\documentclass[12pt,preprint,psfig]{aastex}
\pdfoutput=1

\slugcomment{submitted 2/15/2017, accepted 3/31/2017}

\shorttitle{catalogs of GALEX UV sources}
\shortauthors{Bianchi et al.}

\newcommand{\grado}{\mbox{$^{\circ}$}} 
\newcommand{\degree}{\mbox{$^{\circ}$}} 

\newcommand{\Teff}{\mbox{$T_{\rm eff}$}~}
\newcommand{\teff}{\mbox{$T_{\rm eff}$}}

\newcommand{\as}{\mbox{$^{\prime\prime}$~}}
\newcommand{\am}{\mbox{$^{\prime}$~}}

\newcommand{\beqa}{\begin{eqnarray}} 
\newcommand{\eeqa}{\end{eqnarray}}

\newcommand{\ebv}{\mbox{$E_{B\!-\!V}$}}

\begin{document}


\title{Revised Catalog of GALEX Ultraviolet Sources. I. The All$-$sky Survey: GUVcat\_AIS}


\author{Luciana Bianchi\altaffilmark{1}, Bernie Shiao \altaffilmark{2} and David Thilker\altaffilmark{1}}
\altaffiltext{1}{Dept. of Physics \& Astronomy, The Johns Hopkins University, 3400 N. Charles St.,  Baltimore, MD 21218, USA; 
http://dolomiti.pha.jhu.edu}
\altaffiltext{2}{Space Telescope Science Institute, 3400 San Martin Dr., Baltimore, MD 21210} 
\email{bianchi@jhu.edu}

\begin{abstract}
The Galaxy Evolution Explorer (GALEX) imaged 
the sky in two Ultraviolet (UV) bands, 
 far-UV (FUV, $\lambda$$_{eff}$ $\sim$ 1528\AA) and near-UV (NUV, $\lambda$$_{eff}$ $\sim$ 2310\AA),
 delivering the first comprehensive sky surveys at these wavelengths.  The GALEX database contains FUV and NUV images,  
 $\sim$ 500~million source measurements
and over 100,000 low-resolution UV spectra.   
The UV surveys are a unique resource for statistical
 studies of hot stellar objects, z$\lesssim$2 QSOs, star-forming galaxies, nebulae and the interstellar medium, and 
 provide a  road-map for planning future UV instrumentation
and follow-up observing  programs. 
We present    science-enhanced, ``clean''  
catalogs of GALEX UV sources, with useful tags 
to facilitate scientific investigations. 
The catalogs are an improved and expanded version of our previous  catalogs of  UV sources (Bianchi et al. 2011, 2014: BCScat). 
With respect to BCScat, we have patched 640 fields 
for which  the pipeline had improperly coadded non-overlapping observations, we provide a version with a larger sky coverage (about 10\%) by relaxing the restriction to the central area of the GALEX field to 1.1\degree diameter (GUVcat\_AIS\_fov055), as well as the cleaner, more restrictive version using only the 1\degree~ central portion of each field as in BCScat (GUVcat\_AIS\_fov050). We added new tags to facilitate selection and cleaning of statistical samples for science applications: we flag sources within the footprint of extended objects (nearby galaxies, stellar clusters) so that these regions can be excluded for estimating source density.  
As in our previous catalogs, in $GUVcat$ duplicate measurements of the same source are removed, so that  each astrophysical object has only one entry. Such unique-source catalog 
is needed to study  density and distributions of sources, and to match UV sources with catalogs at other wavelengths. 
The 
catalog includes all observations from the All-Sky Imaging Survey (AIS), the survey with the largest area coverage, with both FUV and NUV detectors exposed: over 28,700 fields, made up of a total of 57,000 observations (``$visit$s''). 
The total area covered, when overlaps are removed and gaps accounted for, is 24,790 (GUVcat\_AIS\_fov055) and 22,125 (GUVcat\_AIS\_fov050) square degrees.  The total number of 
``unique'' AIS sources (eliminating duplicate measurements) is 82,992,086  ($GUVcat\_AIS\_fov055$) and 
69,772,677 ($GUVcat\_AIS\_fov050$). 
 The typical depth of the GUVcat\_AIS catalog is FUV=19.9, NUV=20.8~ABmag. 
\end{abstract}




\section{Introduction.} 
\label{s_intro}

Current observational astrophysics benefits from data mining of modern  sky surveys, enabled by large-format detectors, improved  instrument stability, and computational data-base facilities capable of easily handling large data volumes.  
At optical wavelengths, the relevance and variety of science outcomes from the first modern surveys of the sky, such as the Sloan Digital Sky Survey (SDSS), 
 prompted new, more powerful surveys to be planned and built (e.g., Pan-STARRS, ESO-VISTA,  SkyMAPPER, LSST...). 
At infrared, X-ray and $\gamma$-ray wavelengths, series of surveys with increasing quality, depth, and resolution have progressively advanced our view of 
several classes of sources most prominent in each range.  At UV wavelengths, instead, only the Galaxy Evolution Explorer (GALEX)  performed comprehensive sky surveys, with different coverage and depth 
 \citep{morrissey07, bia09, bia11a, bia14, bia14uvsky}.  GALEX surveys therefore remain the most extensive resource in the UV for planning follow-up observations and missions, and for extracting science from the still largely unexplored database (for earlier, pioneering UV missions see e.g., \citet{bia16a}). 

 This work presents the latest version of the GALEX catalog of UV sources, {\bf GUVcat},  that will facilitate  statistical investigations involving UV measurements, and cross-matching with other samples. It follows, expands and improves the earlier versions by \citet{bia11a, bia11b} and  \citet{bia14uvsky}(BCScat). We present here the catalog from the survey with the largest sky coverage; similar source catalogs from the deeper surveys, more limited in area coverage, will follow, as well as   catalogs of UV variables, and a UV spectroscopic database.   

 The paper is arranged as follows: first we  recall the characteristics of the GALEX instrument  (Section \ref{s_galex}), of the major  surveys performed  (Section \ref{s_surveys}), and of the GALEX data and photometry (Section \ref{s_data})  of relevance for catalog users. In Section \ref{s_catalog} we describe the criteria used for construction of the new catalog and improvements with respect to %
 previous versions,   in Section \ref{s_content} 
we give a statistical overview the catalogs' source content, and provide relevant information for using the catalog, in Section \ref{s_area} we explain the calculation of area coverage, and in Section \ref{s_conclusions} we discuss the distribution of sources across the sky as well as summarize useful caveats and suggestions for using this catalog and GALEX data. A detailed description of the procedure used to identify and remove duplicate measurements of sources is given in Appendix \ref{s_remdup}. A complete list of the tags of catalog sources is given in Table \ref{t_tags} of Appendix \ref{s_tags}.  
Appendix \ref{s_oddfields} illustrates in more detail some caveats and the most relevant artifacts.   

\begin{figure*}
\label{f_figure1}
\vskip -.72cm
\centerline{
\includegraphics[width=8.5cm]{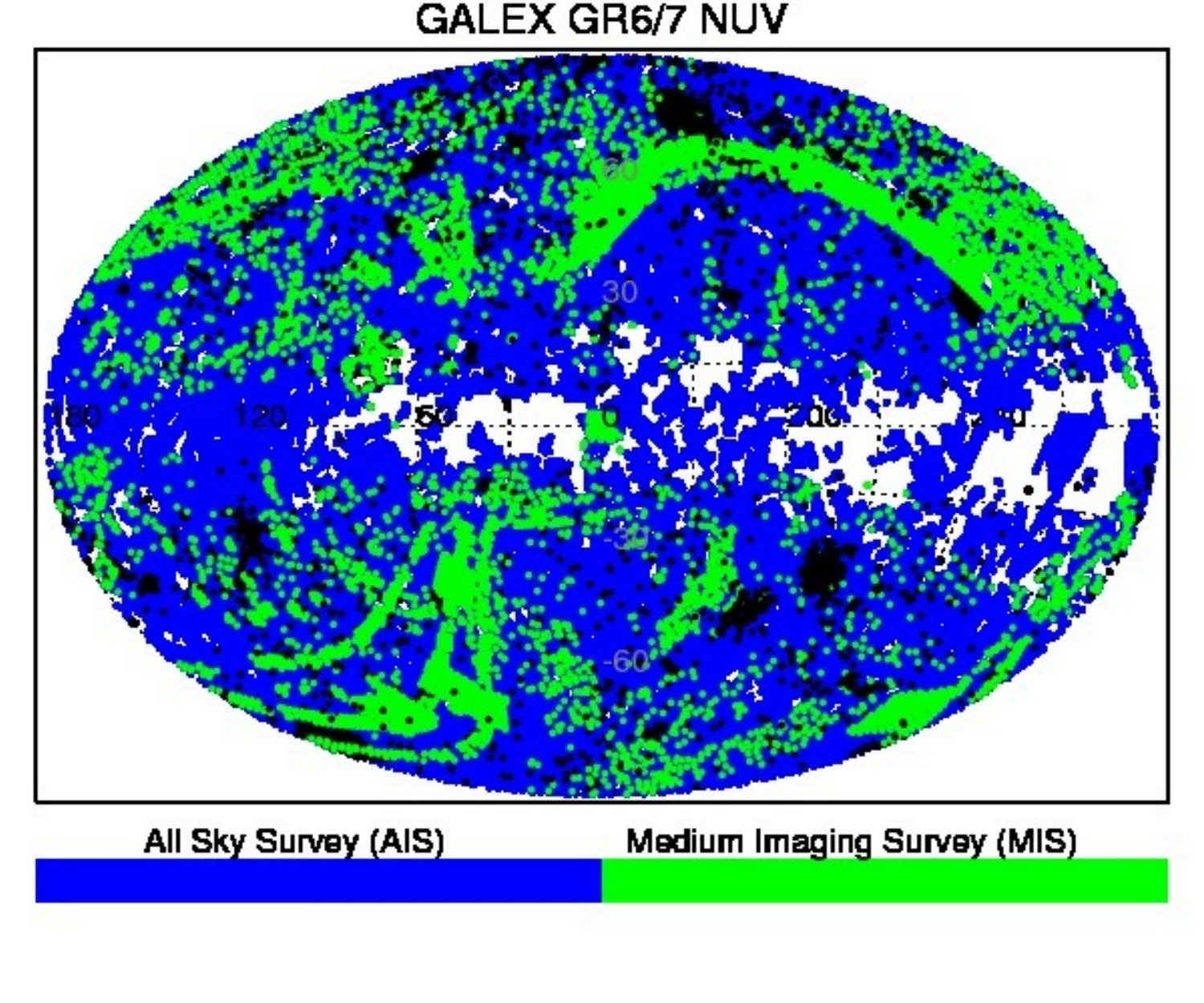}
\includegraphics[width=8.5cm]{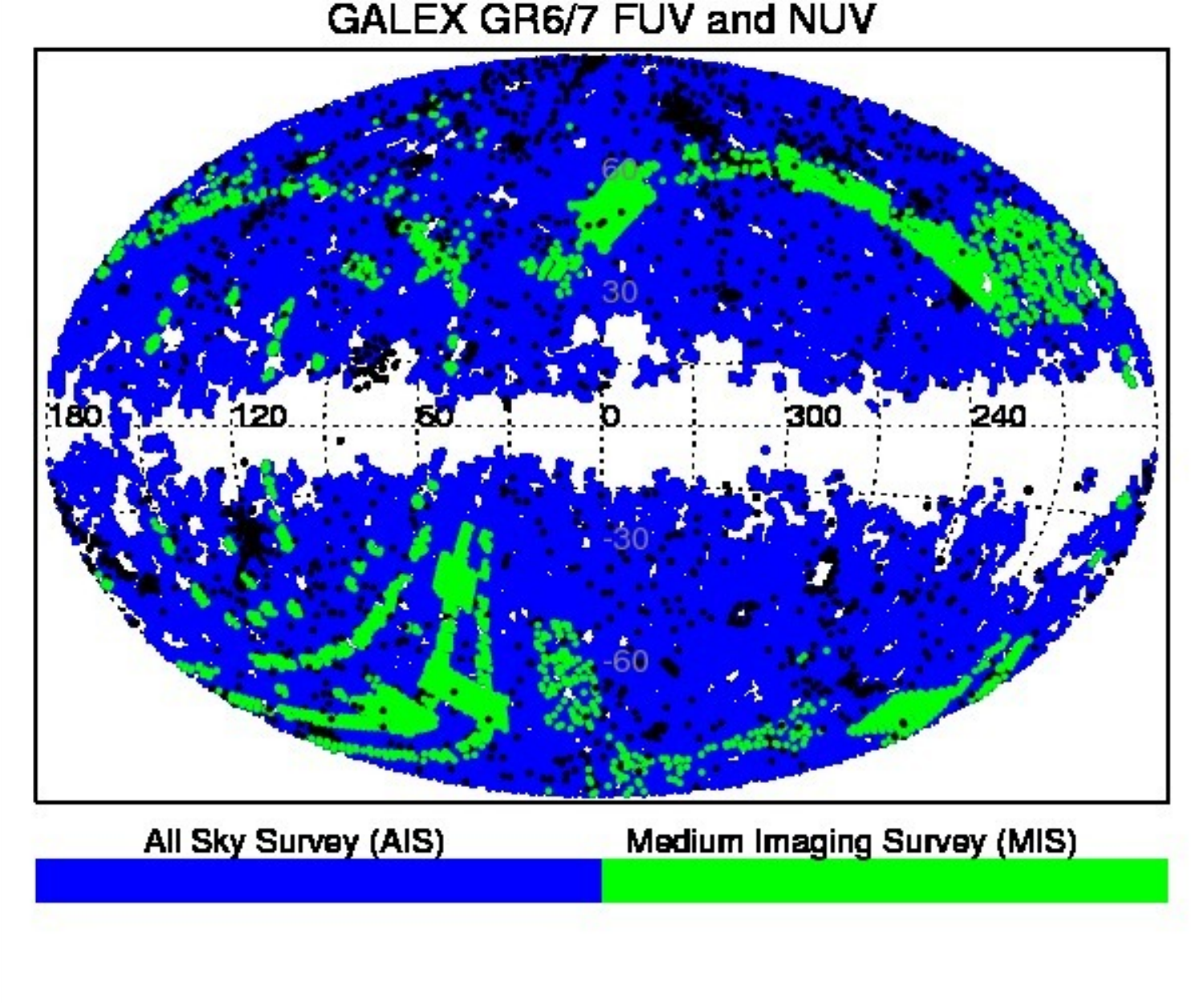}}
\vskip -.62cm 
\caption{Sky coverage, in Galactic coordinates, of the GALEX imaging (GR6plus7 data release). 
The surveys with the largest  area coverage  are 
 AIS (shown in blue) and  MIS (shown in green). Observations from other surveys are shown  in black (figure
adapted from Bianchi et al. 2014a).
Data from the trailed 
CAUSE observations at the end of the mission  are not shown. 
Left: fields observed with 
the NUV detector on, regardless of FUV-detector status; right:
fields observed with both FUV and NUV detectors on. The latter constitute the present catalog.}
\end{figure*}

\section{GALEX instrument and data characteristics}
\label{s_galex}

GALEX \citep{martin05},  a NASA {\it Small Explorer} class mission with contributions from the Centre National d'Etudes Spatiales of France and the Korean Ministry of Science and Technology, performed 
  the first sky-wide Ultraviolet surveys. It was launched on April 28, 2003 and  
 decommissioned by NASA  on June 28, 2013.
GALEX's instrument  consisted of a  Ritchey-Chr\'etien$-$type telescope with a 50~cm primary mirror and  focal length of
299.8cm.  Through a dichroic beam splitter, light was fed  to two detectors simultaneously, yielding observations in  
two broad bands:
 far-UV (FUV, $\lambda$$_{eff}$ $\sim$ 1528\AA, 1344-1786\AA) and  
 near-UV (NUV, $\lambda$$_{eff}$ $\sim$ 2310\AA,  1771-2831\AA). 
GALEX had two observing modes,  direct imaging and 
grism field spectroscopy. 
The FUV detector stopped working in May 2009; subsequent  GALEX observations have only
 NUV data (Figure \ref{f_figure1}).

The GALEX field of view 
is $\approx$1.2$^{\circ}$  diameter (1.28/1.24$^{\circ}$, FUV/NUV),  the spatial resolution is 
 $\approx$ 4.2/5.3\as \citep{morrissey07}.  
 For each observation,  an  FUV and an NUV image, sampled with  virtual pixels of 1.5\as, are reconstructed from the photon list recorded by the two photon-counting micro-channel plate detectors.
From the reconstructed image, the GALEX pipeline then derives a 
sky background image, by interpolating a surface from flux measurements in areas with no detected sources, and performs source photometry in various ways: aperture, psf, Kron-like elliptical (see Appendix \ref{s_tags}). 
Sources detected in the FUV and NUV images of the same observation
 are matched by the pipeline 
 to produce a merged-source list (both bands combined) for 
each observation. We will return to this matching later.

 To  reduce local 
response variations,  in order to  maximize photometric accuracy, 
each observation was carried out in ``AIS mode'' for most AIS data and with a 1$^{\prime}$ spiral dithering pattern for MIS and DIS.\footnote{A trailing mode, rather than the spiral dithering pattern,  was instead used for the latest,  privately-funded observations,
to cover some bright areas near the MW plane.  These latest data currently are not in the public archive. Also, a so-called Petal-mode was used in special cases. The spiral dithering pattern was used for most of the science data, and for all of the data used in this catalog.}
The surveys were accumulated by covering contiguous {\it ``tiles''} in the sky, with series of such observations, sometimes repeated, called ``$visit$s''.

The Galactic plane was largely inaccessible during the prime mission phase because of the 
many bright stars that violated high-countrate safety limits. Such constraints were relaxed at the end of the mission. A survey of 
the Magellanic Clouds (MC), also previously unfeasible due to brightness limits, 
was completed at the end of the mission,  when the initial count-rate saftey threshold (\citet{bia14,simons14,thilker17mc}) was lowered.  Because of the 
FUV detector's failure in 2009, these extensions include only
NUV measurements (Figure \ref{f_figure1}).

\section{The sky surveys}
\label{s_surveys}

GALEX has performed sky surveys with different depth and coverage (\citet{morrissey07},\citet{bia09}). 
The two detectors, FUV and NUV,  observed simultaneously 
as long as the FUV detector
was operational; however,  there are occasional observations in which one of the two detectors
was off (mostly FUV) due to brief shut-down episodes, even in the early part of the mission; in addition, in some
observations the FUV and NUV exposure times differ (see Bianchi et al. 2014a, in 
particular their Table 1 and Fig. 2). 

 The surveys with the largest area coverage are the All-Sky Imaging survey (AIS) 
and the Medium-depth Imaging Survey (MIS): the sky coverage is shown in Figure \ref{f_figure1}. Exposure times slightly vary within each survey,
around the respective 
 nominal exposures of 100~sec for AIS, which corresponds to a detection limit (5~$\sigma$) of FUV$\sim$20/NUV$\sim$21~ABmag,
and 1500~sec for MIS, corresponding to a depth  of $\sim$22.7~ABmag in both FUV and NUV. 
The Deep Imaging Survey (DIS) accumulated exposures of the order of several tens of thousand of seconds in selected fields
(for example, for a 30,000~sec exposure, the depth reached is $\sim$24.8/24.4~ABmag in FUV/NUV).  
In addition, the  ``Nearby Galaxies Survey'' (Bianchi et al. 2003, Gil de Paz et al. 2007), dedicated to mapping large nearby galaxies,  covered initially 436 fields at 
MIS depth, but hundreds of additional nearby galaxies 
 were mapped by GALEX, as part of MIS or other surveys (see also Section \ref{s_inextobj}).
Other observations were obtained during guest investigator (GI) programs, and for other targeted regions  such as, 
for example, the Kepler field (e.g., \citet{myronkepler}).  

 The current GALEX database  (data release GR6plus7)  contains 582,968,330 
source measurements resulting from a total of 100,865 imaging $visit$s;  
most of these source measurements 
are from observations with both  FUV and NUV 
detectors on (64551 $visit$s, 47239 of which from the AIS survey). 
Figure \ref{f_figure1} shows the sky coverage of all GALEX observations 
performed with both FUV and NUV detectors on (right panel), and in NUV regardless of FUV-detector status (left panel). 
The figure does not include the last NUV trailed observations (the privately-funded ``CAUSE'' observing phase, conducted in scan mode). 

\begin{figure*}
\centerline{%
\includegraphics[width=13.5cm]{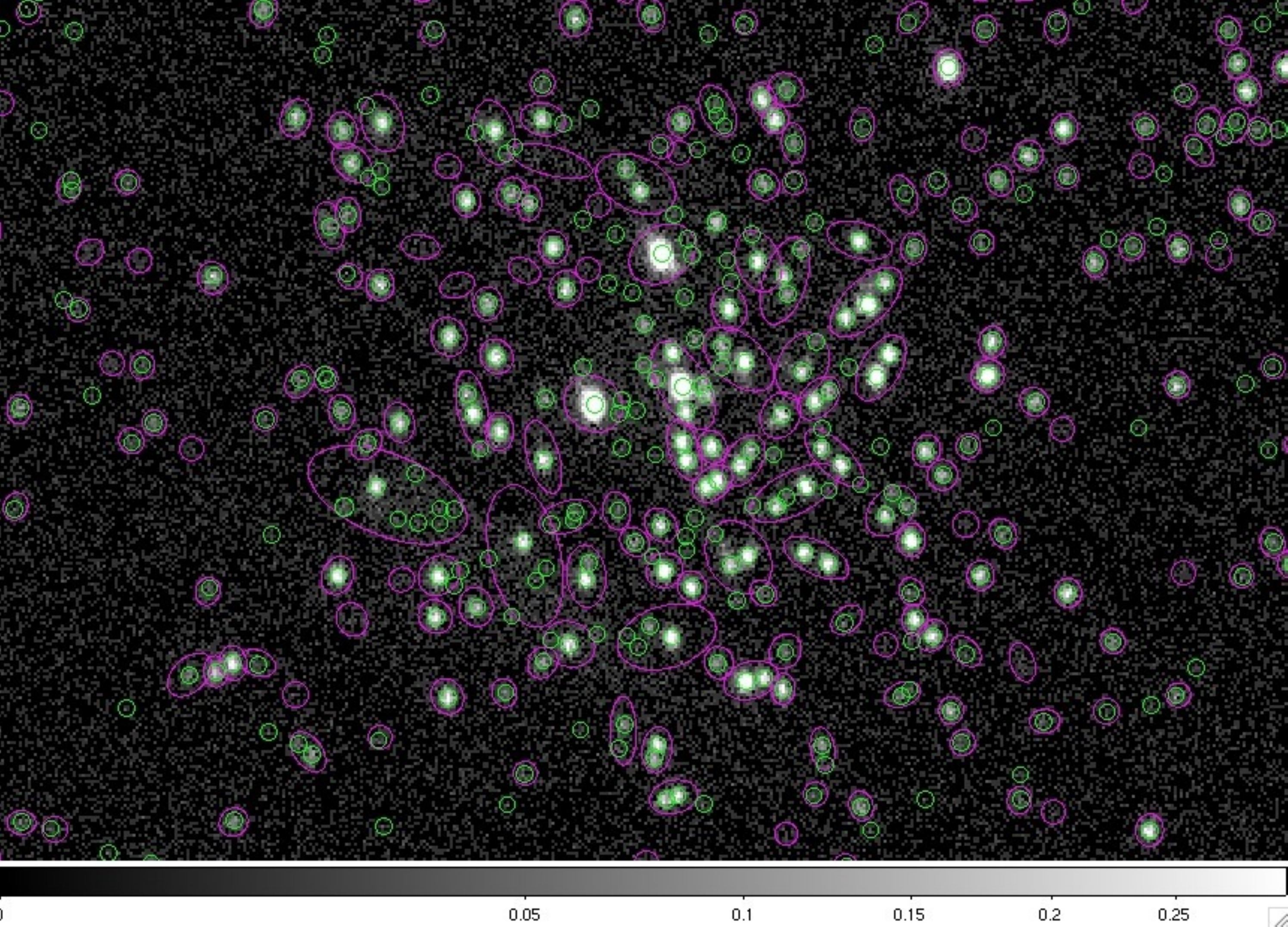}}%
\vskip -.05cm 
\caption{A portion of the Galactic open cluster NGC~2420 imaged by GALEX;  detected sources as defined by the standard pipeline are outlined in pink,  and from our custom-made photometry  \citep{demartino2420} in green.  The example illustrates the case of some crowded pointlike sources being merged by the pipeline into one extended source. In the outskirts of the cluster, less crowded, the pipeline source identification matches ours very well. 
\label{f_crowd} }
\end{figure*}

\section{GALEX data and  photometry}
\label{s_data}

\begin{figure}
\vskip -1.75cm
\centerline{
\includegraphics[width=9.5cm]{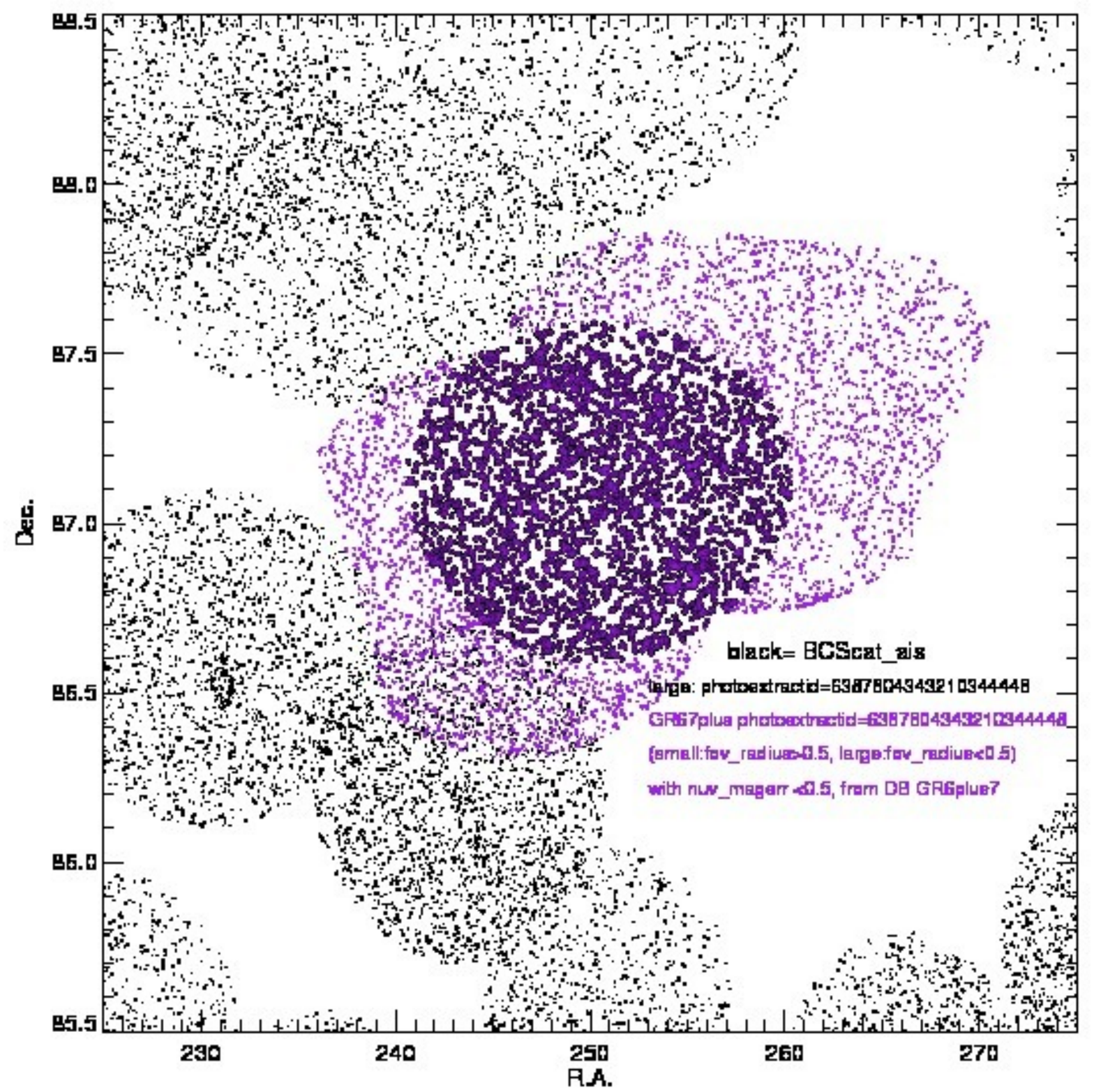}
\includegraphics[width=9.5cm]{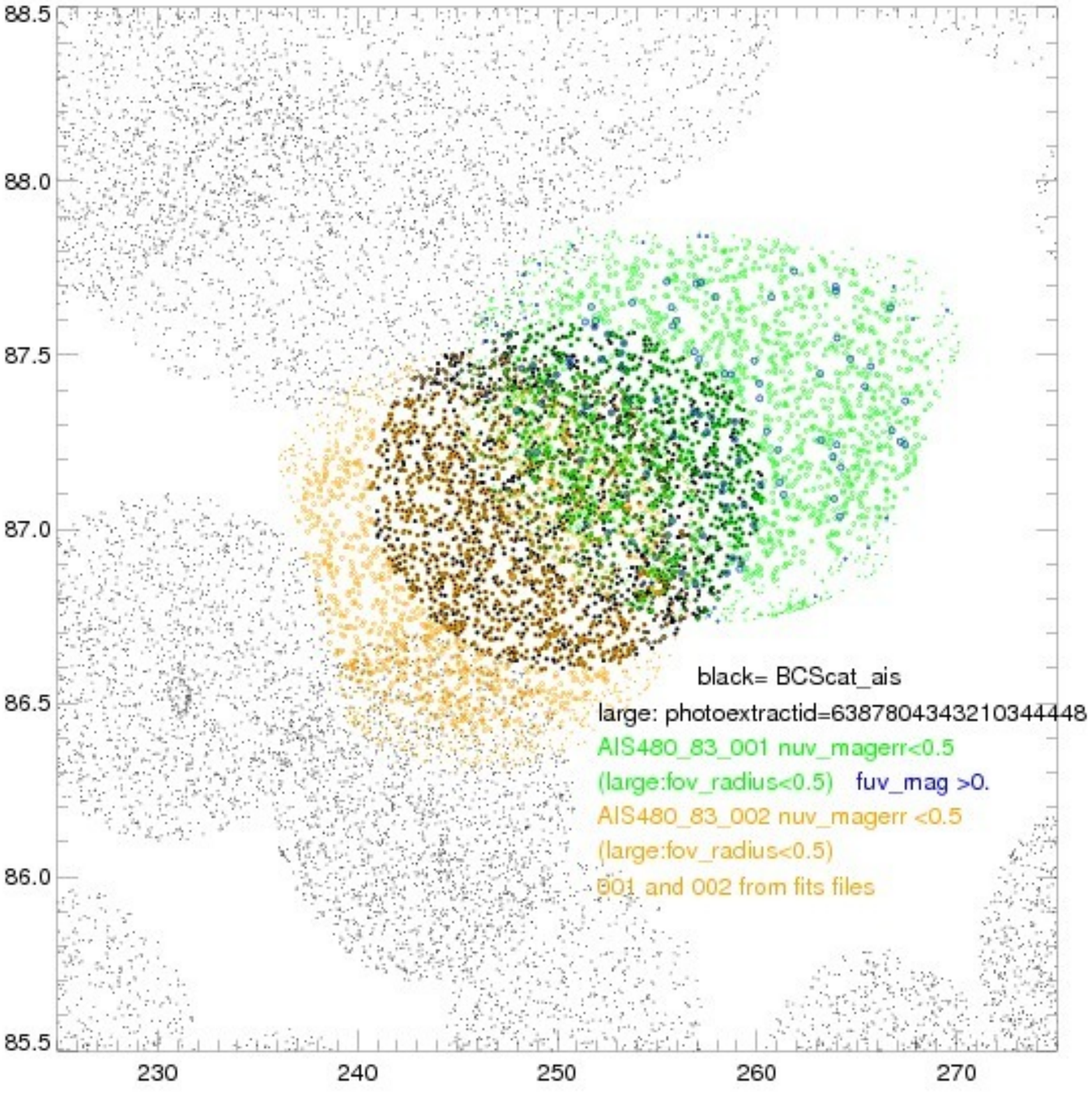}}
\vskip -.65cm 
\caption{ {\bf To be printed in landscape}   {\it Left:}  database sources included in $tile$ AIS\_480 are shown as purple dots. They result from merging two $visit$s (right); the large black dots are sources within 0.5\grado from the center of the $coadd$, thus in principle meeting the criterion for our catalog (and BCScat).  Smaller black dots are sources in nearby fields. 
{\it Right:} Over the pipeline-merged sources with $fov\_radius$ $<$0.5\degree (black dots), the $visit$-level sources are shown in dark yellow ($visit$ with NUV exposure only) and green ($visit$ with both NUV and FUV exposures). On the latter, blue circles mark sources with significant detection also in FUV. We use large/small symbols to indicate sources inside/outside of a 0.5\degree radius from the actual center of each $visit$. In the database $coadd$  all purple sources 
are given  exposure times equal to the sum of the two $visit$s, but this is correct only for those in the small intersection of yellow and green dots. Also, sources from the rim of both $visits$ appear to have a small distance from the $coadd$ center, because the pipeline assigned a value of $fov\_radius$ based on the $coadd$ center, rather than on the center of the parent observation,  and therefore these rim sources will not be discarded by a selection in $fov\_radius$. 
\label{f_badcoadd}  }
\end{figure}

GALEX data include
images through either direct imaging or grism,  and associated photometry from the pipeline or extracted spectra respectively. 
  High-level science products (HLSP) have also been released  \citep{bia11a}, and unique source catalogs (i.e., with no duplicate observations of the same source, \citet{bia11a,bia14uvsky}: BCScat), these are 
 available at MAST and Vizier, and are precursors of the present catalog. 

The photometry calibration for any data release uses the zero points of \citet{morrissey07}, any subsequent pipeline updates were reflected in revised extracted source count-rates (CTR), so that the zero points remained unchanged. On the AB magnitude scale, the GALEX magnitudes are defined as:   
~\\
~\\

UV\_mag= -2.5$\times$ log(CTR) + ZP ~~~~ (AB mag)  \hskip 3.cm (eq. 1) 

~\\
where CTR is the dead-time corrected, flat-fielded count rate (counts s$^{-1}$) and the zero-point values are ZP$_{FUV}$=18.82 and ZP$_{NUV}$=20.08. 

The transformations to Vega magnitudes are \citep{bia11} :\\
~\\
FUV\_mag$_{Vega}$=FUV\_mag$_{AB}$ -2.223   \hskip 7.5cm (eq. 2)\\ 
NUV\_mag$_{Vega}$=NUV\_mag$_{AB}$ -1.699   \hskip 7.5cm (eq. 3) 

In Sections \ref{s_bright} to \ref{s_inextobj} we discuss additional details and relevant caveats for using GALEX data. Practical advice on use of GALEX data and this catalog is summarized in Section \ref{s_advice}.

\subsection{Bright sources} 
\label{s_bright}

 High count-rates from UV-bright  sources  cause non-linearity in the response, or saturation,  
due to the detector's  dead-time correction being overtaken by the photon arrival rate. \cite{morrissey07}  reported  non-linearity  
 at a 10\% rolloff to set in  at 109~counts~s$^{-1}$ for FUV and 311~counts~s$^{-1}$ for NUV.
These countrates correspond to FUV\_mag=13.73~ABmag ($\sim$1.53~10$^{-13}$ erg~s$^{-1}$~cm$^{-2}$~\AA$^{-1}$)
and NUV\_mag=13.85~ABmag ($\sim$6.41~10$^{-14}$ erg~s$^{-1}$~cm$^{-2}$~\AA$^{-1}$). A correction for non-linearity 
is applicable over a limited range, beyond which the measured 
countrate saturates and the true source flux is no longer recoverable (see their Figure 8). 
The bright-object limit during the primary mission was 30,000~counts~s$^{-1}$ per source,
corresponding to $\sim$9$^{th}$ABmag for NUV ($\sim$7~10$^{-12}$ erg~s$^{-1}$~cm$^{-2}$~\AA$^{-1}$)
and 5,000~counts~s$^{-1}$ per source in FUV (ABmag $\sim$ 9.6, 
$\sim$6~10$^{-12}$ erg~s$^{-1}$~cm$^{-2}$~\AA$^{-1}$). 
Such limits were relaxed at the end of the mission. 

 In addition to the non-linearity for sources with high CTR, the total CTR over the entire field affects the stim-pulse correction, which in turn affects the correction for non linearity. We refer to \cite{thilker17mc} for details of the issue, and recipe for correction. 

The calibration of GALEX fluxes is tied to the UV standards used for HST \citep{rcb2001wd}.
However, all but one of the white dwarf (WD) standard stars have GALEX count-rates in the non-linear regime. 
 \cite{camarotaholberg14} derived an empirical correction to the GALEX magnitudes in the non-linear range, 
using a well studied sample of WDs with previous UV spectra and  model atmospheres.  Their correction is valid 
in the bright-flux regime as specified in their work, but would diverge if extrapolated to fainter fluxes. 
Further refinements of the calibration  
have not yet been explored to our knowledge. 
In future works we will examine the stability of the response at very high countrates \citep{bia17galexspec, alexbia}.

\subsection{Crowded fields} 
\label{s_crowd}

Source detection and photometry measurements performed by 
the GALEX pipeline become unreliable where sources are too crowded relative 
 to the instrument's resolution. 
Conspicuous examples include stellar clusters in the Milky Way (Figure \ref{f_crowd}), 
 fields in or near the Magellanic Clouds \citep{simons14,bia14}, 
and nearby extended galaxies (Section \ref{s_inextobj}).  The pipeline, designed for the general purpose of detecting both point-like
and  extended sources (such as galaxies, typically with 
an elliptical shape), sometimes interprets two or more nearby point sources 
 as one extended source; this seems to occur in crowded regions, as Figure \ref{f_crowd} 
shows. 
 Note that, in some crowded fields, at times the pipeline fails to resolve even point-like sources with separation 
comparable to,  or larger than the 
instrumental resolution; see Figure \ref{f_crowd} 
for an example, or Figure 3 of \citet{simons14} 
for a  Magellanic Cloud field. In extended galaxies, the local background of diffuse stellar populations may compound the crowding
 around clustered sources or bright star-forming complexes. 

In extended galaxies, because  UV fluxes are sensitive to the youngest, hottest stars, 
which are typically arranged in  compact groups within star-forming regions, 
  UV-emission peaks 
are identified by the pipeline as individual sources and some star-forming structures may be shredded in individual peaks, or tightly clustered sources may be merged into an extended source. In other cases,  the  extended emission of the central galaxy disk is often interpreted as a single extended source. In many cases, the result from the pipeline is a  single measurement of a large central area and an overdensity of sources in the outer disk. An example is shown in Fig.\ref{f_extgal}.
Custom measurements are needed in extended galaxies, with special care to background subtraction (e.g., 
\citet{kbr09,efremova11,bia14m31,dt07ngc7331,thilker17mc}). 
Useful tags to identify such cases are described in Section \ref{s_inextobj}. 

 For consistency, and completeness, all AIS  measurements from the master database with both FUV and NUV exposures $>$0~seconds
were used to produce our GALEX source catalogs GUVcat\_AIS. Large galaxies, stellar clusters, and MC fields\footnote{We recall that only the periphery of the MC has both FUV and NUV exposures; the coverage of the inner portions has mostly NUV data \citep{bia14, simons14}, and therefore was not included in our catalog; the whole MC catalog, from custom-vetted photometry,  will be published elsewhere \citep{thilker17mc}}   
 were not excluded, to avoid introducing arbitrary gaps in the catalog coverage, because choice of which  regions must be excluded depends
 on the specific science application and the characteristics of the sources to be analyzed (e.g., magnitude range, \citet{bia11b}). As with every large database, it is ultimately the user's choice (and
responsibility) to check crowded or problematic regions or extended objects, and exclude such regions if needed, or carefully check 
the photometry  if these areas cannot be excluded (see Section \ref{s_inextobj}), and use 
specific custom-vetted photometry catalogs for these particular areas when necessary.  For the Magellanic Clouds (MC), initial custom photometry was performed by \citet{simons14}; the final and complete version of the MC catalog is published  by \citet{thilker17mc} and should be used in these regions, instead of GUVcat or the database products.

\section{The UV source catalog.}
\label{s_catalog}

For several sources there are multiple measurements in the GALEX master database, due to repeated  observations of the same
field, or overlap between contiguous fields. For  studies involving UV-source counts, or to match UV samples with catalogs at other wavelengths, 
one needs to eliminate repeats, as well as artifacts.  Therefore, we have  constructed catalogs of 
{\it unique} UV sources, eliminating duplicate measurements of the same object.
Separate catalogs were constructed for AIS and MIS, because of the $\sim$2-3 mag 
difference in depth. 
The catalog presented here is an expanded and improved  version of ``BCScat'' published by Bianchi et al. (2014a), who 
also presented the first 
sky maps showing density of UV sources with various cuts.  
An earlier version, based on the fifth data release 
(GR5), was published by \citet{bia11a}, who extensively discussed the  criteria for constructing  GALEX source catalogs and matched catalogs between GALEX and other surveys. \citet{bia11b}  presented distributions of density of sources as a function of Galactic latitude, magnitude, and colors. 
We refer to these papers for useful presentations of the UV source distributions across the sky, and in magnitudes and colors; such considerations will not be repeated here because the overall statistics will appear very similar, 
but we strongly advise to use the catalog presented here for better quality and completeness.  The improvements with respect to the earlier versions are described in the next section. 
\citet{bia11a} also released matched GALEXxSDSS catalogs, and \citet{bia11b} presented matched GALEXxGSC2 catalogs. Work on source classification from the matched catalogs was presented by \citet{bia09} and \citet{bia11a, bia09qso, bia07apj, bia05apjl}. 
The earlier versions of the unique-source catalogs \citep{bia11a, bia14uvsky} are superseded by  GUVcat presented here. Matched catalogs of GUVcat with SDSS, PanSTARRS, 2MASS, WISE, and Gaia will be released by \citet{bia17guvcatX}.

In the GALEX database, an FUV  magnitude with value of -999 
 means  either that the FUV detector was on and the source was detected in NUV but too faint in FUV to be measured, 
or that the FUV detector was off.  In order to examine and classify sources by color, and the relative 
fraction of sources with different colors, \citet{bia14uvsky, bia11a} 
restricted the catalogs to  those observations in which both detectors were exposed. We do the same here. 
In addition, our previous  catalogs were conservatively restricted to measurements within the central 1\grado~ diameter of the field of view,  to exclude 
the  outer rim, where distortions 
prevent position and photometry of sources to be derived accurately, and counts from rim 
 spikes cause
numerous artifacts to intrude the source list. In the present version we again offer a  catalog restricted to sources within 0.5\grado from the field center, GUVcat\_AIS\_050,  and also a version relaxing this limit to  0.55\grado, GUVcat\_AIS\_055, to  reduce gaps in area coverage, 
as described in Section \ref{s_criteria} and \ref{s_area}.  Sections \ref{s_catalog} and \ref{s_advice} clarify which catalog is preferable depending on the science purpose. 
 The present catalog includes all AIS fields with both FUV and NUV exposed.\footnote{BCScat\_AIS   includes   28,707 AIS fields with both FUV and NUV exposed,
covering a unique area of 22,080 square degrees as they were restricted to sources within the central 1\grado~ of each field,
and BCScat\_MIS  includes 3,008 MIS fields covering a total 
2,251  square degrees. 
The previous catalogs of \citet{bia14uvsky} (``BCScat'' in MAST casjobs: https://archive.stsci.edu/prepds/bcscat/  and Vizier) 
contain $\approx$71~million AIS and $\approx$16.6~million MIS sources.}

\subsection{Patching and updating  BCScat}
\label{s_badcoadds}

 The initial need for patching  $BCScat$ came from the discovery that in some fields the GALEX pipeline had coadded observations from different $visit$s 
 which are largely {\it not} overlapping.  GALEX observed each field (termed ``{\it tile}'' in the database) in  one or more ``$visit$'' (composed of one or more ``subvisit'');  the partial-exposure images ($visit$s) that passed the either automated or manual quality test (``QA'') were coadded, and ``$coadd$'' products (images, photometry) from the pipeline were  entered in the database; the exposure time listed for the $coadd$ is the sum of the partial exposures that were combined. A data-set is listed as ``$coadd$'' in the database even if it consists only of one $visit$. The $coadd$ products are the default data level accessed by browsing the GALEX database with $GALEXview$ (galex.stsci.edu/galexview). 
 For constructing the catalog, and for most other purposes, using the $coadd$s as a starting point is the best option since they provide the total exposure available for each field, with all $visit$s already coadded. Our previous catalogs were therefore constructed combining sources from the $coadd$s, and so is the catalog being released with this paper, with the exceptions described below.

 The UV source catalog BCScat\_AIS \citep{bia14uvsky} was constructed from the 28,707 AIS $coadd$s with both FUV and NUV {\it total} exposures $>$0. These $coadd$s are made up of 57,000 $visit$s (47,239 of which have both detectors exposed).
 We  discovered however that in some GALEX AIS fields  the pipeline had  coadded %
$visit$s centered at significantly differing positions, up to 26.8$^{\prime}$ apart (which means, in this extreme case, almost no overlap).  The pipeline then places the nominal center of the resulting $coadd$ in between the centers of the merged $visit$s, compounding the problem and making some critical tags useless (mis-leading). $Coadd$s made of non-overlapping $visit$s  cause three potential problems, affecting any analysis,  and all previous catalogs. To illustrate these  problems we show  an example, $tile$ AIS\_480, in Fig. \ref{f_badcoadd}. In this case the database has merged two $visit$s: one with both detectors exposed (shown as green dots in  Fig. \ref{f_badcoadd}) %
and one with only the NUV-detector exposed (yellow dots). %
 The database  sources  associated with this $tile$, i.e. the $coadd$, are shown in purple. %
The total exposure time given in the database is the sum of the exposures of the two $visit$s, hence  all the AIS\_480 sources (the purple dots) appear to have  FUV exposure equal to that of the first $visit$, and  NUV exposure equal to the sum of the two $visit$s. 
But this is only true in the area of overlap of the two $visit$s, which is very small in this case. In the yellow-dot-only area (portion of $visit$ 2 not overlapping with $visit$~1), sources  have FUV\_mag=-999 (i.e., non-detection), but they appear to have an FUV exposure $>$0 (as the same exposure is given for the entire $coadd$), therefore they would be erroneously interpreted as having FUV flux below the detection threshold, while in fact they have no FUV data.  In the green-dot-only area, sources appear to have an NUV exposure equal to the sum of the two $visit$s, while they only have the exposure time of $visit$~1.  Such improper $coadd$s then introduce two biases
  when one selects - as we do 
in our previous, and current, catalogs - only fields with both detectors exposed, and  we include for each field only sources within 
a certain radius from the field center  
to avoid rim artifacts and poor source photometry in the outer edge of the field of view (f.o.v.).    
First,  the green-only sources, that would meet our catalog selection criteria (both detectors exposed), are not included in the catalog, because the center of the tile is the center of the $coadd$ (in Figure  \ref{f_badcoadd}-top black dots are the sources within 0.5\grado from the $coadd$'s field center). Second, some yellow-only sources (within the 0.5\grado circle from the coadd center) intrude the sample in spite they actually have no FUV exposure. The consequences will be for example that the ratio of FUV detections over NUV detections will be incorrect, and so any interpretation of UV color. 
In sum, these  ``bad $coadd$s'' cause: {\it (i)} loss of sources that should have been included, {\it (ii)} intrusion of sources not meeting the criteria, and {\it (iii)} misleading exposure times for the included sources. In addition, and worst 
 of all, {\it (iv)} our criterion of limiting the catalog to sources within 0.5\grado from the field center, intended to exclude the numerous rim artifacts and distorted sources along the edge of the fields, is nullified by the $fov\_radius$ value being assigned by the pipeline with respect to the centering of the $coadd$: Fig.  \ref{f_badcoadd} shows that the merged sources within 0.5\grado ~from the $coadd$ center include part of the rim of both $visit$s.  In fact, by imposing a limit of $fov\_radius$$\leq$0.5\grado, we would expect no sources with rim artifact flag in the catalog; instead, 
there are 116,530 sources with $fuv\_artifact$=32 and 74,579 with  $nuv\_artifact$=32 in BCScat\_AIS.   These were introduced by the $coadds$ in which non-overlapping $visit$s had been merged by the pipeline (hereafter $bad~coadds$).   This problem had never been reported previously to our knowledge.   When we discovered it, we undertook an effort to identify all  the $bad~coadds$ in the database, and patch the catalogs. The result is the GUVcat\_AIS presented here. 

 The first step for constructing a revised catalog was therefore to identify the $bad~coadds$, and  
to use the data from the corresponding  individual $visits$ instead of the $coadd$ in such cases. 
 To identify the $bad~coadds$,  
we compared the center of each of the  28,707 AIS $tiles$ with the centers of their associated $visit$s (i.e., the $visit$s used by the pipeline to build each $coadd$). 
For all cases where the center of one or more of the associated  $visit$s differs by more than 5\am from the center of the coadded $tile$, we discarded the $coadd$  and ingested in the catalog the corresponding $visit$s (those that satisfy the criteria of both detectors being exposed). In this way we ensure that an FUV non-detection in the catalog is an actual non-detection and not a non-exposure, that the exposure times are correct, and that the centers correspond, within a given tolerance,  to the actual centers of the observation ($visit$) so there is no loss of good sources, and no inclusion of rim artifacts (see also next section).  We chose a tolerance of 5\am between $visit$ centers, as a good compromise to use as many as possible of the $coadds$ (which offer the most exposure available in each field) without introducing the negative effects described above. 

Out of a total 28,707 AIS fields with both FUV and NUV exposure $>$0, made up of 57,000 $visit$s, there are 640 
  $bad~coadds$\footnote{we  term {\it bad~coadd} a  field in the database which was made combining $visit$s of which at least one has its center  $>$5\am away from the $coadd$'s center}, made up of 1195 $visit$s. Of these $visit$s,  886  have both FUV and NUV exposed: these have been used to construct the new catalog, in place of their corresponding 640 bad~$coadd$s. 
The $bad~coadds$ identified in this way  are spread all across the sky, 
therefore it was not possible to simply patch a subset of the previous (BCScat) catalog by removing the $bad~coadds$ and replacing them with data from  the individual $visit$s, because to construct the unique-source catalog duplicate measurements of the same source had been identified and removed. Some of the $bad~coadds$ overlap  
with other (good) fields, and the procedure constructing the catalogs eliminates duplicate measurements 
 from overlapping fields.   

We therefore constructed a new catalog, GUVcat\_AIS, using all the ``good'' $coadd$s (28067, with both FUV and NUV exposed, $visit$ positions within each $coadd$ differing by no more than 5\am), and for the $bad~coadds$, the individual $visit$s of that $tile$ instead.   Table \ref{t_CoaddsVisits} (electronic only) lists the centers of the $tiles$ used to construct the new catalog, and specifies whether $coadd$ (``C'') or $visit$ (``V'') photometry was used.
The 640 $bad~coadds$ are listed in Table \ref{t_badcoadds} (electronic only); we release this list too, because it may be of general interest, in providing to users of the GALEX database a quick quality check of the data they use.  Because of its  potential more general use,  in Table \ref{t_badcoadds} we include all AIS $coadd$s and $visit$s regardless of exposure, although in our catalog we only retain observations with both detectors exposed. These are easy to identify, having both exposure times $>$0, and are indicated as 'G'  in the last column of the table (they were included in BCScat), 'N' indicates those not included.

 In the next section we describe the criteria used to construct the new catalog, which largely follows our previous recipe \citep{bia11a, bia14uvsky}, with several improvements. 
Five of the $coadd$s, which appear to have both FUV and NUV exposure,  were not replaced by their individual $visit$s because each one consists only of two $visit$s, non overlapping, one exposed only in NUV and one exposed only in FUV. These $coadd$s  were included in BCScat, but are excluded in the present catalog, and not replaced by $visit$s; They have the following $photoextractid$: 
6385728408348786688, %
   6385728422307430400, 
   6386256187888762880,
   6386748750844395520, 
   6386748759434330112.  
These entries are marked with 'N' in the last column of Table \ref{t_badcoadds}. 
For one of these fields the difference between the center of the two visits 
is only 5.4\am: this implies that most of their sources (except 
for an outer annulus) 
 may have good measurements in both filters. We had nonetheless to apply a consistent criterion to discard $bad~coadds$, therefore these data are not included in GUVcat\_AIS.  

To summarize: in the GALEX database there are 28,707 AIS fields ($coadd$s) that appear to  have both FUV and NUV exposure $>$ 0;  we examined the distance between the center of each $coadd$ and the center of the $visit$s which were combined to produce it, and found 28,067 $good$ $coadds$ (distance between all $visit$s of the same $coadd$ $<$5\am) resulting from 54,996 $visit$s, and 1,195 $visit$s whose centers differ $>$ 5\am from the center of their $coadd$, affecting 640 $coadds$. These 640 bad $coadds$ are made up of 2,400 $visits$ in total;  we discarded these $coadds$, and  used only $visits$ with both FUV and NUV exposure $>$0 to replace them: 1,468 $visits$.

\subsection{Criteria for constructing the UV source catalog}
\label{s_criteria} 

 The catalog was constructed from the database source photometry with the  criteria given below, following the recipe of 
 \citet{bia11a,bia11b,bia14uvsky}, where other details can be found, and of which the present catalogs
represent the updated and  expanded version.  We used the photometry from 
28,067 AIS good $coadds$, plus 
1,468 $visits$ that replaced 635 
 of the 640 $bad~coadds$ as described in the previous section; the ensemble of these datasets includes the whole AIS coverage with both FUV and NUV detectors exposed. 

 The catalog includes sources:

\begin{itemize}
\item{{\bf from observations with both FUV and NUV detectors on.} This restriction is useful for  science applications in which
 the fraction of sources with a given FUV-NUV color is of interest, or to estimate the fraction of sources with significant detection in  FUV  over the total %
 NUV detections (e.g., \citet{bia14uvsky}, and Sections \ref{s_content} and \ref{s_conclusions}). More observations, 
taken with one of the two detectors turned  off (mostly FUV), exist in the MAST database. 
Including in our catalog observations where one detector was not exposed would bias any statistical analysis,
since the FUV magnitude of a NUV-detected source 
appears in the database as a non-detection (FUV$_{mag}$=-999) either because the FUV detector was turned off, 
or the FUV detector was on but the FUV flux of that source was actually below detection threshold. In some cases the exposure is not the same in both detectors (exposure times are also  given in Table \ref{t_CoaddsVisits}). We used all AIS data in which both detectors' exposures were $>$0.  }
\item{{\bf within the central 0.55\grado (GUVcat\_AIS\_055) or 0.50\grado (GUVcat\_AIS\_050)  radius of the field-of-view} ({\it fov\_radius} $\leq$ 0.55\degree ~ or 0.50\degree, respectively), to avoid sources with poor photometry and astrometry near the edge of the field, and rim artifacts. This restriction yields source samples  with overall homogeneous quality, and minimize artifacts, without great loss of area coverage. Users interested in a particular source that falls on the outermost edge of a GALEX field should obtain the measurements from the GALEX database and carefully examine the quality.  The less conservative $fov\_radius$ $\leq$ 0.55\degree~ limit  %
reduces gaps between fields and increases total area coverage (see Section \ref{s_area}), while still excluding the outermost rim in nearly all data (Section \ref{s_artifacts}).}
\item{{\bf with NUV magnitude errors $\leq$ 0.5mag}; that is,  all sources with  NUV detections are retained, regardless of detection in the FUV filter.
 Typically, about 10\% of the NUV-detected sources are also detected in FUV \citep{bia11b}. Effects of error cuts on the resulting samples can be seen from 
Figure 4 of \citet{bia11a}, and Figures 2$-$4 of \citet{bia11b}. Sources in the database having a FUV detection with no NUV counterpart  will not make it into the catalog:  these cases are very rare, and are either mismatches or artifacts (see later), or cases where the pipeline resolves individual sources in FUV but merges them into one extended source in NUV, such as for example in the center of globular clusters (Fig. 5.c). }
\item{{\bf Unique,} i.e. duplicate measurements of the same source are identified and removed: each object is counted only once in the GUVcat catalog. The procedure for defining duplicates is fully described in Appendix \ref{s_remdup}, as it involves often neglected complexities.  The unique-source catalog is useful for most  science applications, such as  examining density of sources, and for cross-matching with other catalogs.  We provide online also a master catalog (GUVcat\_plus) in which duplicate measurements are identified and flagged but not removed. Details can be found in Appendix \ref{s_remdup}. 
The identified repeated measurements could be used in principle for serendipitous variability searches; 
 we provide tags giving magnitude difference between ``primary'' and ``secondary'' sources, but mainly for the purpose of checking consistency between repeated measurements. Because our catalog made use of $coadd$s as much as possible, variability searches will be more productive on catalogs extracted at $visit$ level, or better yet with sub-$visit$ integrations, which we will present in  follow-up works \citep{bia17var,million17var}.  }
\end{itemize} 

 There are five  AIS fields ({\it photoextractid} = 6379923033125027840, 6381259965176217600, 6379711852804308992, 
6372041728408420352 and 6379571150749433856)  where both FUV and NUV detectors were exposed, but 
 NUV and FUV sources do not match:  
all sources with NUV measurements  show no FUV detection (FUV\_mag=-999), and viceversa
all FUV sources have NUV\_mag=-999.  These fields are nonetheless included in the catalog because they satisfy all the defined criteria, however users must keep in mind that such mismatch would cause a false statistics of FUV-NUV colors in these fields. 
In Appendix \ref{s_oddfields} we show one of these fields, and also use it as example to illustrate the main artifact flags of the GALEX sources. 

\section{Content and Structure of the Catalog} %
\label{s_content}

The catalog includes  82,992,086 unique sources (GUVcat\_AIS\_055), from a total of 86,632,284 AIS measurements (GUVcat\_AIS\_plus, before duplicates are removed). The version restricted to sources within the central 1\grado~ of the GALEX field, GUVcat\_AIS\_050, contains 69,772,677 sources.  Note that the majority of these measurements are from $coadd$s (Section \ref{s_badcoadds}), therefore duplicate measurements only occur in field overlaps or repetitions. These fields are the result of over 56,000 $visit$s,  many repeats at $visit$ level were already merged in the $good$ $coadd$s we used. 

  Tables \ref{t_stats} and \ref{t_statscolors} give 
the number of sources included in GUVcat\_AIS, at different galactocentric latitudes, the fraction which have 
 multiple measurements, those affected by artifacts, and  samples with  magnitude and color selections. %
Whole-sky maps of the density of UV sources and their characteristics across the sky were shown by \citet{bia14uvsky}, which highlighted interesting distributions of hot stars in the Milky Way, among other trends.

The catalog gives several tags for each source, including position
(R.A., Dec., Galactic $l,b$), photometry measurements in FUV and NUV and their errors (``nuv\_mag'' and ``fuv\_mag'' are the
``best'' measurements as chosen by the pipeline, and preferable in most cases; other measurements are also
included, such as  PSF photometry, aperture photometry with different apertures, 
and Kron-like elliptical aperture magnitudes), other
 parameters useful to retrieve the original image from which the photometry was extracted (tag {\it photoextractid}),
as well as artifact flags and extraction flags that can be used to eliminate spurious sources (see Section \ref{s_artifacts} below).
In addition to these astrometry and photometry tags, propagated from the GALEX pipeline processing, we include new tags informative of the existence of duplicate [AIS] measurements or nearby sources (described in Appendix \ref{s_remdup}), and tags indicating whether the source falls within the footprint of a large object such as galaxy or Milky Way stellar cluster. These added tags facilitate extraction of clean samples for science applications of the catalog. 
The complete list of tags and their description is given in Appendix \ref{s_tags}. 

The catalogs can be downloaded from the author's web site: \\
\url{http://dolomiti.pha.jhu.edu/uvsky/\#GUVcat}, and will be also  available from the MAST casjobs 
  web site (\url{http://mastweb.stsci.edu/gcasjobs})
and from the SIMBAD Vizier  database,
which allows VO-type queries including  cross-correlation with other catalogs in the
same database. \footnote{
The earlier version of these catalogs (with less data coverage and fewer parameters) is  
accessible with Vizier 
at: 
\url{http://vizier.u-strasbg.fr/viz-bin/VizieR-3?-source=II/312},  and from MAST at:
\url{http://archive.stsci.edu/prepds/bianchi\_gr5xdr7/} . 
}

\begin{figure}
\centerline{
\includegraphics[width=14.5cm]{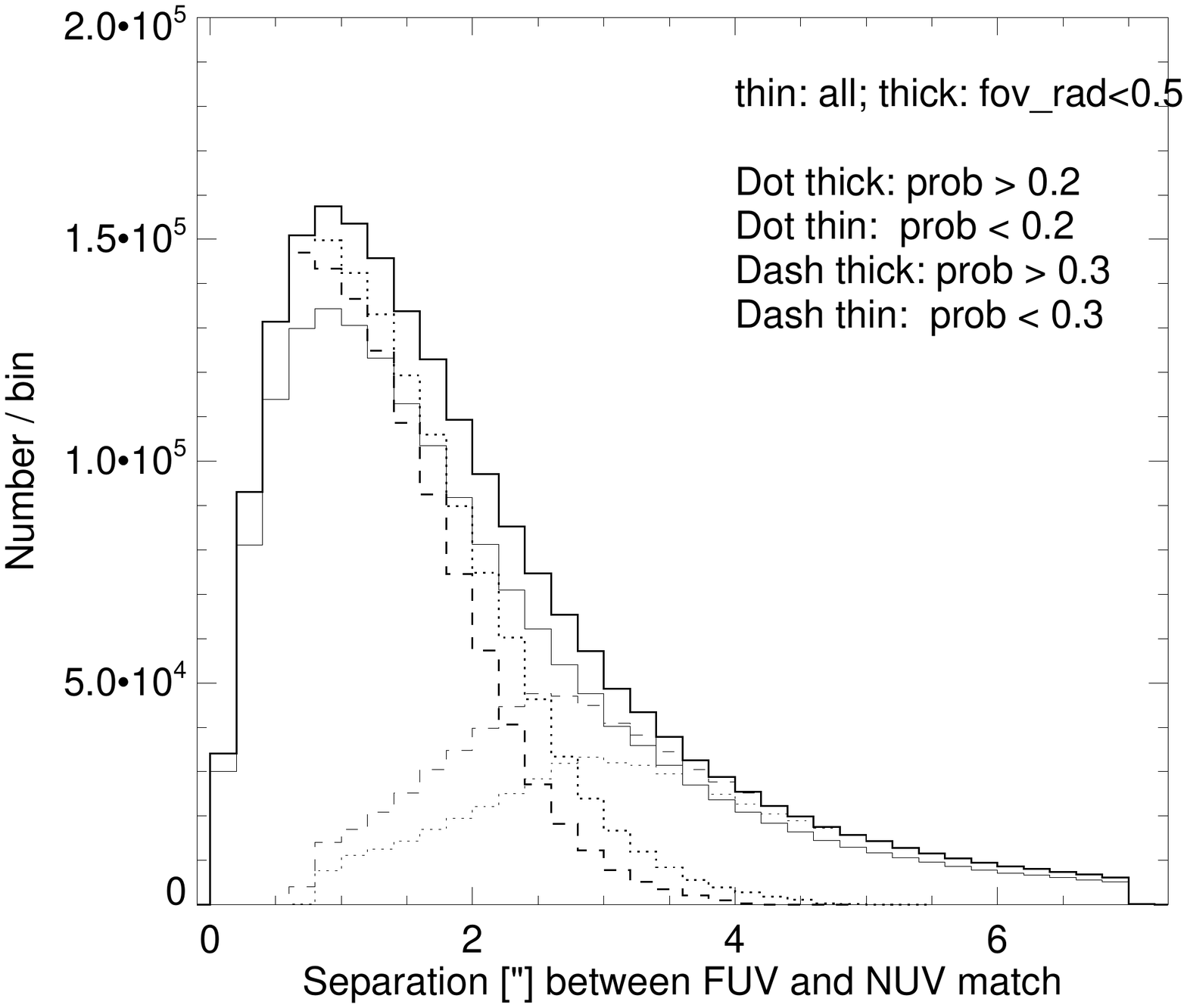}}
\caption{Distribution of the separation between FUV and NUV position of the sources in the GALEX database. A representative sample of 2 million sources is shown. Subsamples with cuts in the tag $probability$ [that the match is real] are shown.
\label{f_fuvnuvsep} }
\end{figure}

\subsection{Sources in Extended Clusters or Galaxies}
\label{s_inextobj}

 While we cannot and should not exclude from the catalog the  sources (as measured by the pipeline) in extended galaxies or crowded fields, for convenience of catalog's users we flagged all sources that fall within the footprint Galactic stellar clusters or galaxies larger than 1\am.  We added a tag $inlargeobj$ which contains the identifier of the large object prefixed by {\bf ``GA:''} for galaxies (e.g., GA:M33 ),  {\bf ``GC:''} or {\bf ``OC:''} for globular clusters and open clusters respectively (e.g., GC:NGC5272), {\bf ``SC:''} for less well defined cluster types. We also added a tag $largeobjsize$ which gives the D$_{25}$ diameter for galaxies, or 
twice the radius for stellar clusters. Note that 1\am is a very conservative limit, for the purpose of eliminating crowded regions, but a user can choose to worry only about larger objects by using a combination of these two tags, which we highly recommend. 
We provide finding charts for all of the extended objects ($>$1\am) in the footprint of GUVcat\_AIS. These can be found in the GUVcat tools on the author's web site 
\url{http://dolomiti.pha.jhu.edu/uvsky/\#GUVcat}.

\begin{figure}
\vskip -.75cm
\centerline{
\includegraphics[width=14.cm]{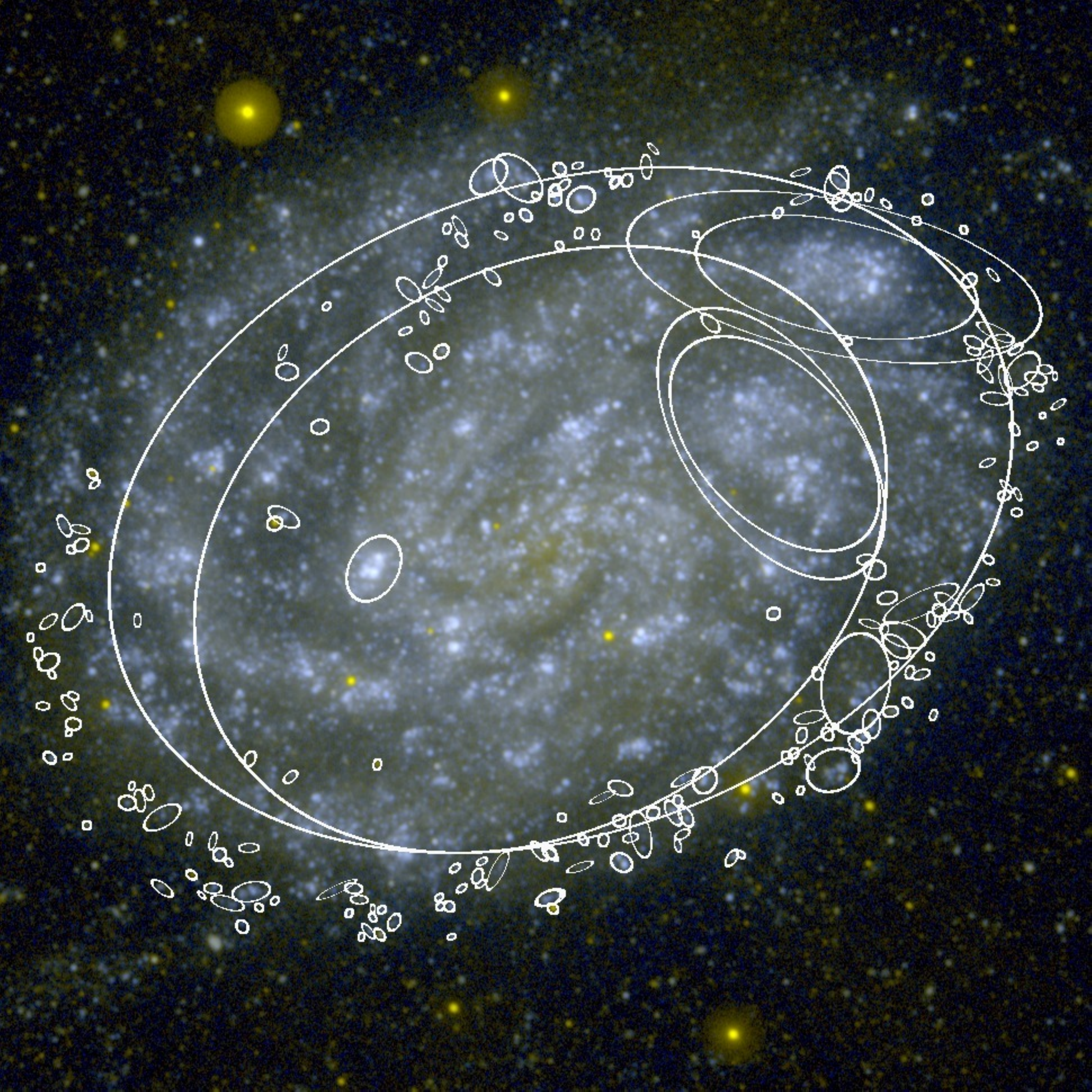}}
\caption{Example of pipeline photometry for an extended disk galaxy, NGC300 (D$_{25}$ $\approx$ 0.2\grado).   The central parts of the disk are measured by the pipeline  as  unresolved extended sources; in the periphery and less dense regions, where individual peaks are resolved, source density is much higher than in the surrounding field.  Therefore, density counts of foreground stars or background AGNs for example will be highly biased if sources in this region were not excluded.  
We marked all  sources retained in GUVcat  (duplicate measurements are removed) and associated to NGC~300 by our $inlargeobj$ flag 
(within 1.25$X$ D$_{25}$).  The  source shape is drawn, with an ellipse based on the pipeline-derived 
2.35$\times$$nuv\_a\_world$, 2.35$\times$$nuv\_b\_world$ (this choice is to match the pipeline .ds9reg file), $nuv\_theta$ (position angle). 
 They may appear different using $kron\_radius$$\times$$nuv\_a\_world$, which would show the area where the $mag\_auto$ are integrated.  Aside from details and differences among various  magnitude extraction options, which can be examined in the catalog, the figure illustrates convincingly that pipeline photometry in very extended galaxies must not be used for source counts.  The GUVcat tag $inlargeobj$  allows  sources in these areas to be excluded.  Note that some large sources have two measurements: these come from two overlapping AIS observations, that placed the centers of the big ellipses more that 2.5\as apart from each other, therefore they were not eliminated as duplicates in GUVcat. The image is 1455\as on a side. 
\label{f_extgal} }
\end{figure}

 The stellar clusters included for flagging were taken from the compilation available at \url{https://heasarc.gsfc.nasa.gov/W3Browse/all/mwsc.htm}, which basically includes all globular clusters from \citet{harrisGC}, which are all confirmed objects, and includes as ``open clusters'' confirmed, candidate or doubtful clusters, or spurious objects such as OB associations and large nebulae.  Of course the definition of open clusters is less specific than is possible for globular clusters, and their stellar density also varies more widely.  As pointed out in Section \ref{s_crowd}, only 
in the most crowded regions of clusters the source extraction would fail. In the dense central regions of globular clusters, the pipeline sometimes integrates a large area as one extended source. This may happen both for galaxies and for crowded stellar clusters: examples are shown  in Figure \ref{f_extgal}. 

   The $heasarc$ catalog gives three values of radius: r$_o$ (radius of the cluster core in the visible, corresponding to the distance from the center where the radial density profile becomes flatter), r$_1$ (where the radial density profile abruptly stops decreasing) and r$_2$ (where the surface density of stars equals the average density of the surrounding field); it also gives the  number of (optical) sources within these radii. In order to select the most appropriate value of cluster radius for our purpose, i.e.  to exclude only sources which would very likely introduce statistical biases, we examined two classical examples, NGC188 and NGC2420. By combining the number of sources  with the cluster sizes, we concluded that  r$_1$ is a good compromise, although somewhat conservative. OB associations are interesting objects $per se$,  but are sparse and are much less likely to suffer from crowding problems, and to introduce significant overdensities in  global source counts.  Therefore, we restrict the 'open cluster' list to only $confirmed$ clusters, and we further restricted these by combining the criteria of $cluster\_status$ not ``C'' (candidate) and $cluster\_type$ neither ``DUB'' nor ``NON''.   In total,  48  GC and 324 OC are included, entirely or partly, in the GUVcat footprint, all are shown in our $uvsky$ web pages.
Table \ref{t_clustersGC} (electronic only) lists  centers, size and other parameters for Galactic clusters.

Table \ref{t_extendedGA} (electronic only) gives a list of 
 centers, major and minor axis and position angle (p.a.) and other basic parameters for extended galaxies with major axis D$_{25}$ $\geq$ 1\am.  The galaxies (22,037) were selected from the hyperleda database, with no other restriction than the size, D$_{25}$ $\geq$ 1\am.   
 In total,  15,659 of these galaxies with D$_{25}$ $\geq$ 1\am are included (at least partly) in the GUVcat\_AIS footprint. We flagged sources out to 1.25$\times$D$_{25}$, a choice based on inspection of several maps, available on our web site\footnote{\url{http://dolomiti.pha.jhu.edu/uvsky/\#GUVcat} }, of which Figure \ref{f_extgal} shows an example.   Note that most galaxies with size $\approx$1\am are probably detected as a single (extended) source, or a few sources, in the GALEX data.  Therefore, while the 1\am size limit provides a very comprehensive flagging, for statistical analyses of large samples of sources a much larger radius can be used to exclude only galaxies for which the pipeline photometry is misleading.  

 For many science applications, such as statistical studies of source densities and luminosity functions, the area covered by the catalog must be calculated.  Portions optionally excluded (because in the footprint of a cluster or galaxy) must be taken into account in the area calculation. Our interactive  area calculation tools will offer some options (Section \ref{s_area}) for area estimate in the cases where large object footprints are excluded from the samples. We stress that, when catalogs over large areas are used, removing very small footprints may introduce additional un-necessary uncertainties in area calculations, depending on the tessellation steps of the sky grid used for area calculation relative to size of the areas being excluded.  More details are provided with the area calculation tools \citep{bia17area}.

\renewcommand{\thefigure}{\arabic{figure} (Cont.)}
\addtocounter{figure}{-1}
\begin{figure}
\centerline{
\includegraphics[width=8.5cm]{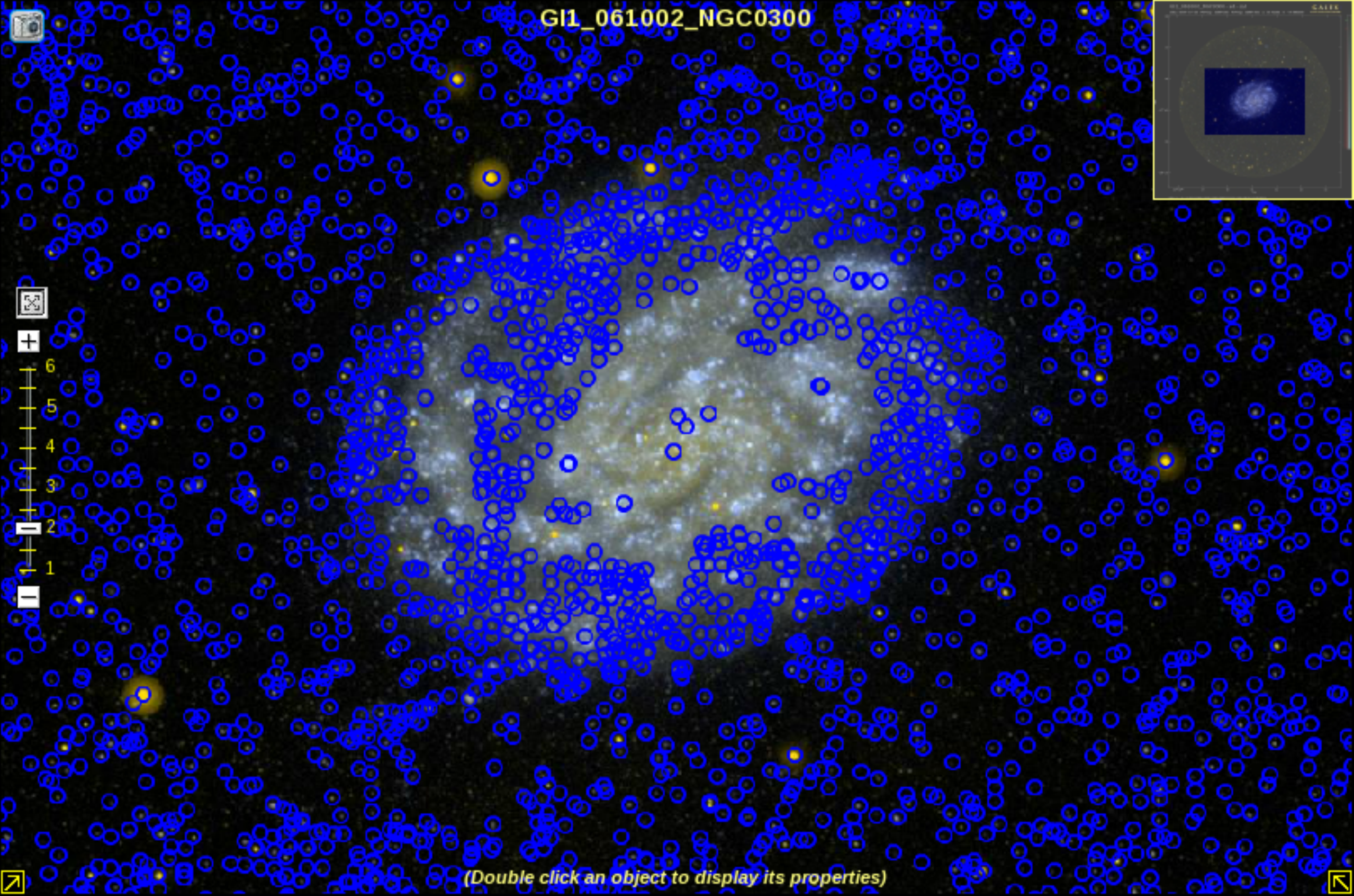}%
\includegraphics[width=8.5cm]{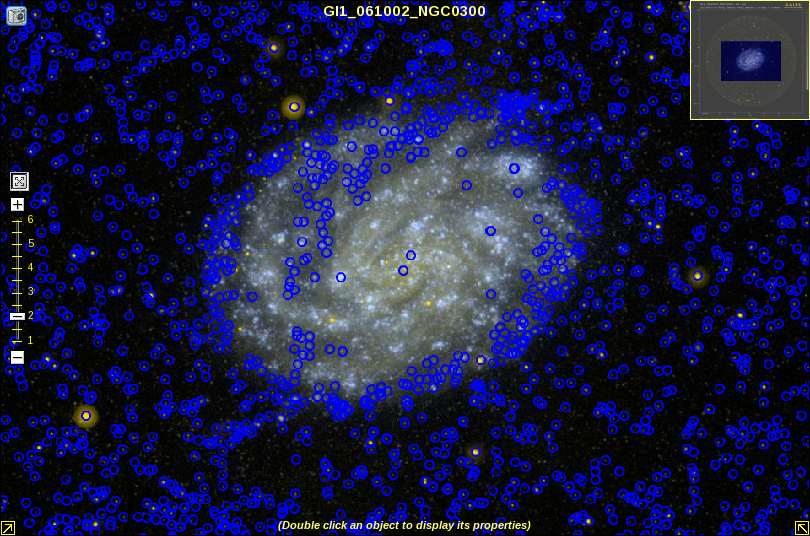}}
\vskip 0.1cm
\centerline{
\includegraphics[width=8.5cm]{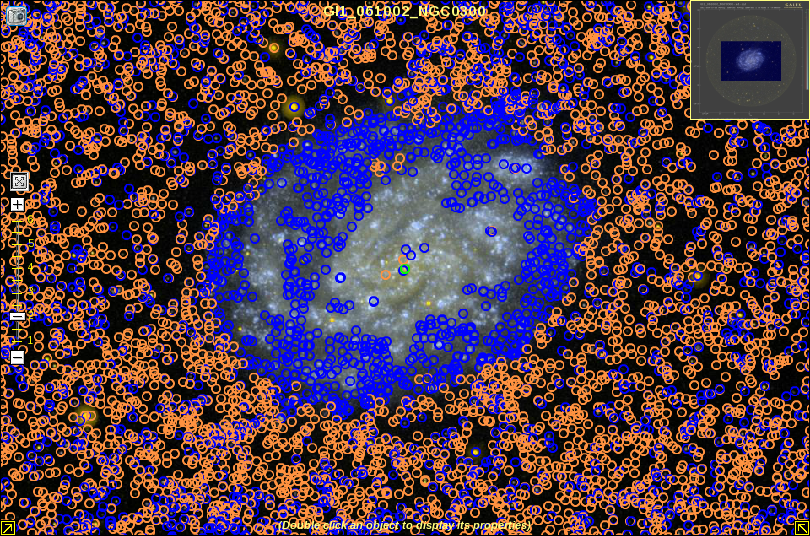}
\includegraphics[width=8.5cm]{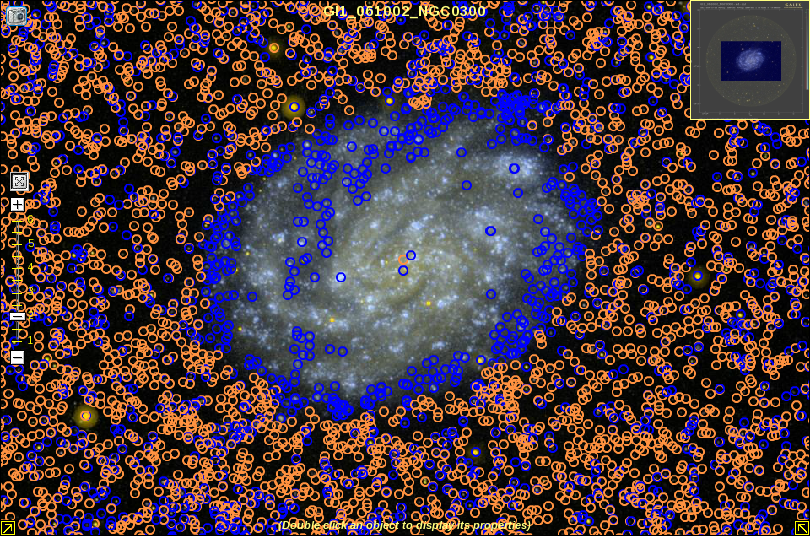}}
\caption{GALEX pipeline sources in the master database around NGC300: AIS as blue circles, NGS (about two mag deeper than AIS, see \citet{bia09}) as orange circles. Note that here duplicates have not been removed, all measurements are shown, making the sources appear more numerous than in GUVcat (previous figure). There is an even deeper GI observation, not shown for clarity. The left panels show all entries in the master database, the right panels only those with %
NUV\_err$\leq$0.5 (as in GUVcat, which eliminates some spurious sources and many artifacts).
}
\end{figure}
\renewcommand{\thefigure}{\arabic{figure} (Cont.)}
\addtocounter{figure}{-1}
\begin{figure}
\centerline{
\includegraphics[width=15.8cm]{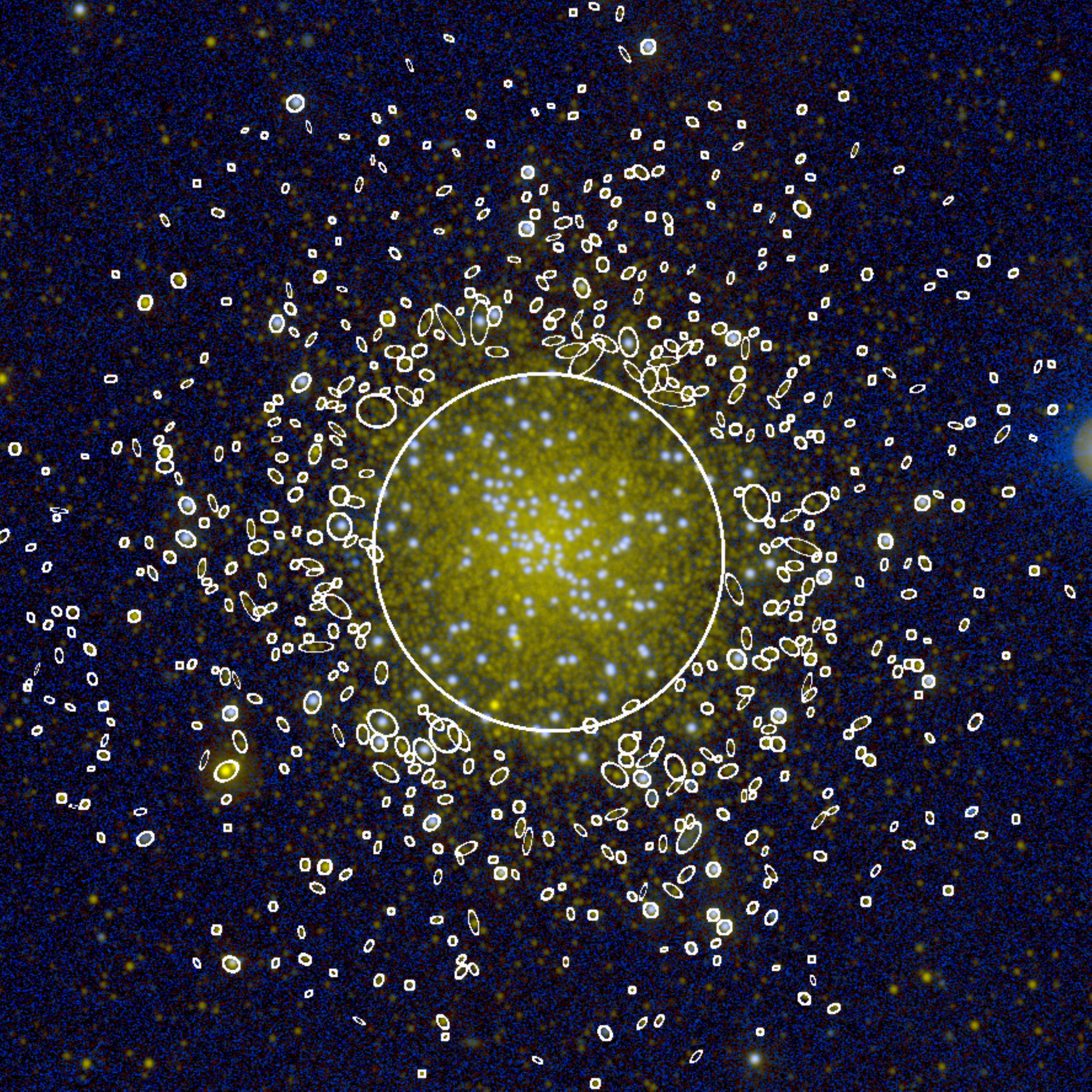}}
\caption{Example  of pipeline photometry for a crowded stellar cluster, NGC6218. Sources retained in GUVcat are drawn, as in the Figure 5a, according to their database photometry extraction parameters. The color image was constructed for all available imaging for the field, including deeper exposures.  Clearly visible blue sources in the cluster are only detected by the pipeline in FUV (therefore not retained in GUVcat, which uses as starting point the NUV-source detections), see next figure. The large circle is showing the pipeline aperture of the central source (drawn as explained in the previous figure), 
 taken from NUV, showing how the pipeline neither provides accurate integrated measurements nor robust crowded-field measurements of resolved stars in the cluster.   The image is 1332\as on a side. 
}
\end{figure}
\renewcommand{\thefigure}{\arabic{figure} (Cont.)}
\addtocounter{figure}{-1}
\begin{figure}
\centerline{
\includegraphics[width=8.5cm]{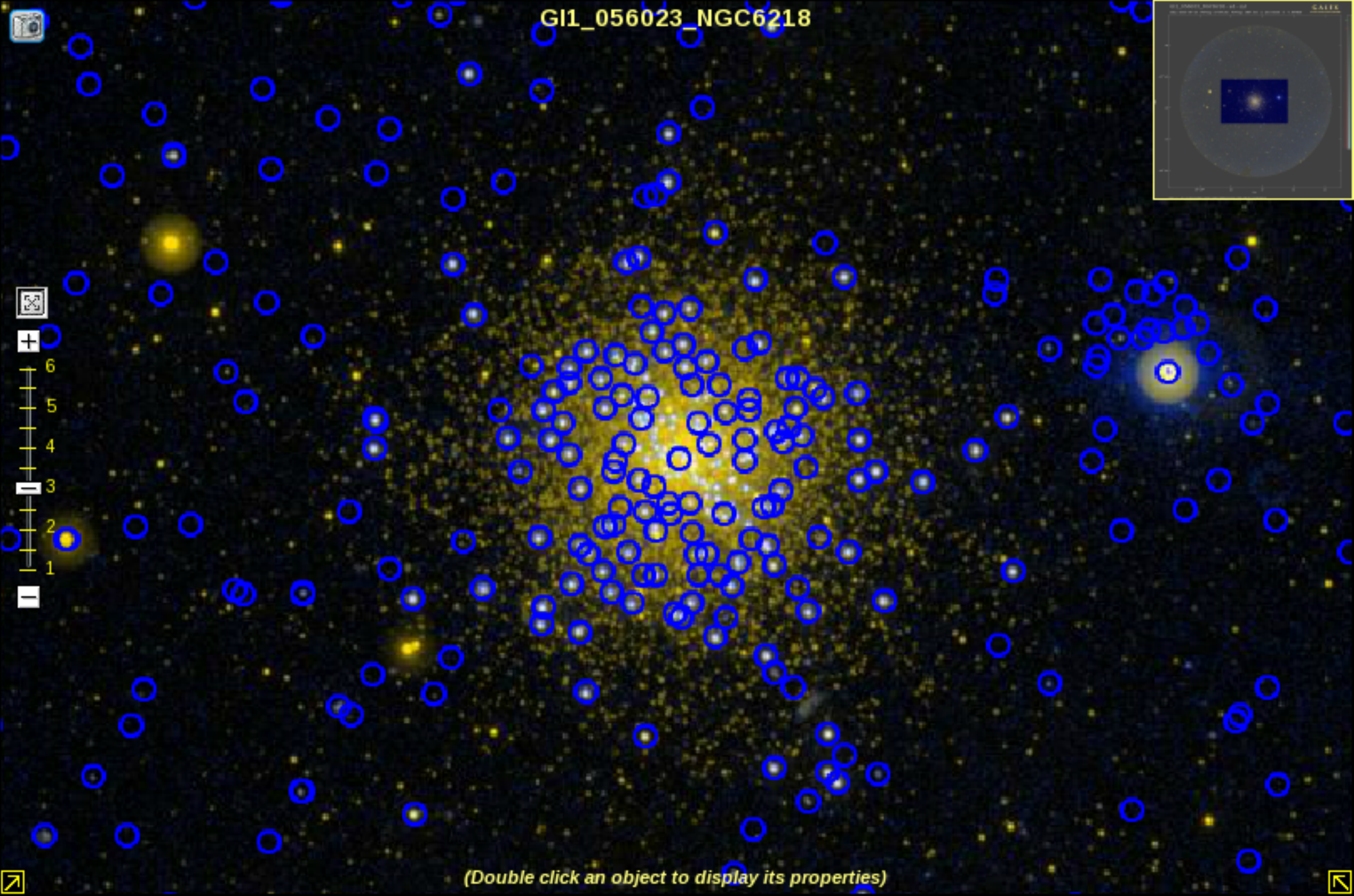}
\includegraphics[width=8.5cm]{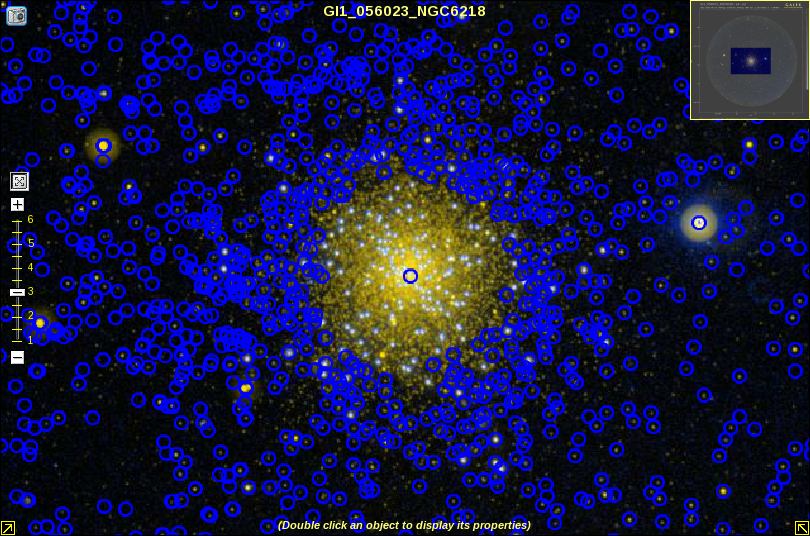}}
\centerline{
\includegraphics[width=12.5cm]{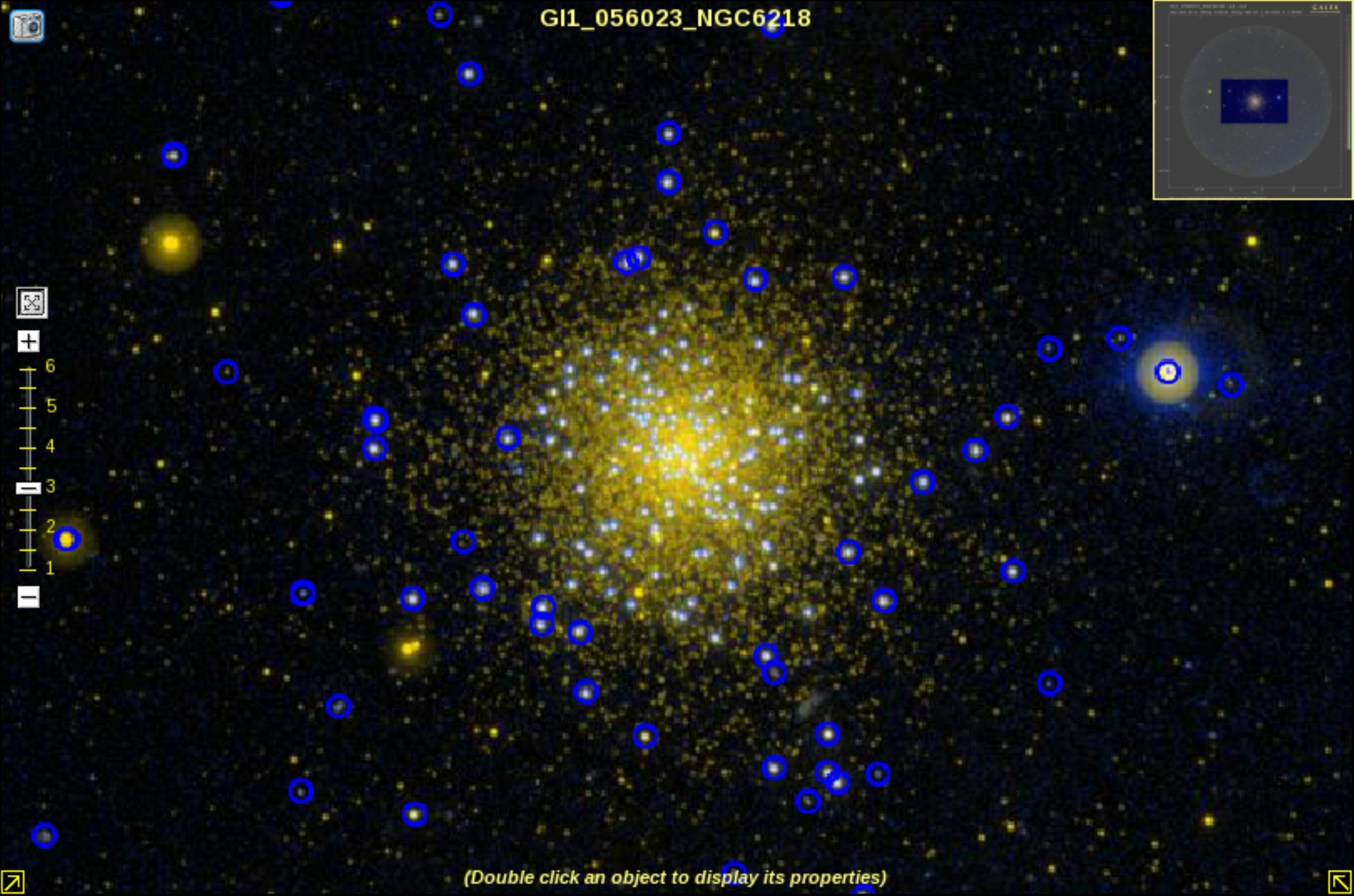}}
\caption{AIS detections in the master database for NGC~6218 (only source centers shown, not source shapes).  Top:  FUV detections (left) and NUV detection (right); bottom: sources detected in  both FUV and NUV.   Note, from the shape of the pipeline  sources shown in the previous figure, that matching FUV and NUV colors in the central region would not be correct, even for sources where a match exists.  }
\end{figure}
\renewcommand{\thefigure}{\arabic{figure}}

\subsection{Flagged artifacts} 
\label{s_artifacts}

 Table 2 of the GALEX GR6 documentation (galex.stsci.edu/GR6/?page=ddfaq\#6) lists the value of the $artifact$ flags ($FUV\_artifact$ and $NUV\_artifact$ in the catalog), and suggests that the only artifact flags causing real concern are the $Dichroic ~ reflection$ ($artifact$=4, base 10 value, or $artifact$=64 when a $coadd$ has enough $visit$s at different position angles  that masking the Dichroic reflection does not decrease the flux by more than 1/3rd)  and $Window~ reflection$ (applicable to the  NUV detector only: $NUV\_artifact$=2). 
 Most of the artifacts in  the original database are caused by the detector rim ($artifact$=32), or reflections around the edge: these do not affect our catalog since we exclude the outer edge  of the field of view (Figure \ref{f_oddfields}). In more detail: the version which retains sources within 0.55\degree~ from the field centers, GUVcat\_AIS\_055, excludes a 0.06\degree-wide outer ring; this  is sufficient to eliminate rim artifacts, except in a few cases because in GUVcat we retained $coadd$s of $visit$s with a tolerance of up to 5\am between the pointings of the individual $visit$s. In the worse case of two coadded $visit$s having centers 5\am apart, the $fov\_radius$ of the $coadd$ sources may also differ by up to 5\am from the actual distance of the source from the center of its $visit$, therefore a few rim artifacts may be included.  Such tolerance of 5\am centreing difference  between $visit$s of $coadd$s was chosen to maximize the area coverage of the catalog, and to avoid throwing away much data or much exposure depth. As a consequence, in GUVcat\_AIS\_055 there remain  23,218 sources with either FUV or NUV rim artifact flag set, out of the $\sim$93~million catalog sources. These sources have $fov\_radius$ (distance from the $coadd$ center) between 0.5125 and 0.55\degree, and all come from $coadds$ as expected. 
 By comparison, there are 31,184,260 sources  with either $fuv\_artifact$ or $nuv\_artifact$ rim flag set
 (6,765,612 with $fuv\_artifact$ flag set)  in the whole $visitphotoobjall$ GALEX database, and 
25,259,384 sources with $fuv\_artifact$ or $nuv\_artifact$ rim flag set (25,221,382 NUV; 18,592,421 FUV)  in the whole $photoobjall$ GALEX database of 292,296,119 entries. 
Note that the GALEX field has a diameter of $\approx$1.2\degree, therefore the actual fov\_radius of any source should always be $\leq$0.6\degree, 
and rim sources should have $fov\_radius$$\sim$0.6, but in the MAST GALEX database the sources with ``rim'' artifact flag set have values of  $fov\_radius$ between 0 and $>$1\degree, an effect of the improper $coadd$ described in Section \ref{s_badcoadds}, where the rim artifact has been propagated from the $visit$-level processing, while $fov\_radius$ has been recalculated using the center of the $coadd$, therefore an actual rim source may end up having apparent $fov\_radius$ near zero (near the center of the $coadd$) or a value almost twice the GALEX f.o.v. radius. This is illustrated in Figure \ref{f_badcoadd} and was explained in Section \ref{s_badcoadds}. This problem is cured in GUVcat.

 In the  GUVcat\_AIS\_050 catalog there are no rim or edge  artifacts, 
since we only retained sources within 0.5\degree from the field center, which leaves out, even with a 5\am tolerance for $coadds$,  an outer ring of $\geq$0.1\degree~ width. This restriction comes at the price of a $\sim$ 10.7\% decrease in area coverage,  as explained in Section \ref{s_area}, introducing occasional gaps between adjacent fields. 

 Masked variable pixels ($artifact$=128) and masked detector hotspots ($artifact$=256) may degrade the quality of a photometric measurement but would not introduce spurious sources, and they are rare, therefore they are not relevant for the purpose of source counts.  What does introduce a high number of spurious source detections (once the rim is excluded) are reflections and 'ghosts' near very bright sources. We show examples in Appendix~C.  A conservative recommendation 
 is to eliminate sources with $artifact$=4 or 2.  Note that if more than one artifact is deemed to be present, the flag value is the sum of all the artifacts affecting the source.   Table \ref{t_stats} gives also the fraction of sources with different artifact flags in the GUVcat\_AIS catalog, and report the artifact definitions in the table's footnote.

\section{Area Coverage of the Catalogs}
\label{s_area}

For studies involving density of sources (number per unit area), the exact area coverage of the 
catalog  must be known. 
As we removed duplicate measurements of the same source, we  must  calculate the  area 
covered by the surveys accounting for 
overlaps. We must also account 
for possible gaps between fields; these may occur because of the tiling strategy (for example, to avoid bright stars that would damage the detectors), 
or because the actual pointing of an observation is slightly off from the planned position, 
and  because we limited our catalogs to sources within the central 1.1$^{\circ}$ (or 1.0\grado) diameter of the GALEX field. 

We calculated the total actual area covered by GUVcat\_AIS with the method 
of  \citet{bia11a}: we divided the sky in small tesserae, 
 and added the 
areas of all tesserae which fall within 0.55$^{\circ}$ (or 0.50$^{\circ}$) from the center of  every field  used in the catalog, ensuring that each tessera
is counted only once. 
The  total area covered  is 24,790  square degrees for GUVcat\_AIS\_055, and 22,125 square degrees for GUVcat\_AIS\_050. 
This  area of ``unique-coverage''  is $\approx$95\% (with $fov\_radius$$\leq$0.55) and 88\% (with $fov\_radius$$\leq$0.5) of the sum of  areas of the fields used (if there was no overlap between observations), implying an overall overlap of $\approx$ 11.7\% and 4.6\%  respectively  among the AIS fields used. Area coverage of 5-degrees latitude slices for the catalog are given in Table \ref{t_stats}.

 Because both gaps and overlaps between fields occur,  the actual area coverage 
must be computed for each region of the sky where one desires to extract a sample, 
if the density of sources has to be estimated.  An online interactive tool will be presented elsewhere, 
for  area calculations of custom-chosen regions, for the GUVcat and BSCcat catalogs, and for matched GUVcat$-$optical catalogs \citep{bia17area}. 

\section{Conclusions and Summary}
\label{s_conclusions}

\subsection{The UV sources across the sky}
\label{s_stats}

\begin{figure}
\vskip -.75cm
\centerline{
\includegraphics[width=14.5cm]{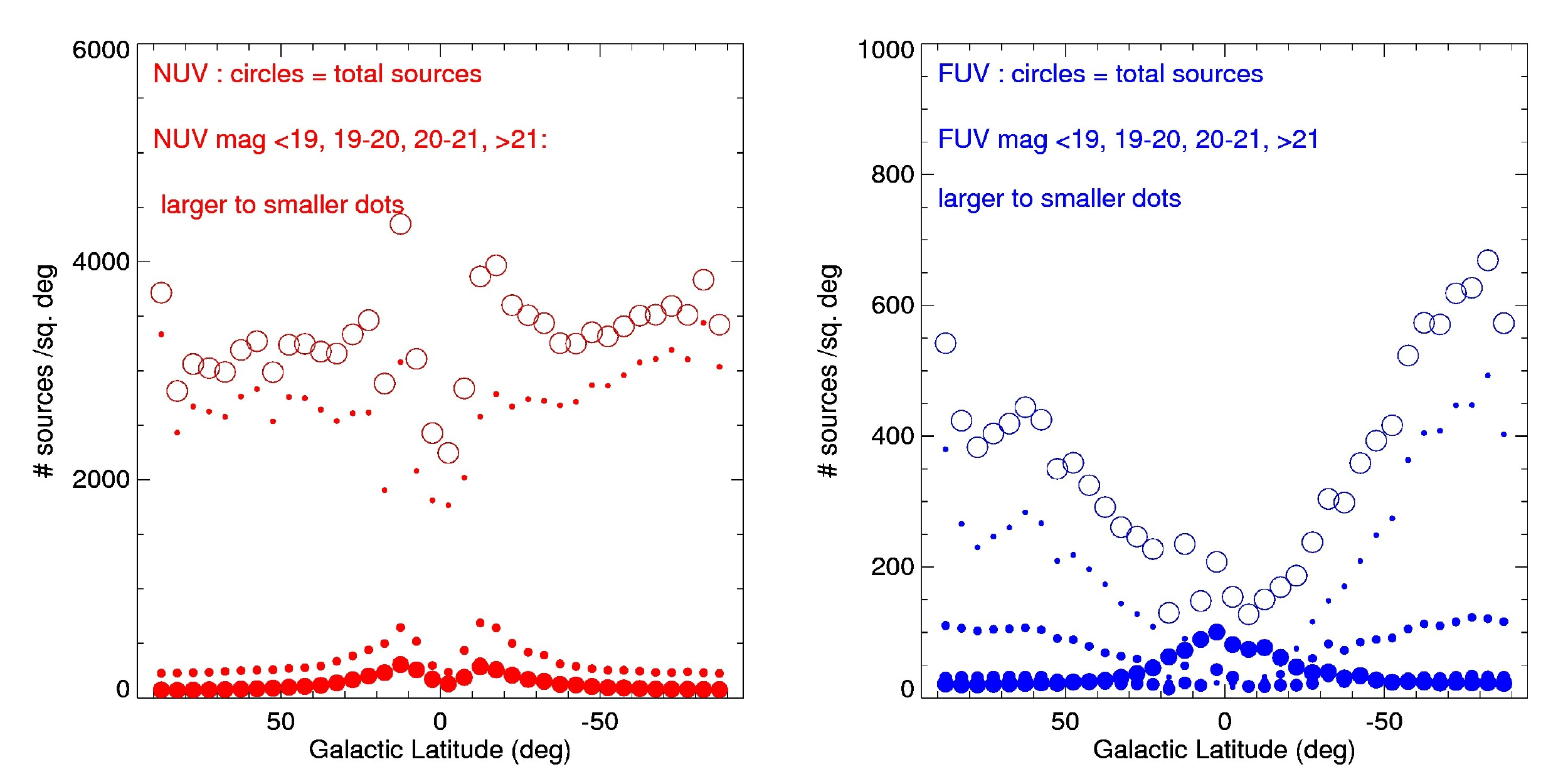}}
\centerline{
\includegraphics[width=16.5cm]{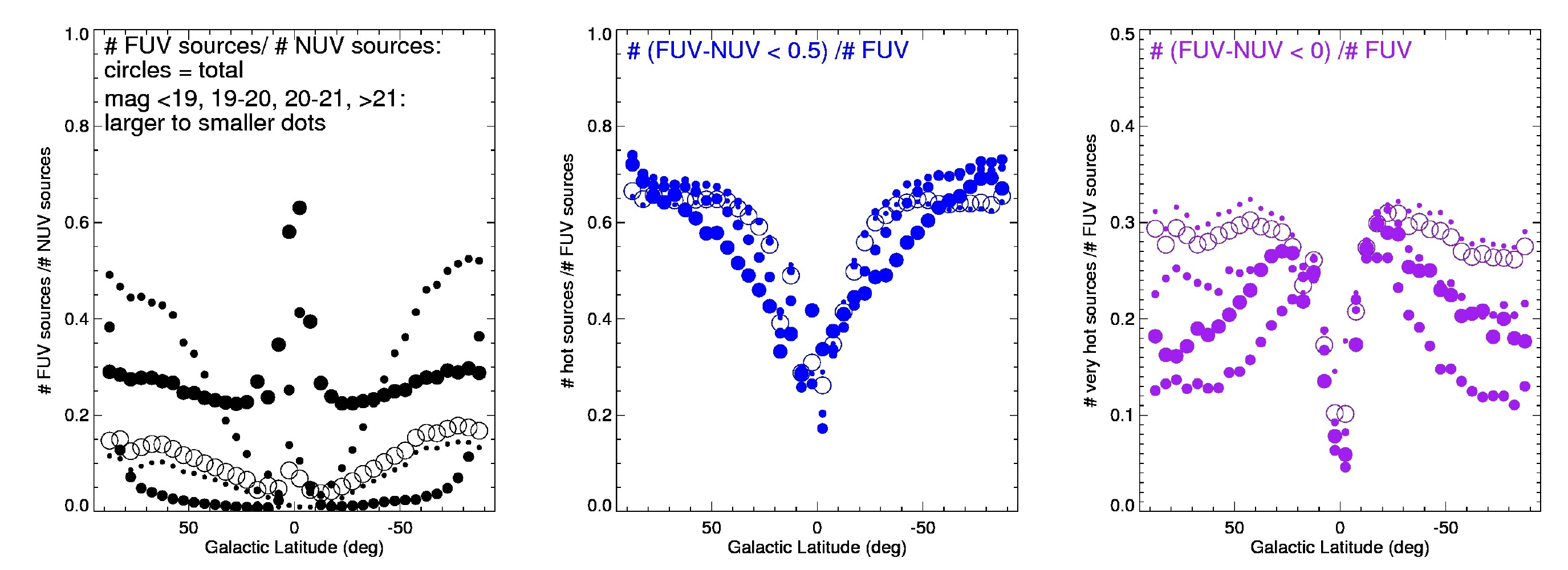}}
\caption{{\it Top:} number of sources per square degree detected in NUV (left) and  in FUV (right), as a whole (circles) or by magnitude ranges (dots). Values are  shown for every 5-degree latitude strip.  {\it Bottom: } Fraction of FUV detections over NUV detections (left), fraction of sources with FUV-NUV $<$0.5 (middle) and $<$0.0 (right) among the FUV detections (no error cuts, but sources in the footprint of extended objects have been excluded from the counts).  
While the faint sources (largely extragalactic) dominate the total samples, the FUV-detected sources are mostly bright stars. 
 We recall that, for average Galactic dust, the UV extinction is similar in  FUV and NUV, and much higher in both bands than at  optical wavelengths (Table \ref{t_ext} and \citet{bia11}).   \label{f_stats} }
\end{figure}

\citet{bia14uvsky} published  several maps showing the  distributions of UV sources in the sky, for both the AIS and the deeper MIS survey. 
In Figure \ref{f_stats} we show the density of sources (number per square degree) detected  in the NUV and FUV bands; as discussed extensively by \citet{bia11a, bia11b, bia14uvsky} the number of FUV detections is typically ten times less than the NUV detections overall; this happens because hot stars, and blue galaxies, are much more rare than cooler (redder) objects. More specifically, the fraction depends  on Galactic latitude and on the magnitude depth considered, because the number of extragalactic sources with respect to Galactic stars increases rapidly towards fainter magnitudes. The relative fractions are a combinations of the intrinsic distribution of different types of sources, whereby the density of Galactic stars increases towards the disk of the Milky Way, while the distribution of  extragalactic sources does not depend on the Milky Way structure, but all sources are affected by the Milky Way dust, which is mostly confined to a thin disk. The reddening depends therefore on the line of sight towards the sources going through more or less of the dust disk. This effect was dramatically illustrated by Figure 2 (bottom) of \citet{bia11a}: the ``V-shape'' region devoided of UV-source counts in their figure is essentially a direct image of the dust disk. It is also visible, though less evident,  in Figure \ref{f_stats}.

In Figure \ref{f_stats} we plot the density of NUV and of FUV sources in GUVcat\_AIS (top plot), as a whole and divided by NUV magnitude ranges: the plots show that the sources fainter than NUV\_mag=21mag dominate the sample, in spite the AIS is the shallowest survey, and especially so in the NUV where extra-galactic objects are more prominent.   The bottom panels show the fraction of FUV detections over NUV detections, again as a function of Galactic latitude, and among these, the hot and very hot sources (FUV\_mag-NUV\_mag $\leq$ 0.5 and $\leq$0.0 respectively). Such UV color cuts correspond to different stellar \Teff for different types of stars \citep{bia09}, but roughly hotter than $\sim$15,000K.   Some QSOs may intrude these FUV-NUV  color cuts, as shown by  \citet{bia09qso}: these affect the faint sources most.   The different behaviour of relative source densities in Figure \ref{f_stats} reflects the fact that brighter samples (and hotter samples)  are dominated by Galactic stars, which are more numerous in the Milky Way  disk (see e.g., \citet{bia11a}). 

\subsection{Summary of Suggestions and Caveats for Using the Catalog}
\label{s_advice}
To conclude, we distill here, in terms of practical advice for users, the relevant information on the %
 catalogs presented in this paper. 

\begin{itemize}

\item
{\bf GUVcat\_AIS contains unique measurements} of all sources from AIS observations  with both FUV and NUV detectors exposed. 
Duplicates measurements are removed in the main catalog, however a version GUVcat\_AISplus is accessible where duplicate measurements are flagged but not removed.

\item
{\bf GUVcat is available} from \url{http://dolomiti.pha.jhu.edu/uvsky/\#GUVcat}   as well as MAST casjobs (\url{http://mastweb.stsci.edu/gcasjobs}), and SIMBAD Vizier.

\item  {\bf Sources near the field's edge} have been excluded, because they are mostly artifacts and have poor photometry.   
GUVcat\_AIS\_0.55 contains sources within 0.55\grado of the field's center, GUVcat\_AIS\_0.50 only sources within 0.50\grado. The first has a larger area coverage, fewer gaps, at the expense of a few rim artifacts intruding the catalog, these must be sieved from samples by using the artifact=32 flag (Section \ref{s_artifacts}).
Tables \ref{t_stats} and \ref{t_statscolors} give statistical information on the number of sources, and the fraction of sources affected by artifacts, or in given UV-color ranges, in total and divided by Galactic latitude. 

\item
{\bf area coverage} of our catalog  GUVcat, BCScat, and of overlap of  these catalogs  with optical databases, can be calculated for any desired region of the sky with the tool of \citet{bia17area}. See also Column 9 of Table \ref{t_stats}.

\item
{\bf Extended Objects:}
beware of sources in the footprint of large galaxies or crowded stellar clusters (Section \ref{s_inextobj}). These can be identified and eliminated with the two tags $inlargeobj$ and $largeobjsize$, provided in GUVcat. The web site \url{http://dolomiti.pha.jhu.edu/\#GUVcat} gives also 
 finding charts and information on all large objects included entirely or partly in GUVcat. 

The size limit  of the extended objects that one should  eliminate from the catalog depends on the specific objectives and sample size; if one needs to compute area coverage of the extracted sample, the excluded footprints can be accounted for with our area calculation tool (Bianchi \& de la Vega 2017), however the interactive public version currently uses a sky tessellation with a grid step of 0.1\grado (so that a computation over the whole sky can be accomplished in a few tens seconds); excluding any area smaller than, or comparable to the grid tesserae will introduce uncertainties in the area estimate.

\item 
{\bf  Magellanic Clouds: } only the periphery of the Magellanic Clouds is included in GUVcat, because the central regions are only observed in NUV. Even the peripheral fields are crowded enough to pose a challenge to the pipeline photometric procedures: for point sources within a 15\grado radial distance from the center of the LMC, and a 10\grado radial distance from the SMC, it is preferable to use the custom-made catalog of Thilker et al. (2017) and to avoid using this catalog or the master database. In Table \ref{t_statscolors} we count sources within 15\grado / 10\grado radial distance from LMC / SMC, but for consistency with other galaxies, the flag $inlargeobj$ is set only for GUVcat sources within 1.25$\times$ the Hyperleda D$_{25}$ size, which is much smaller; these sources have flag GA:ESO056-115 and GA:NGC0292 for LMC and SMC respectively.  For the statistical overview in Figure \ref{f_stats} we conservatively excluded the  15\grado / 10\grado degree areas, since we noted an overdensity of sources even in the outermost periphery of the Clouds.

\item
{\bf Reddening correction}:  Table \ref{t_ext} gives extinction coefficients in the GALEX FUV and NUV bands for representative known types of interstellar dust; these coefficients can be used to correct the UV magnitudes for reddening. In the GUVcat catalog, an  \ebv~ value is given for each source, based on the extinction maps of   \citet{Schlegeletal98};  this value is approximate as it represents an interpolation from low-resolution maps at the source position, and as such it is also an upper limit (a Galactic source very close will only suffer the absorption by the local component of the dust along the line of sight), anyway it is a convenient indication of reddening.  As noted by \citet{bia11a, bia11b, bia11}, the GALEX FUV-NUV color is almost reddening-free, for Milky-Way typical dust (see also Table \ref{t_ext}), and therefore it could be used  to select hot stellar sources, almost independently of reddening,  by \citet{bia11a}. 

\end{itemize}

\subsection{Is a source not detected, or not observed?}
\label{s_inorout}

When one matches a source list to the GALEX catalog, if a source is not found (either in the entire database, or in GUVcat\_AIS, or in any other catalog), one needs to know whether the source was observed but too faint to be detected in a given filter, or it was not in the footprint of any actual observation. This holds for GALEX, SDSS, and any database which does not have a complete coverage of the sky, due to the nature of the survey or because there are some gaps or unusable portions of data. 

The easiest and safest way  to find out whether a given celestial position is within the footprint of any  GALEX observation is to match the source coordinates to the list of $visit$ centers ($visitphotoextract$ in MAST casjob ) and check if the source position  is within the f.o.v. radius from the  center of any observation ($navaspra$, $navaspde$ for NUV; $favaspra$, $favaspde$ for FUV;   $avaspra$, $favaspdec$ for the combined FUV+NUV source list).
This test should be done at $visit$ level, because of the issue of $bad$ $coadds$ described in Section \ref{s_badcoadds}. For the ``good $coadds$'' (only) one could use the $photoextract$ values.

For other surveys, such as for example  SDSS, where gaps among fields or failed observations are not always mapped %
consistently into the footprint tool (e.g., \citet{bia11a}), one has to search for sources in a wider  area around the source of interest, and if other sources are found around the position, a negative detection for the source of interest will imply that the source was observed but its flux is below detection threshold. In this case, one could derive an upper limit from the exposure time of the observations in the area.   This procedure works in any case, but it is more cumbersome, and it may not be entirely safe: if the catalog sources are sparse, one would need to probe farly large portions of sky around the source of interest, to avoid false negatives, but in this way a 'positive' detection will mean that some wide area around the desired position has some sources: if that happens to be near a field edge and the desired  position is just outside the edge, the ``poor resolution'' sampling of the surroundings may give a false positive. 

\acknowledgments
 We are very grateful to Imant Platais for helpful suggestions concerning the selection of stellar cluster parameters, to Chase Million for always providing expert advice on GALEX data issues and clarifications on the GALEX pipeline, and to Scott Fleming for useful discussions on GALEX science projects. 
This work was supported by NASA ADAP grant NNX14AF88G. 
We made use of the GALEX database in the MAST archive, which is funded by the NASA Office of Space Science.

{}
\clearpage 
\begin{table*}
\caption{Broad-band reddening corrections for different types of interstellar dust \label{t_ext} }
\begin{tabular}{@{}lllll}
\hline
                    &  \multicolumn{4}{c}{Type of selective extinction\tablenotemark{a}} \\ 
\hline
                    &  MW     &    LMC   & LMC~2   &  SMC \\
\hline
\hline
E$_{FUV-NUV}$/\ebv & 0.11 &  1.08 &  2.00 &  4.60  \\
A$_{FUV}$/\ebv    & 8.06 &  8.57 &  9.02 & 12.68 \\
A$_{NUV}$/\ebv    & 7.95 &  7.49 &  7.02 &  8.08 \\
A$_{U}$/\ebv      & 4.72 &  3.96 &  4.11 &  4.61 \\
A$_{B}$/\ebv      & 4.02 &  3.26 &  3.46 &  3.85 \\
A$_{V}$/\ebv      & 3.08 &  2.34 &  2.54 &  2.93 \\
E$_{U-B}$/\ebv    & 0.70 &  0.70 &  0.66 &  0.76 \\
\hline
\tableline
\end{tabular}
\tablenotetext{a} 
{ values derived using the standard  Milky Way extinction curve from Cardelli et al. (1989): ``MW'', 
  using the curve of Misselt et al. (1999) for sightlines in LMC2 (``LMC2''), and using the average LMC extinction curve outside the LMC~2 region  (``LMC'') and 
the UV-steep extinction curves for SMC sightlines (``SMC'') by Gordon \& Clayton (1998). 
The quantities for each broad-band 
are derived by applying the filter passbands
 to progressively reddened model atmospheres for stars with \teff~ between 30,000K and 15,000K,
 and comparing unreddened and reddened model colors with \ebv=0.4.
The mean values are given, the dispersion is always less than 1\% within this  \teff~ range. }
\end{table*}

\clearpage 
\begin{deluxetable}{ccccccc}
\tabletypesize{\scriptsize}
\tablecaption{List of clusters included in the GUVcat footprint (electronic only, sample shown here) \label{t_clustersGC} }
\tablehead{
\colhead{name} & \colhead{ra} & \colhead{dec} & \colhead{central radius} & \colhead{broad type} & \colhead{cluster status} & \colhead{cluster type} 
}
\startdata
IC4499 &  225.076996 &    -82.213997 &      0.085000 &      G &     O &    GLO\\
\enddata
\end{deluxetable}

\begin{deluxetable}{cccccccccccc}
\tabletypesize{\scriptsize}
\rotate
\tablecaption{List of galaxies larger than 1\am included in the GUVcat footprint (electronic only, sample shown here) \label{t_extendedGA} }
\tablehead{
\colhead{name} &\colhead{type}& \colhead{ra} & \colhead{dec} & \colhead{v3k} & \colhead{vlg} & \colhead{P.A.} & \colhead{inclination} &  \colhead{log(D$_{25}$)} & \colhead{log(D$_{25}$ error)} & \colhead{log(R25)} & \colhead{log(R25 error)}\\
\colhead{}  &\colhead{ }& \colhead{degrees}& \colhead{degrees} & \colhead{km/sec} & \colhead{km/sec}& \colhead{degrees} &  \colhead{degrees} & \colhead{(0.1 arcmin)}& \colhead{(0.1 arcmin)} & \colhead{D$_{25}$/d$_{25}$} & \colhead{}
}
\startdata
UGC12889       &       G  &   0.0070005 & 47.27450 &    4751 &     5319 & 163.5 &   53.4 &     1.27 &      0.03 &     0.20 &     0.03\\
\enddata
\end{deluxetable}
\newpage
\begin{deluxetable}{cccccccccccccc}
\tabletypesize{\scriptsize}
\rotate
\tablecaption{List of AIS GALEX fields ($Coadd$s and $Visit$s) used to construct the GUVcat\_AIS catalog \label{t_CoaddsVisits}}
\tablewidth{0pt}
\tablehead{
\colhead{Field }           & \colhead{RA center}    & \colhead{Dec center} & \colhead{{\it l} center} & \colhead{{\it b} center} &
\colhead{FUV exp.time}     & \colhead{NUV exp.time} & \colhead{C or V} \\%
\colhead{} & \colhead{degrees} & \colhead{degrees} &  \colhead{degrees} &  \colhead{degrees} & 
\colhead{(seconds)}        & \colhead{(seconds)}    & \colhead{} \\   
\colhead{\it (photoextractID)} & \colhead{\it (avaspra)} & \colhead{\it (avaspdec)} &  \colhead{} &  \colhead{} & 
\colhead{\it (fexptime )}        & \colhead{\it (nexptime)}    & \colhead{} 
}
\startdata
   6370915756560875520 &      291.449459 &       75.146935 &      106.972770 &       23.963759 &   219.050 &   219.050 &   $coadd$ \\
   6370915757634617344 &      273.010299 &       80.131423 &      111.833168 &       28.552309 &   213.000 &   213.000 &   $coadd$ \\
   6370915758708359168 &      282.810539 &       78.033136 &      109.575259 &       26.360347 &   251.000 &   251.000 &   $coadd$ \\
   6370915759782100992 &      278.835416 &       78.547001 &      110.147665 &       27.387182 &   216.000 &   216.000 &   $coadd$ \\
   6370915760855842816 &      273.760205 &       79.023568 &      110.744419 &       28.396397 &   198.000 &   198.000 &   $coadd$ \\
 ... &  ... \\
 6370915708343156736&     257.1537508&      71.986534&     103.457631&      33.558659&  109.000&  109.000& $visit$\\
  ... &  ... 
\enddata
\tablecomments{Table \ref{t_CoaddsVisits} is published in its entirety in the electronic edition.
 A portion is 
shown here for guidance regarding its form and content. In total, 28067 $coadd$s and 1468 $visit$s are used to build GUVcat\_AIS. }
\end{deluxetable}
\newpage
\begin{deluxetable}{lrrrrrrrrrrrrr}
\tabletypesize{\scriptsize}
\rotate
\tablecaption{List of AIS eliminated $bad~coadds$, and their associated visits \label{t_badcoadds}} 
\tablewidth{0pt}
\tablehead{
\colhead{ $Coadd$ ID} & \colhead{ $Coadd$ RA} & \colhead{ $Coadd$ Dec}  & \colhead{ $Coadd$ lon}   & \colhead{$Coadd$ lat}  &  \multicolumn{2}{c}{$Coadd$  exp.time (sec)} 
 &  \colhead{$Visit$ ID }  & \colhead{$Visit$ RA}  & \colhead{ $Visit$ DEC} & \multicolumn{2}{c}{ $Visit$  exp.time}  &  \colhead{ Distance $Coadd$-$Visit$} & \colhead{ in GUVcat?}\\
\colhead{(photoextractID)}   &\colhead{ (degrees)} & \colhead{(degrees)} &  \colhead{(degrees)}   & \colhead{(degrees)}  &  \colhead{FUV }  & \colhead{NUV}  & \colhead{(photoextractID)}   & \colhead{(degrees)}  & \colhead{ (degrees)} & \colhead{ FUV (sec.)}  &\colhead{ NUV (sec.)} &  \colhead{ (arcmin)} &\colhead{ (Y/N)}
} 
\startdata
  6370915845681446912&     257.195457&      71.984578&     103.457631&      33.558659&  109.000&  267.000&   6370915708276047872&     257.490785&      71.946748&    0.000&   62.000& 5.937&N\\
   6370915845681446912&    257.195457&      71.984578&     103.457631&      33.558659&  109.000&  267.000&   6370915708309602304&     257.051516&      72.006587&    0.000&   96.000& 2.978&N\\
   6370915845681446912&    257.195457&      71.984578&     103.457631&      33.558659&  109.000&  267.000&   6370915708343156736&     257.153751&      71.986534&  109.000&  109.000& 0.783&G\\
   6370950948449157120&     250.541348&      70.399218&     102.543859&      36.089296&  171.000&  350.050&   6370950811043758080&     250.893395&      70.436709&    0.000&   75.050& 7.428&N\\
   6370950948449157120&     250.541348&      70.399218&     102.543859&      36.089296&  171.000&  350.050&   6370950811077312512&     250.614435&      70.469918&    0.000&  104.000& 4.489&N\\
   6370950948449157120&     250.541348&      70.399218&     102.543859&      36.089296&  171.000&  350.050&   6370950811110866944&     250.340422&      70.356973&   91.000&   91.000& 4.776&G\\
   6370950948449157120&     250.541348&      70.399218&     102.543859&      36.089296&  171.000&  350.050&   6370950811144421376&     250.346796&      70.319619&   80.000&   80.000& 6.181&G\\
...  & ... & ... & ... & ... & \\
\enddata
\tablecomments{Table \ref{t_badcoadds} is published as online data only. 
Note: the 640 $bad coadds$  were included in BCScat, prior to our discovery of the database improper coadding of non-overlapping $visit$s; \\ they are not used in GUVcat, and their corresponding $visit$s with both FUV and NUV exposure times $>$ 0 are used instead (1468 out of 2004 total).}
\end{deluxetable}

\newpage 

\begin{deluxetable}{lrrrrrrrrlrrrrrrrrrrrr} 
\tabletypesize{\tiny}
\rotate
\tablewidth{0pt}
\tablecaption{Catalog Source Statistics. 
 \label{t_stats}}
\tablehead{ 
 \colhead{latitude}  &  \colhead{\#sources} & \multicolumn{3}{c}{ \#sources with grank } 
 &  \multicolumn{3}{c}{ \% sources with grank}  & \colhead{ area}& \colhead{band}
 &  \multicolumn{10}{c}{ \#sources with artifact =  } \\
   \colhead{range}  &  \colhead{ total} & \colhead{=0}  &    \colhead{=1} &  \colhead{=-1}& \colhead{=0}&\colhead{=1}&\colhead{=-1} & \colhead{(deg$^2$)}
  & & \colhead{none}  & \colhead{1}& \colhead{2}&\colhead{4}& \colhead{8}&\colhead{16}& \colhead{32}&\colhead{64}& \colhead{128}& \colhead{256}&\colhead{512}  
}
\startdata
85\_90N&      249745&      240811&        8934&           0& 96.42 &   3.58 &   0.000&   67.2 & FUV: &      230090&           0&           0&         101&           5&           0&           0&           0&           4&       19549&           0\\
                     &            &            &            &            &      &      &    &    & NUV: &       196416&       43063&        1646&        3147&         738&       13147&          28&           0&        1573&        6777&           0\\
80\_85N&      575444&      554881&       20563&           0& 96.43 &   3.57 &   0.000&  204.4 & FUV: &      540343&           0&           0&         204&          20&           0&           0&           0&          17&       34883&           0\\
                     &            &            &            &            &      &      &    &    & NUV: &       469695&       88308&        4071&        4757&        1759&       25779&           0&           0&        2763&       10800&           0\\
75\_80N&     1044253&     1007254&       36999&           0& 96.46 &   3.54 &   0.000&  341.1 & FUV: &      985099&           0&           0&         303&          77&           0&           0&           0&          36&       58785&           0\\
                     &            &            &            &            &      &      &    &    & NUV: &       831957&      176601&        6871&        9833&        3134&       51158&         127&           0&        4992&       24194&           0\\
70\_75N&     1465351&     1411115&       54229&           7& 96.30 &   3.70 &   0.000&  484.4 & FUV: &     1377058&           0&           0&         499&          67&           0&           4&           0&          21&       87745&           0\\
                     &            &            &            &            &      &      &    &    & NUV: &      1167688&      249353&        9686&       14211&        4157&       74265&          32&           0&        6962&       31158&           0\\
65\_70N&     1703241&     1640819&       62371&          51& 96.34 &   3.66 &   0.003&  569.8 & FUV: &     1593211&           0&           0&         664&          96&           0&          59&           0&          52&      109232&           0\\
                     &            &            &            &            &      &      &    &    & NUV: &      1357300&      291364&       11658&       16035&        4601&       84784&         269&           0&        7579&       34633&           0\\
60\_65N&     2098214&     2020029&       78030&         155& 96.27 &   3.72 &   0.007&  657.5 & FUV: &     1947288&           0&           0&        1069&         153&           0&          60&           0&          45&      149689&           0\\
                     &            &            &            &            &      &      &    &    & NUV: &      1620388&      400773&       16315&       24681&        7068&      118614&         169&           0&       11204&       50461&           0\\
55\_60N&     2593208&     2498594&       94480&         134& 96.35 &   3.64 &   0.005&  792.6 & FUV: &     2411132&           0&           0&        1410&         161&           0&         398&           0&          93&      180186&           0\\
                     &            &            &            &            &      &      &    &    & NUV: &      1972090&      522630&       19991&       31454&        8072&      154025&         513&           0&       13731&       65505&           0\\
50\_55N&     2684044&     2582141&      101731&         172& 96.20 &   3.79 &   0.006&  898.6 & FUV: &     2511586&           0&           0&        1518&         166&           0&         581&           0&         100&      170299&           0\\
                     &            &            &            &            &      &      &    &    & NUV: &      2011991&      578042&       22768&       32662&        9454&      181656&         773&           0&       15296&       58839&           2\\
45\_50N&     3235658&     3109755&      125810&          93& 96.11 &   3.89 &   0.003&  999.1 & FUV: &     3013291&           0&           0&        1841&         193&           0&         154&           0&         124&      220300&           0\\
                     &            &            &            &            &      &      &    &    & NUV: &      2338774&      769016&       28932&       46392&       14503&      241091&         567&           0&       23295&       82319&           9\\
40\_45N&     3621419&     3472606&      148525&         288& 95.89 &   4.10 &   0.008& 1115.3 & FUV: &     3365832&           0&           0&        2546&         267&           0&         949&           0&         196&      252037&           0\\
                     &            &            &            &            &      &      &    &    & NUV: &      2548161&      926429&       34956&       58091&       16539&      295057&        1468&           0&       25753&       93417&          13\\
35\_40N&     3724140&     3573912&      150061&         167& 95.97 &   4.03 &   0.004& 1172.7 & FUV: &     3442159&           0&           0&        3185&         170&           0&         681&           0&         165&      278265&           0\\
                     &            &            &            &            &      &      &    &    & NUV: &      2553402&     1018776&       38696&       64992&       21180&      332445&        1033&           0&       32668&       92204&          26\\
30\_35N&     3887581&     3730652&      156816&         113& 95.96 &   4.03 &   0.003& 1230.8 & FUV: &     3600303&           0&           0&        3701&         265&           0&         377&           0&         329&      283109&           0\\
                     &            &            &            &            &      &      &    &    & NUV: &      2593126&     1133700&       47398&       75458&       22874&      382694&         701&           0&       35456&       89550&         113\\
25\_30N&     3923424&     3759335&      163994&          95& 95.82 &   4.18 &   0.002& 1176.8 & FUV: &     3620962&           0&           0&        5153&         356&           0&         379&           0&         398&      296849&           0\\
                     &            &            &            &            &      &      &    &    & NUV: &      2477770&     1268424&       55885&       92317&       28419&      460009&         584&           0&       44238&       96318&         273\\
20\_25N&     3830858&     3667481&      163192&         185& 95.74 &   4.26 &   0.005& 1105.7 & FUV: &     3517568&           0&           0&        6657&         543&           0&         234&           0&         486&      306230&           0\\
                     &            &            &            &            &      &      &    &    & NUV: &      2327595&     1326992&       64264&      101293&       27587&      522211&         548&           0&       43320&       90249&         174\\
15\_20N&     2813302&     2708785&      104400&         117& 96.28 &   3.71 &   0.004&  975.9 & FUV: &     2568553&           0&           0&        9333&         537&           0&         247&           0&         677&      235196&           0\\
                     &            &            &            &            &      &      &    &    & NUV: &      1543588&     1111672&       73814&      109084&       31465&      546654&         351&           0&       46558&       66419&         325\\
10\_15N&     3417929&     3274835&      142954&         140& 95.81 &   4.18 &   0.004&  786.4 & FUV: &     3134410&           0&           0&        8604&         627&           0&          42&           0&         717&      274696&           0\\
                     &            &            &            &            &      &      &    &    & NUV: &      1977151&     1271433&       69731&      109821&       28675&      553273&         110&           0&       43687&       78262&         343\\
05\_10N&     1129788&     1093910&       35866&          12& 96.82 &   3.17 &   0.001&  363.4 & FUV: &     1034186&           0&           0&        3932&         296&           0&          74&           0&         312&       91513&           0\\
                     &            &            &            &            &      &      &    &    & NUV: &       616241&      445946&       36739&       49952&       11801&      223662&          89&           0&       17880&       25910&         304\\
00\_05N&      137477&      134551&        2926&           0& 97.87 &   2.13 &   0.000&   56.6 & FUV: &      124470&           0&           0&         229&          34&           0&          26&           0&          49&       12719&           0\\
                     &            &            &            &            &      &      &    &    & NUV: &        80621&       48640&        3789&        4975&        1347&       19017&          26&           0&        2034&        3147&          17\\
05\_00S&       98225&       96655&        1570&           0& 98.40 &   1.60 &   0.000&   43.7 & FUV: &       88557&           0&           0&          98&          12&           0&           0&           0&          18&        9558&           0\\
                     &            &            &            &            &      &      &    &    & NUV: &        57906&       34686&        2037&        2752&        1987&       12999&           0&           0&        2522&        2519&           3\\
10\_05S&      576514&      562934&       13580&           0& 97.64 &   2.36 &   0.000&  203.1 & FUV: &      523808&           0&           0&        1989&          94&           0&          12&           0&         130&       50713&           0\\
                     &            &            &            &            &      &      &    &    & NUV: &       321858&      223024&       15458&       22510&        5781&      104650&          12&           0&        8329&       12724&          70\\
15\_10S&     1885871&     1826551&       59315&           5& 96.85 &   3.15 &   0.000&  487.9 & FUV: &     1723937&           0&           0&        6317&         327&           0&         163&           0&         538&      155435&           0\\
                     &            &            &            &            &      &      &    &    & NUV: &      1021945&      765421&       46088&       73137&       15308&      375815&         273&           0&       22996&       44205&         433\\
20\_15S&     2980587&     2873863&      106665&          59& 96.42 &   3.58 &   0.002&  751.3 & FUV: &     2729970&           0&           0&        7385&         442&           0&         331&           0&         626&      242850&           0\\
                     &            &            &            &            &      &      &    &    & NUV: &      1684070&     1154229&       59825&       94801&       20667&      539053&         378&           0&       31628&       70174&         399\\
25\_20S&     3296612&     3159205&      137154&         253& 95.83 &   4.16 &   0.008&  915.2 & FUV: &     3020972&           0&           0&        6150&         348&           0&         590&           0&         506&      268930&           0\\
                     &            &            &            &            &      &      &    &    & NUV: &      1950745&     1194417&       55483&       89115&       22519&      503868&         743&           0&       34918&       77904&         167\\
30\_25S&     3662226&     3502796&      159010&         420& 95.65 &   4.34 &   0.011& 1043.4 & FUV: &     3347437&           0&           0&        5286&         351&           0&         852&           0&         511&      308595&           0\\
                     &            &            &            &            &      &      &    &    & NUV: &      2290154&     1210376&       50534&       88398&       21007&      465364&        1490&           0&       33336&       91010&         162\\
35\_30S&     3872172&     3702613&      169181&         378& 95.62 &   4.37 &   0.010& 1126.6 & FUV: &     3556663&           0&           0&        4797&         358&           0&         911&           0&         454&      309746&           0\\
                     &            &            &            &            &      &      &    &    & NUV: &      2512707&     1190891&       51189&       86347&       22718&      444204&        1173&           0&       35154&       92925&         124\\
40\_35S&     3628761&     3456338&      171585&         838& 95.25 &   4.73 &   0.023& 1115.4 & FUV: &     3328898&           0&           0&        3615&         224&           0&        1383&           0&         262&      294907&           0\\
                     &            &            &            &            &      &      &    &    & NUV: &      2435514&     1034355&       42202&       78427&       17853&      372865&        2093&           0&       28284&       95120&          51\\
45\_40S&     3438189&     3287766&      149958&         465& 95.62 &   4.36 &   0.014& 1058.0 & FUV: &     3144125&           0&           0&        3379&         387&           0&        2256&           0&         265&      288448&           0\\
                     &            &            &            &            &      &      &    &    & NUV: &      2366844&      929025&       37294&       65412&       17132&      338423&        2530&           0&       27296&       83680&          89\\
50\_45S&     3210828&     3054911&      154880&        1037& 95.14 &   4.82 &   0.032&  957.5 & FUV: &     2932586&           0&           0&        2629&         267&           0&        1054&           0&         160&      274610&           0\\
                     &            &            &            &            &      &      &    &    & NUV: &      2252162&      826472&       30788&       58111&       14007&      303229&        1437&           0&       22500&       83051&          18\\
55\_50S&     2936108&     2786459&      147733&        1916& 94.90 &   5.03 &   0.065&  885.6 & FUV: &     2696422&           0&           0&        2172&         205&           0&        2546&           0&         162&      235095&           0\\
                     &            &            &            &            &      &      &    &    & NUV: &      2108224&      709768&       27170&       48545&       14823&      260102&        3318&           0&       22550&       71974&          32\\
60\_55S&     2825703&     2694479&      130287&         937& 95.36 &   4.61 &   0.033&  829.0 & FUV: &     2578445&           0&           0&        1991&         167&           0&         656&           0&         113&      244689&           0\\
                     &            &            &            &            &      &      &    &    & NUV: &      2098459&      616894&       24339&       43856&       12815&      217084&         843&           0&       19843&       64026&          19\\
65\_60S&     2358898&     2260825&       97496&         577& 95.84 &   4.13 &   0.024&  672.8 & FUV: &     2146308&           0&           0&        1754&         169&           0&         101&           0&          46&      210770&           0\\
                     &            &            &            &            &      &      &    &    & NUV: &      1777379&      489657&       18926&       35212&        8410&      167698&         153&           0&       13456&       56966&          31\\
70\_65S&     2181934&     2082443&       99008&         483& 95.44 &   4.54 &   0.022&  620.9 & FUV: &     1987819&           0&           0&        1263&         127&           0&         160&           0&          76&      192733&           0\\
                     &            &            &            &            &      &      &    &    & NUV: &      1655414&      439389&       17260&       31733&        8414&      151401&         192&           0&       13195&       55032&          15\\
75\_70S&     1695985&     1624317&       71484&         184& 95.77 &   4.21 &   0.011&  471.7 & FUV: &     1539154&           0&           0&         979&          79&           0&         136&           0&          59&      155726&           0\\
                     &            &            &            &            &      &      &    &    & NUV: &      1292244&      336714&       12705&       24818&        5251&      115650&         249&           0&        8421&       44232&           3\\
80\_75S&     1169851&     1111405&       57573&         873& 95.00 &   4.92 &   0.075&  333.0 & FUV: &     1060612&           0&           0&         651&          41&           0&         223&           0&          48&      108422&           0\\
                     &            &            &            &            &      &      &    &    & NUV: &       903018&      221460&        8616&       16997&        3331&       78678&         225&           0&        5719&       29585&           4\\
85\_80S&      811212&      767146&       43119&         947& 94.57 &   5.32 &   0.117&  211.5 & FUV: &      728465&           0&           0&         540&          31&           0&         198&           0&           7&       82066&           0\\
                     &            &            &            &            &      &      &    &    & NUV: &       623816&      151325&        5570&       13579&        3201&       51800&         454&           0&        4989&       23363&          18\\
90\_85S&      227334&      217689&        9574&          71& 95.76 &   4.21 &   0.031&   66.4 & FUV: &      207506&           0&           0&         111&           3&           0&           0&           0&           3&       19720&           0\\
                     &            &            &            &            &      &      &    &    & NUV: &       178466&       39773&        1681&        3415&         509&       13344&          25&           0&         902&        5436&           0\\
Total  &    82992086&    79549861&     3431053&       11172& 95.85 &   4.13 &   0.01 & 24790.3&   FUV:  &76359225&           0&           0&         111&           3&           0&           0&           0&           3&       19720&           0\\
                     &            &            &            &            &       &       &  &     &NUV:  &   56214879&       39773&        1681&        3415&         509&       13344&          25&           0&         902&        5436&           0\\
\enddata
\tablecomments{GALEX artifact flags :\\
 Artifact~~1(~~1):(edge) Detector bevel edge reflection (NUV only).\\
 Artifact~~2(~~2):(window) Detector window reflection (NUV only).\\
 Artifact~~3(~~4):(dichroic) Dichroic reflection.\\
 Artifact~~4(~~8):(varpix) Variable pixel based on time slices.\\
 Artifact~~5(~16):(brtedge) Bright star near field edge (NUV only).\\
 Artifact~~6(~32):Detector rim (annulus) proximity ($>$0.6 deg from field center).\\
 Artifact~~7(~64):(dimask) dichroic reflection artifact mask flag.\\
 Artifact~~8(128):(varmask) Masked pixel determined by varpix.\\
 Artifact~~9(256):(hotmask) Detector hot spots.\\
 Artifact~10(512):(yaghost) Possible ghost image from YA slope.\\
}
\end{deluxetable}
\newpage
\begin{deluxetable}{crrrrrrrrrrr} %
\tabletypesize{\scriptsize}
\rotate
\tablewidth{0pt}
\tablecaption{Catalog Source Statistics. UV colors   \label{t_statscolors}}
\tablehead{ 
 \colhead{latitude}  &  \colhead{\#sources} & \colhead{density} & \multicolumn{1}{c}{ \# FUV } & \colhead{fraction     }& \colhead{\#sources}& \colhead{\#sources}& \colhead{\#sources}& \colhead{\#sources} &  \colhead{\#sources} &  \colhead{\#sources}\\
                     &                      &  \colhead{(\# /deg$^2$)} &                       & \colhead{(\#FUV/\#NUV)}& \colhead{FUV-NUV$\leq$0}& \colhead{FUV-NUV$\leq$0.5}& \colhead{in galaxies}& \colhead{in clusters}  &  \colhead{not MC}&  \colhead{$\leq$15\grado from LMC}&  \colhead{$\leq$10\grado from SMC}
}
\startdata
85\_90N&      249745&3716.58&       36434&   0.15&       10751&       24240&         316&        8643&      249745&           0&
           0\\
80\_85N&      575444&2814.77&       86642&   0.15&       23838&       56095&         375&       21553&      575444&           0&
           0\\
75\_80N&     1044253&3061.59&      130726&   0.13&       38373&       85756&        1386&        2607&     1044253&           0&
           0\\
70\_75N&     1465351&3024.87&      195794&   0.13&       56013&      127660&        2175&         305&     1465351&           0&
           0\\
65\_70N&     1703241&2988.98&      238919&   0.14&       66078&      154227&        1351&         280&     1703241&           0&
           0\\
60\_65N&     2098214&3191.04&      292232&   0.14&       81716&      188597&        1347&           0&     2098214&           0&
           0\\
55\_60N&     2593208&3271.72&      337085&   0.13&       96573&      218088&        2233&           0&     2593208&           0&
           0\\
50\_55N&     2684044&2986.87&      314592&   0.12&       91570&      203731&        1425&           0&     2684044&           0&
           0\\
45\_50N&     3235658&3238.55&      359081&   0.11&      106576&      232645&        1409&         976&     3235658&           0&
           0\\
40\_45N&     3621419&3247.02&      362529&   0.10&      109455&      233388&        2258&         708&     3621419&           0&
           0\\
35\_40N&     3724140&3175.61&      341953&   0.09&      100999&      215049&        1601&        1363&     3724140&           0&
           0\\
30\_35N&     3887581&3158.53&      321160&   0.08&       93943&      196661&        2799&       19836&     3887581&           0&
           0\\
25\_30N&     3923424&3334.07&      290238&   0.07&       84044&      171420&        1621&        1876&     3923424&           0&
           0\\
20\_25N&     3830858&3464.68&      251682&   0.07&       69040&      139172&        1398&        4614&     3830858&           0&
           0\\
15\_20N&     2813302&2882.87&      126863&   0.05&       29719&       49576&         854&        7159&     2813302&           0&
           0\\
10\_15N&     3417929&4346.37&      185024&   0.05&       48188&       90471&        2078&       15187&     3417929&           0&
           0\\
05\_10N&     1129788&3109.13&       53793&   0.05&        9270&       15451&         147&        3323&     1129788&           0&
           0\\
00\_05N&      137477&2429.08&       11771&   0.09&        1199&        3619&           0&        1473&      137477&           0&
           0\\
05\_00S&       98225&2246.47&        6756&   0.07&         681&        1762&           0&         629&       98225&           0&
           0\\
10\_05S&      576514&2838.62&       25939&   0.04&        5375&        8977&          33&        2630&      576514&           0&
           0\\
15\_10S&     1885871&3865.60&       73420&   0.04&       20202&       30698&       48693&        2576&     1885871&           0&
           0\\
20\_15S&     2980587&3967.39&      127231&   0.04&       38069&       63299&       74678&       24730&     2954498&       26089&
           0\\
25\_20S&     3296612&3602.23&      170977&   0.05&       53134&       95767&        5211&        9308&     3130822&      165790&
           0\\
30\_25S&     3662226&3509.82&      248345&   0.07&       73965&      148739&       86561&        5099&     3326385&      335841&
           0\\
35\_30S&     3872172&3436.90&      342725&   0.09&       95991&      210826&      217832&        7626&     3391602&      472916&
        7688\\
40\_35S&     3628761&3253.33&      333036&   0.09&       99801&      213257&       78466&        1054&     3150549&      337319&
      181294\\
45\_40S&     3438189&3249.72&      379787&   0.11&      111606&      248892&       47417&        4920&     2842873&      303802&
      355523\\
50\_45S&     3210828&3353.38&      376489&   0.12&      109139&      244447&       27606&        3436&     2831925&       67392&
      312887\\
55\_50S&     2936108&3315.46&      369087&   0.13&      105669&      238995&        1637&       12324&     2797762&           0&
      138346\\
60\_55S&     2825703&3408.57&      433883&   0.15&      117038&      277066&        1713&       33317&     2825703&           0&
           0\\
65\_60S&     2358898&3506.30&      385773&   0.16&      101852&      245762&        1833&          82&     2358898&           0&
           0\\
70\_65S&     2181934&3514.37&      354422&   0.16&       94703&      226806&        1163&         993&     2181934&           0&
           0\\
75\_70S&     1695985&3595.28&      291753&   0.17&       76879&      186544&         658&           0&     1695985&           0&
           0\\
80\_75S&     1169851&3512.67&      208792&   0.18&       54909&      133938&         694&        1728&     1169851&           0&
           0\\
85\_80S&      811212&3835.03&      141506&   0.17&       36981&       90073&         321&        1240&      811212&           0&
           0\\
90\_85S&      227334&3423.16&       38041&   0.17&       10449&       24905&         162&         582&      227334&           0&
           0\\
Total: &    82992088&1673.85&     8244480&   0.10&     2323788&     5096599&      619451&      202177&    80393016&     1709149&      995738\\
\enddata
\end{deluxetable}

\clearpage
\appendix

\section{Appendix A. Criteria for Identifying and  Removing Duplicate Measurements} 
\label{s_remdup}
The  GALEX database
in the MAST archive  contains all existing measurements. For sources with repeated AIS observations (e.g. where different fields overlap, or the same field was repeated), we removed duplicate measurements as follows, to produce a unique source catalog. GALEX sources {\it within 2.5\as of each other but from different  observations} were considered duplicates. In such cases, the measurement from the observation with the longest exposure time (sum of FUV and NUV exposures) was retained, and - in cases of equal exposure time - the one closest to the center of the field  in its parent image. 

 The choice of a 2.5\as match radius was based on several considerations.   According to the archive documentation, the accuracy of GALEX source positions  is of 0.32/0.34\as (NUV/FUV)  in the GR7 data\footnote{the current GALEX database is called ``GR6plus7'' because not all data have been reprocessed yet, the latest version of the pipeline was used only for the GR7 addition} and slightly worse, especially in FUV, in previous data. 
 The GALEX pipeline uses a complicated probability algorithm to match FUV sources to the NUV sources of the same observation. In short, NUV and FUV matches are allowed up to 7\as.  We have examined statistically the tags informative of the FUVxNUV match process. Fig. \ref{f_fuvnuvsep}  shows the distribution of the 
 separation between  FUV and NUV position for a 2million source subsample of the database, as a whole and divided by cuts in match $probability$ (as defined by the pipeline algorithm).  
Our choice of a 2.5\as match radius to define duplicates corresponds statistically to a FUVxNUV match $probability$ $>$0.3, and is consistent with 
 the early versions of our catalogs (\citet{bia11a, bia14uvsky}), where it was found from other tests to be a good compromise between not excluding real sources 
and not retaining duplicate measurements. GALEX astrometry is  more accurate than 2.5\as, but the deblending of sources closer than 
this separation is not always robust due to the instrument resolution ($\approx$4.2/5.3\as, FUV/NUV). This was discussed in Section \ref{s_crowd}.  In general, we expect most science applications of this catalog to be restricted to point sources, for which the chosen limit is appropriate. 

\begin{figure*}
\centerline{
\includegraphics[width=15.cm]{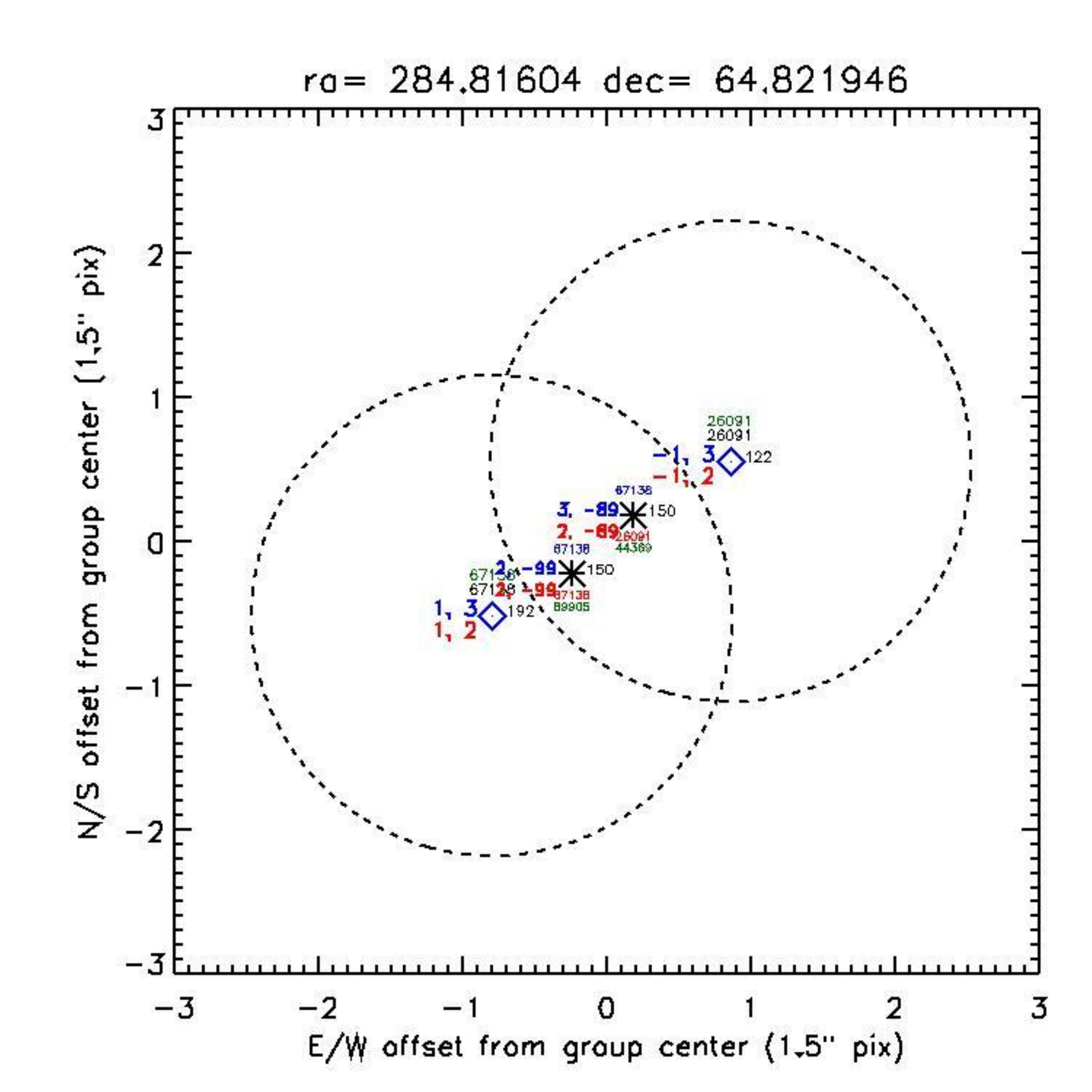}}
\vskip -.62cm 
\caption{\label{f_grank1} Examples of multiple observations for the same source.  The source at the center of the lower-left circle (blue diamond, id=67138) has the best measurement out of three within 2.5\as from its position (dashed circle); therefore it is assigned $grank$=1, $ngrank$=3 (blue numbers  at the left). Its closest neighbor %
has $grank$=2, $ngrank$=-99 (because it is not a primary). The second closest, with id = 44309, has $grank$=3, but  $ngrank$=-89 because it also falls within 2.5\as of another source (id=26091) which is further than 2.5\as from the first primary and has exposure time shorter than source 44309. Object id=26091 is therefore assigned $grank$=-1, because it cannot be discarded as duplicate of the first primary, but has a nearby source which has a better measurement but cannot be ``primary'' because it is secondary with respect to a better primary.  The black numbers to the right of the sources are exposure time in seconds. The $grank$ tag is used to eliminate secondaries in the unique-source catalog.  The red numbers show the values of the same tags if we used instead a distance criterion and associate secondaries to the closest primary rather than to the ``best'' primary.  In that case, each of these two primaries would get one secondary. Note that  $grank$ for primaries would not change.}
\end{figure*}

 While the criterion is simple in principle, we add here some important clarifications which were not described in the previous versions of the catalog, and are relevant for any work requiring merging of overlapping observations. To identify duplicates, and eventually remove them, we search the master catalog around the position of each source, within the chosen match radius. If there is no other entry within the match radius of 2.5\as, we assign to the source ``$grank$=0''. Thus, all sources with $grank=0$ are unique (i.e., have only one measurement) also in the original AIS database.\footnote{we recall that GUVcat\_AIS only includes the AIS exposures, for homogeneity of exposure depth across the catalog. Some regions were observed repeatedly with deeper exposures, see \citet{bia14}, therefore some  AIS sources may have additional observations in other surveys, with longer exposure times. 
A unique-source catalog at MIS-depth was published by \citet{bia14uvsky}. Deeper exposures will be addressed in a future work.} %
 If  within the match radius around a source ``i'' we find other sources, measured in a different observation, we assign $grank$=1 to the best measurement of this group, 
the ``primary'' (which will be retained in the final catalog when duplicates are removed); the best measurement is the one with longest exposure, or - for equal exposure - closer to the field center in its parent observation;  we assign $grank$=2,3,... to other sources within 2.5\as of the primary, ranked in order of distance from the primary. %
 To keep track of  %
duplicate measurements, since only the primary is retained in the end, we added 
 a tag $ngrank$,  indicating the number of matches to the primary (including the primary itself), 
and $primgid$, the identifier of the primary (the source  with $grank$=1) to which the sources  with $grank$$>$1 are associated. 
This basic definition is simple. 
However, there  may happen to be 
 sources - let's say, a source "j" - farther than 2.5\as from the  primary "i", therefore not included in its group, 
 but closer than 2.5\as to  a source with $grank$$>$1  in the  group of the primary ``i''. If the source previously assigned  $grank$$>$1 (with respect to source "i"), has better exposure time than its neighbor "j", 
its $grank$ cannot be reclassified to =1 because it does not satisfy the primary  criterion with respect to the (better) primary "i". 
Therefore,  the new source "j" must be retained in the catalog because it's farther than 2.5\as from  "i", 
but we set its $grank=-1$ (instead of =1), to  indicate that another source within 2.5\as would have been a primary with respect to "j", according to our "best measurement" criteria, if it were not a secondary with respect to another, better primary.  The $grank$$>$1 neighbor in our example is given $ngrank$=-89, so it can be identified in the master catalog as a potential primary (with respect to source "j") which could not be retained in the unique-source catalog because it was a secondary with respect to source "i".  If, instead, source "j" has longer exposure than its neighbor with previous $grank$$>$1 
but shorter exposure than source "i", for the  source with  $grank$$>$1 
which is a secondary associated to "i" (its $primgid$ tag indicates the $objid$ of its primary "i"), we still want to retain the information that there is another source ("j") within the match radius from it.  This information is given in tag $groupgid$, where all $objid$'s of sources within the match radius are concatenated. Also, tag $ngrank$ for source "j"  indicates the number of all sources within the match radius (including the $grank$$>$1 secondary associated to source "i") but fewer secondaries than its $ngrank$ will have $primgid$ equal to $objid$ of source "j".  This is easier to understand from some  examples, shown in Figure \ref{f_grank1}. 
Such variety of cases may seem  irrelevant subtleties for users of the final catalog, but is worth mentioning; in fact, any code performing associations of repeated measurements must include provisions for such cases, and more odd (and rare) situations, otherwise more sources will be eliminated  than it is necessary, or wrong associations will result.  A code simply performing rank assignment looking for neighbors of each source sequentially, and not accounting for intersecting groups, would eliminate duplicates inconsistently among the sample. 

 With these -or any other - criteria to define duplicates, there may be sources which are within the match radius of more than one primary, the primaries being more distant than 2.5\as from each other. The assigned primary to each secondary, according to our standard recipe, is the one with the longest exposure time (best measurement) as explained above.  For completeness, we also include in the master catalog the tags $grankdist$, and $primgiddist$, the latter indicates the closest primary to the source, and $grankdist$  is its ranking with respect to the closest primary. This may be different from the "best-measurement" primary ($primgid$). With the distance-criterion tags, a secondary may be reassigned from the original primary (best measurement, $primgid$) to the closest primary ($primgiddist$) and therefore the number of secondaries for each primary may differ from $ngrank$; we record this number in the tag    $nkgrank$. These details only concern users who wish to delve in the master catalog GUVcat\_AISplus, where we include all AIS measurements from the archive, and create these tags so that one can chose the primary sources only ($grank$=0, 1 or -1), i.e. removing all duplicates  at once, and obtain a catalog where each source is counted only once, or viceversa examine repeated measurements of AIS sources. GUVcat\_AISplus is also available from MAST's casjobs. 
 
 However,  for most purposes only the primary sources are needed, and it is not convenient for a user to download all measurements and having to apply cuts later using our $grank$ tags described above. GUVcat\_AIS contains only 'unique sources', with duplicate measurements removed. This is extracted from the master catalog GUVcat\_AISplus by retaining only sources with grank=0, 1 or -1. 

\renewcommand{\thefigure}{\arabic{figure} (Cont.)}
\addtocounter{figure}{-1}
\begin{figure*}
\centerline{
\includegraphics[width=15.cm]{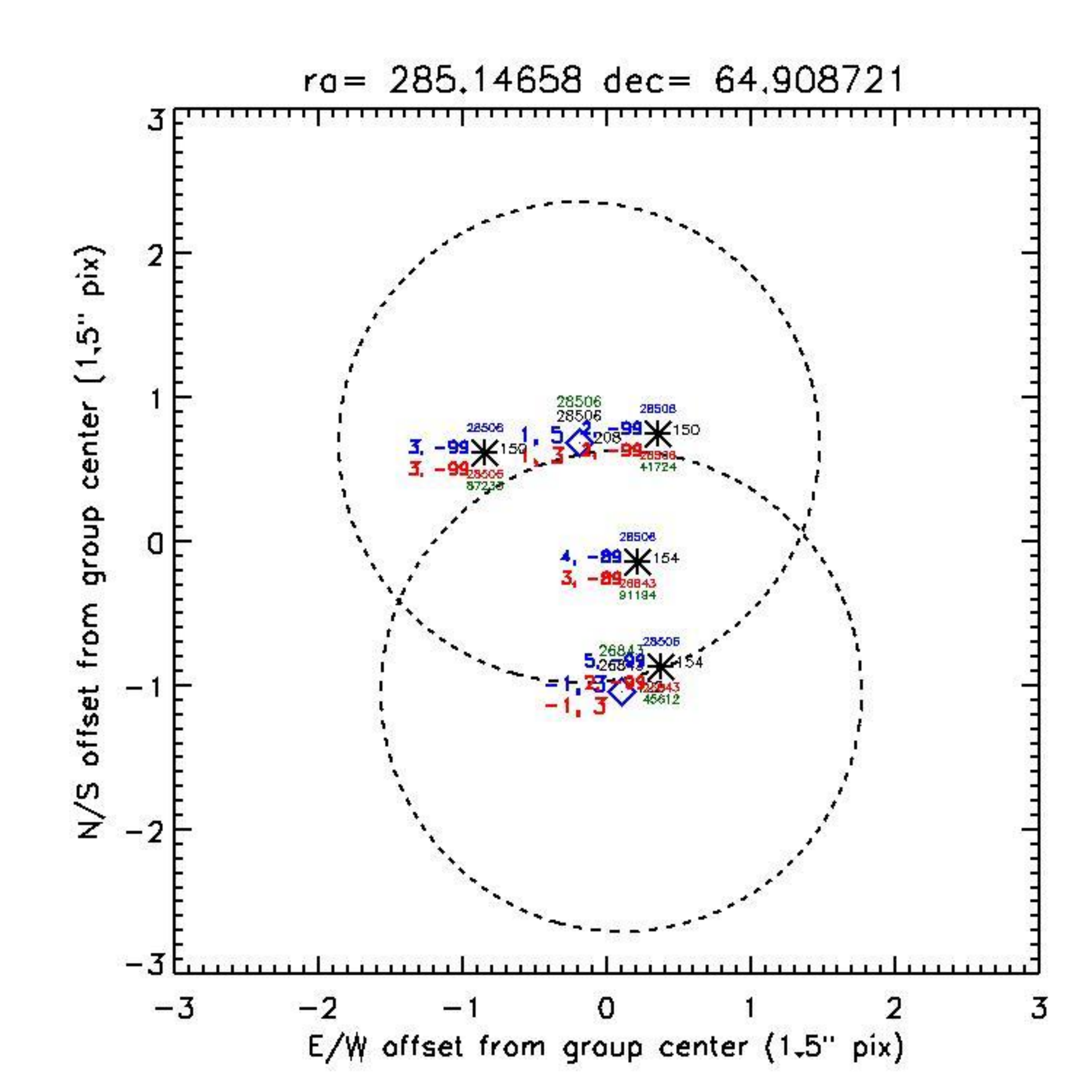}}
\vskip -.62cm 
\caption{Another example illustrating the definition of multiple observations for the same source, showing a different combination; the tag coding is the same as in Figure \ref{f_grank1}-a. }
\end{figure*}
\renewcommand{\thefigure}{\arabic{figure}}

\newpage

\section{Appendix B. Description of The Catalogs' Columns}
\label{s_tags}
Below we list the tags included in the online catalogs presented in this paper, and 
available at \url{http://dolomiti.pha.jhu.edu/uvsky/\#GUVcat}  , as well as from MAST casjobs  and SIMBAD/Vizier.
 The columns of greatest interest in most cases are in bold in the Table below. 
The first sets of tags are propagated from the pipeline database, and give information on the source photometry;  tags $CORV$ and beyond, indicated in italics,  are generated by us and described in this paper; some indicate whether the source has duplicate (AIS) measurements, that have been removed (Section \ref{s_remdup}, or flagged if one uses the 'plus' catalog). The last two tags indicate whether the source is in the footprint of a large ($>$1\am) object (Section \ref{s_inextobj}). 
\begin{deluxetable}{ll} 
\tabletypesize{\scriptsize}
\tablecaption{Catalog Columns \label{t_tags} }
\tablehead{ 
\colhead{ Tag } & \colhead{ Description} }
\startdata
photoextractid   &  Pointer to photoExtract Table (identifier of original observation \\  & on which the measurement was taken)\\
mpstype	   & which survey  (e.g, "MIS", or ``AIS'', ...)\\
avaspra	   & R.A. of center of field where object was measured\\
avaspdec   & Decl. of center of field where object was measured\\	
{\bf objid}	   & GALEX identifier for the source\\
{\bf ra	  } & source's  Right Ascension (degrees).\\
{\bf dec }	   & source's Declination  (degrees)\\
{\bf glon }   & source's Galactic longitude  (degrees)\\
{\bf glat }	   & source's Galactic latitude (degrees)\\
tilenum	   & ``tile'' number \\
img	   & image number (exposure \# for \_$visit$s) \\
subvisit   & number of subvisit if exposure was divided\\	
fov\_radius & distance of source from center of the field in which it was measured\\	
type	   & Obs.type (0=single,1=multi)\\
band	   & Band number (1=nuv,2=fuv,3=both)\\
{\bf e\_bv	 }  & E(B-V) Galactic Reddening (from Schlegel et al. 1998 maps)\\
istherespectrum	   & Does this object have a (GALEX) spectrum? Yes (1), No (0) \\
chkobj\_type	   & Astrometry check type\\
{\bf fuv\_mag	}  & FUV calibrated magnitude\\
{\bf fuv\_magerr	 } & FUV calibrated magnitude error \\
{\bf nuv\_mag	 } & NUV calibrated magnitude\\
{\bf nuv\_magerr	}  & FUV calibrated magnitude error \\
fuv\_mag\_auto	   & FUV Kron-like elliptical aperture magnitude\\
fuv\_magerr\_auto	   & FUV RMS error for AUTO magnitude\\
nuv\_mag\_auto	   & NUV Kron-like elliptical aperture magnitude\\
nuv\_magerr\_auto	   & NUV RMS error for AUTO magnitude\\	
fuv\_mag\_aper\_4	   & FUV Magnitude aperture ( 8 pxl ) \\
fuv\_magerr\_aper\_4  & FUV Magnitude aperture error ( 8 pxl ) \\	
nuv\_mag\_aper\_4	   & NUV Magnitude aperture ( 8 pxl )\\
nuv\_magerr\_aper\_4   &NUV  Magnitude aperture ( 8 pxl ) error\\	
fuv\_mag\_aper\_6	   & FUV  Magnitude aperture ( 17 pxl )\\
fuv\_magerr\_aper\_6  & FUV  Magnitude aperture ( 17 pxl ) error \\	
nuv\_mag\_aper\_6	   & NUV Magnitude aperture ( 17 pxl )\\
nuv\_magerr\_aper\_6  & NUV Magnitude aperture ( 17 pxl ) error\\
{\bf fuv\_artifact }	   & FUV artifact flag (logical OR near source)\\
{\bf nuv\_artifact}	   & NUV artifact flag (logical OR near source)\\
fuv\_flags	   & Extraction flags\\
nuv\_flags	   & Extraction flags\\
fuv\_flux	   & FUV calibrated flux (micro Jansky)\\
fuv\_fluxerr	   & FUV calibrated flux (micro Jansky) error \\
nuv\_flux	   & NUV calibrated flux (micro Jansky)\\
nuv\_fluxerr	   & NUV calibrated flux (micro Jansky) error \\
fuv\_x\_image	   & Object position along x\\
fuv\_y\_image	        & Object position along y\\
nuv\_x\_image   	& Object position along x\\
nuv\_y\_image   	& Object position along y\\
fuv\_fwhm\_image	& FUV FWHM assuming a gaussian core\\
nuv\_fwhm\_image	& NUV FWHM assuming a gaussian core\\
fuv\_fwhm\_world	& FUV FWHM assuming a gaussian core (WORLD units) \\
nuv\_fwhm\_world	& NUV FWHM assuming a gaussian core (WORLD units)\\
nuv\_class\_star	& S/G classifier output \\
fuv\_class\_star	& S/G classifier output \\
{\bf nuv\_ellipticity}	& 1 - B\_IMAGE/A\_IMAGE\\
{\bf fuv\_ellipticity}	& 1 - B\_IMAGE/A\_IMAGE\\
nuv\_theta\_J2000	& Position angle (east of north) (J2000) \\
nuv\_errtheta\_J2000    & Position angle error (east of north) (J2000)\\	
fuv\_theta\_J2000	 & Position angle (east of north) (J2000)\\
fuv\_errtheta\_J2000    & Position angle error (east of north) (J2000)\\	
fuv\_ncat\_fwhm\_image & FUV FWHM\_IMAGE value from -fd-ncat.fits (px)\\	
fuv\_ncat\_flux\_radius\_3	& FUV FLUX\_RADIUS \#3 (-fd-ncat)(px)[0.80]\\
nuv\_kron\_radius& Kron apertures in units of A or B\\	
nuv\_a\_world & Profile RMS along major axis (world units)\\
fuv\_kron\_radius& Kron apertures in units of A or B\\	
fuv\_b\_world & Profile RMS along major axis (world units)\\
nuv\_weight  & NUV effective exposure (flat-field response value) in seconds at the source position (center pixel) given alpha\_j2000, delta\_j2000 \\
fuv\_weight  & FUV effective exposure \\
prob         & probability of the FUV x NUV match \\
sep          & separation between FUV and NUV position of the source  in the same observation \\
nuv\_poserr  & [arcseconds] position error of the source in the NUV image\\
fuv\_poserr  & [arcseconds] position error of the source in the FUV image \\
IB\_POSERR  & [arcseconds] inter-band position error in arcseconds \\
NUV\_PPERR  & [arcseconds] NUV Poisson position error (the part of the position error due to counting statistics) \\
FUV\_PPERR  &  [arcseconds] FUV Poisson position error (the part of the position error due to counting statistics)\\
{\it  CORV}  & whether the source comes from a $Coadd$ or $Visit$\\
{\it GRANK}   & grank=0  if the are no other sources (from different observations) within 2.5\as \\
              & grank=1 if this is the best (see text) source of $>$1 sources  within 2.5\as \\
              & grank=-1 if this is a primary but has a better source within  2.5\as \\
              & grank =n (n$>$1) is this is the n$^{th}$ source   within  2.5\as of the primary \\
{\it NGRANK} & if this is a primary, number of sources within 2.5\as (otherwise, 99 or 89, see text)\\
{\it PRIMGID} & objid of the primary (only of use for the 'plus' catalog)  \\
{\it GROUPGID} & objid's of all sources (AIS) within 2.5\as, concatenated by ``+''\\
{\it GRANKDIST} & as for grank, but based on distance criterion \\
{\it NGRANKDIST} &  as for ngrank, but based on distance criterion \\
{\it PRIMGIDDIST} & as for primgid, but based on distance criterion (objid of the closest primary \\ & rather than the best primary)(only of use for the 'plus' catalog)  \\
{\it GROUPGIDDIST} &  as GROUPGID,  but based on distance criterion \\
{\it GROUPGIDTOT} & objid's of all sources within 2.5\as \\
{\it DIFFFUV} & mag difference between primary and secondary (only of use for the 'plus' catalog) \\
{\it DIFFNUV} &  mag difference between primary and secondary (only of use for the 'plus' catalog)  \\
{\it DIFFFUVDIST} &  mag difference between closest primary and secondary (only of use for the 'plus' catalog)  \\
{\it DIFFNUVDIST} &   mag difference between  closest and secondary (only of use for the 'plus' catalog) \\
{\it SEPAS} & separation (arcsec) between primary and secondary  \\
{\it  SEPASDIST} &  separation (arcsec) between primary (distance criterion) and secondary  \\
{\it {\bf INLARGEOBJ}} &  is the source in the footprint of an extended object?  if not, INLARGEOBJ=N\\
                 & if yes, INLARGEOBJ= XX:name-of-the-extended-object ; where XX=GA (galaxy), \\ &  GC (globular cluster),  OC (open cluster), SC (other stellar clusters)\\
{\it {\bf LARGEOBJSIZE} } & size of the extended object;  LARGEOBJSIZE = 0. if INLARGEOBJ=N, \\ & otherwise LARGEOBJSIZE= D25 for galaxies and 2xR$_1$ for stellar clusters  \\
\enddata
\end{deluxetable}

\newpage
\section{Appendix C. Odd Fields and Artifacts} 
\label{s_oddfields}

In Section \ref{s_artifacts} we mentioned the different artifacts flagged by the GALEX source extraction pipeline, and  Table \ref{t_stats} gives the statistics of sources with artifact flags.  Fig. \ref{f_oddfields}  showd examples of ghost reflections from bright sources,  in FUV and NUV, and of other types of artifacts. 
We use as example one of the fields where there is also an apparent mismatch in coordinates between FUV and NUV detections (Section \ref{s_criteria}).
 
The definition of artifacts can be found in the documentation\footnote{http://www.galex.caltech.edu/wiki/Public:Documentation/Chapter\_8\#Artifact\_Flags}
and is reported in the footnote of Table \ref{t_stats}.

\begin{figure*}
\centerline{
\includegraphics[height=6.cm]{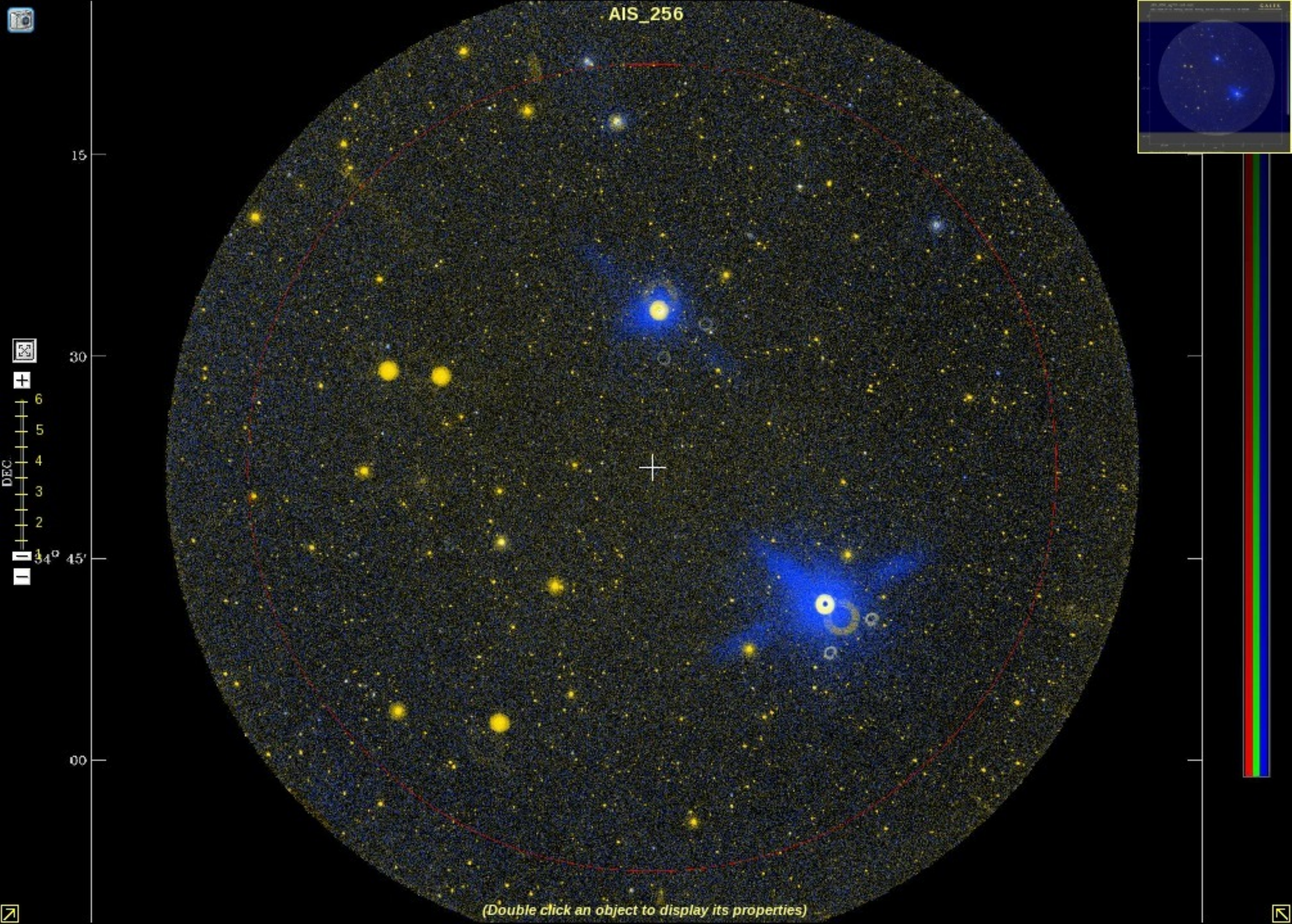}
\includegraphics[height=6.cm]{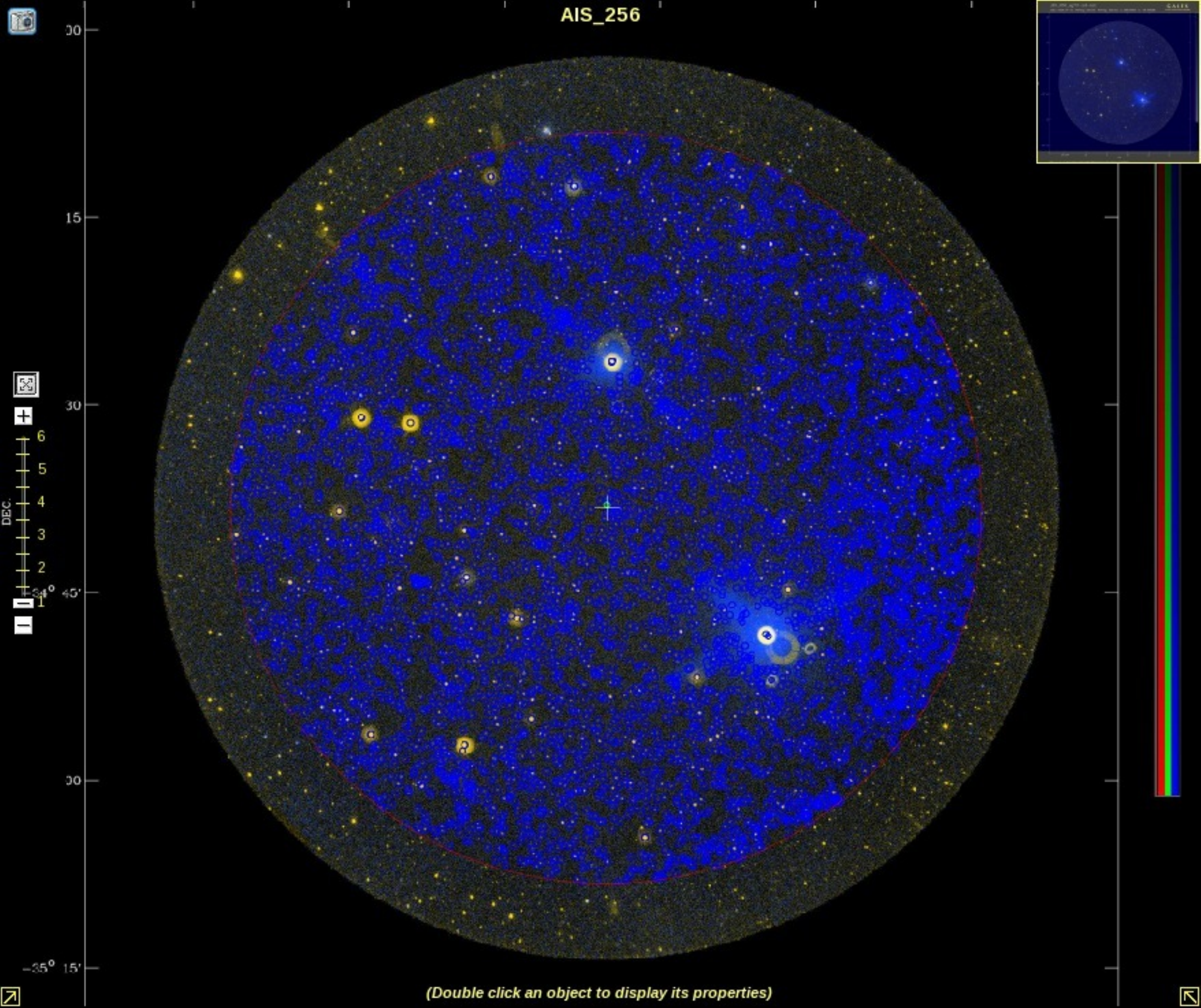}}
\vskip0.2cm
\centerline{
\includegraphics[width=6.5cm]{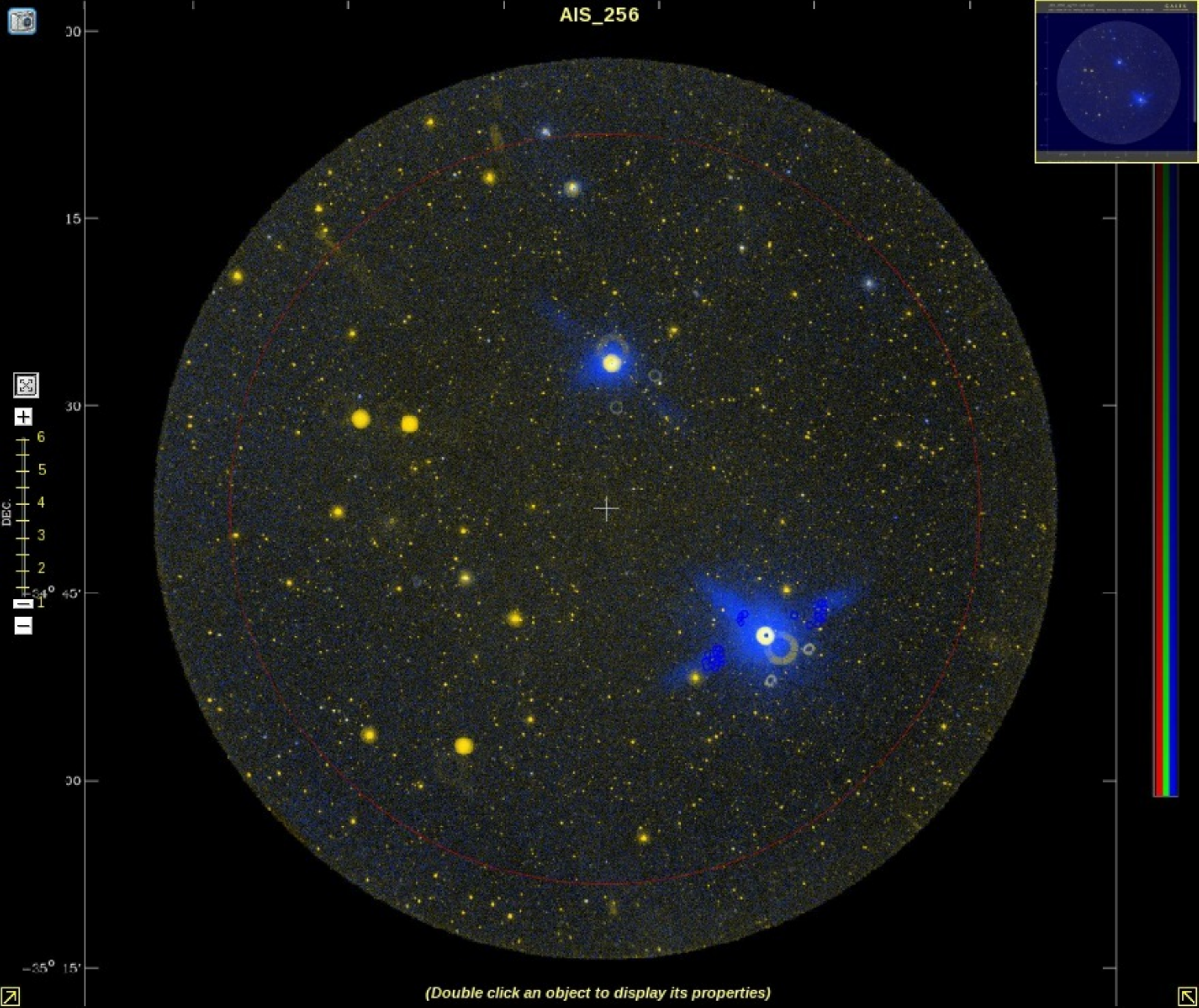}
\includegraphics[width=6.5cm]{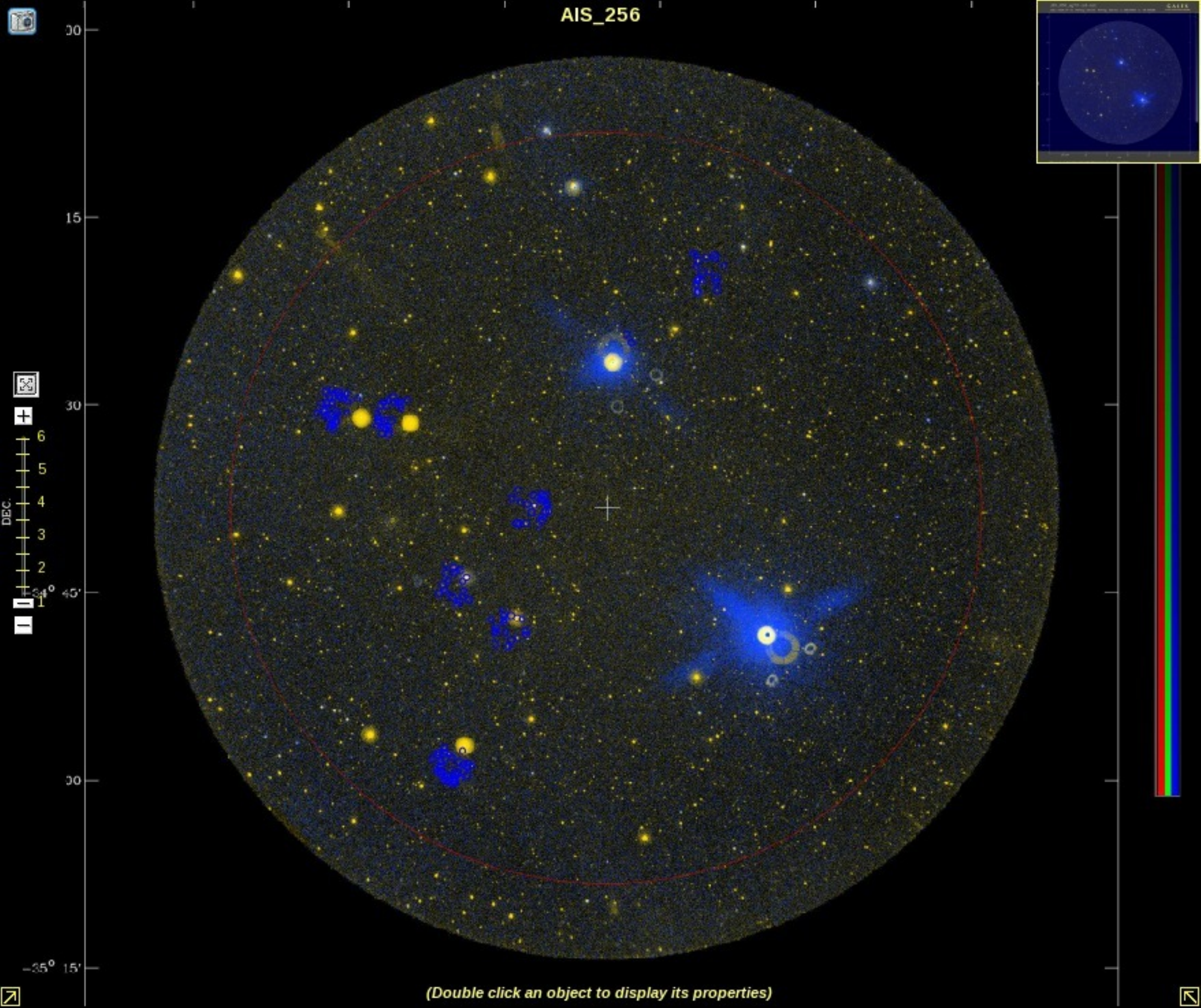}
}
\caption{\label{f_oddfields} GALEX Field AIS\_256: the top panels show the entire field with and without pipeline sources  overlaid (blue circles, all detections in the central 1\grado), combining the FUV (blue) and NUV (yellow) images. The red circle indicates a diameter of 1\grado.  Ghosts from very bright sources are clearly visible as yellow rings offset from the source position for NUV, and as extended streaks in FUV.  The bottom panels show only the sources with FUV\_artifact=128 (left, ghosts from the bright sources), and with NUV\_artifact=2 (right, ghost rings around bright sources; note that the sources are always plotted as blue circles. 
}
\end{figure*}

\renewcommand{\thefigure}{\arabic{figure} (Cont.)}
\addtocounter{figure}{-1}
\begin{figure*}
\centerline{
\includegraphics[width=6.cm]{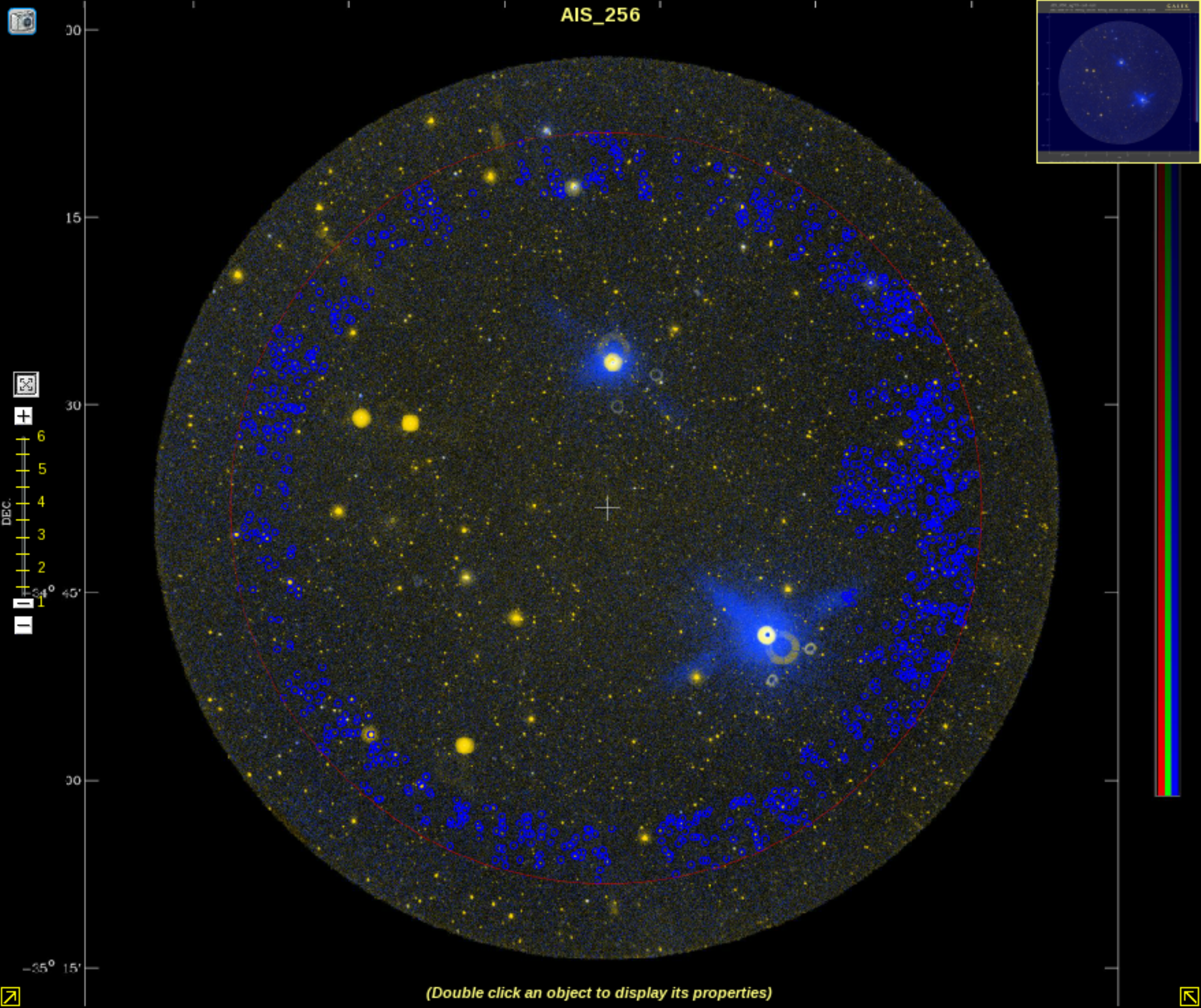}
\includegraphics[width=6.cm]{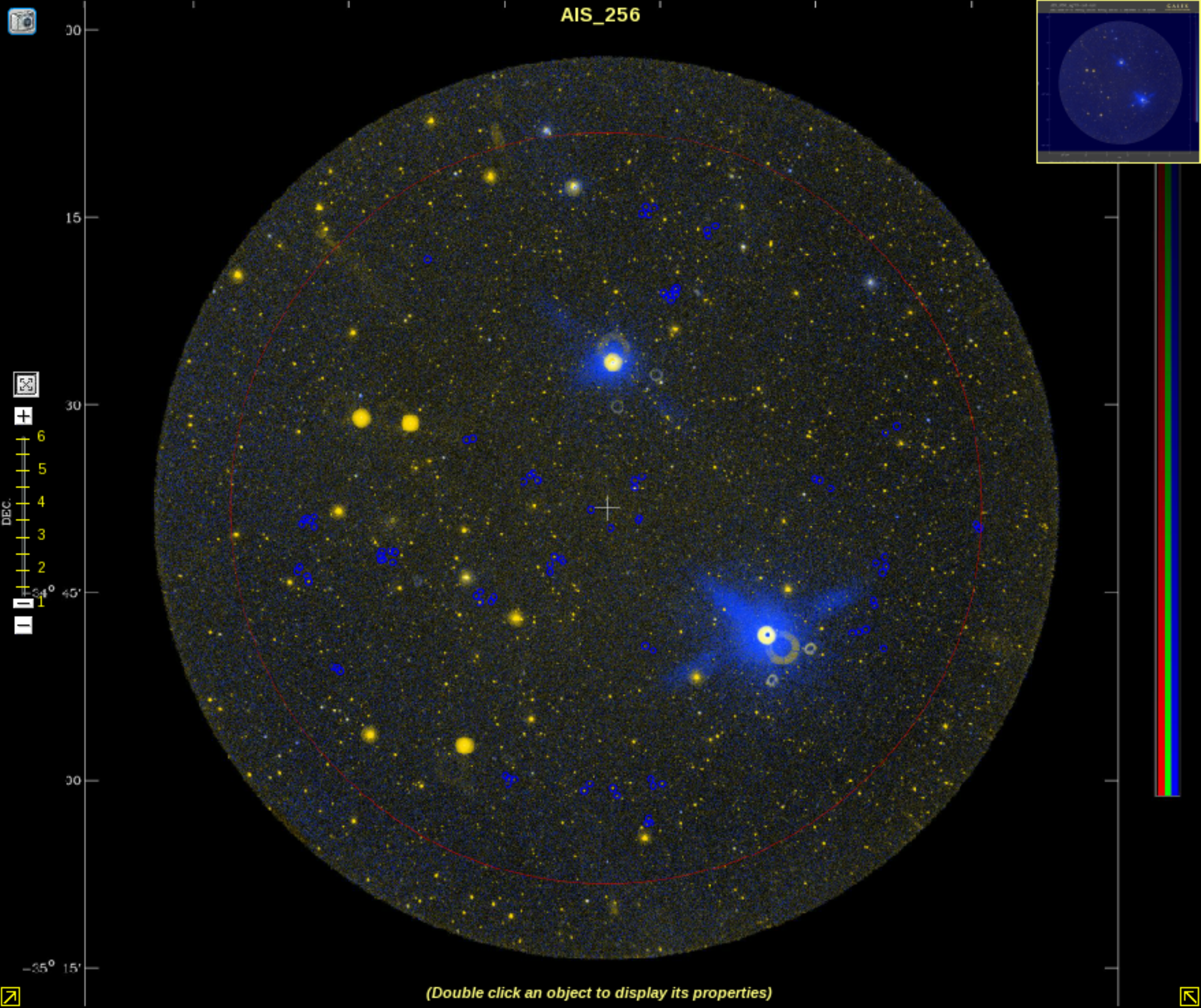}}
\vskip0.2cm
\centerline{
\includegraphics[width=6.cm]{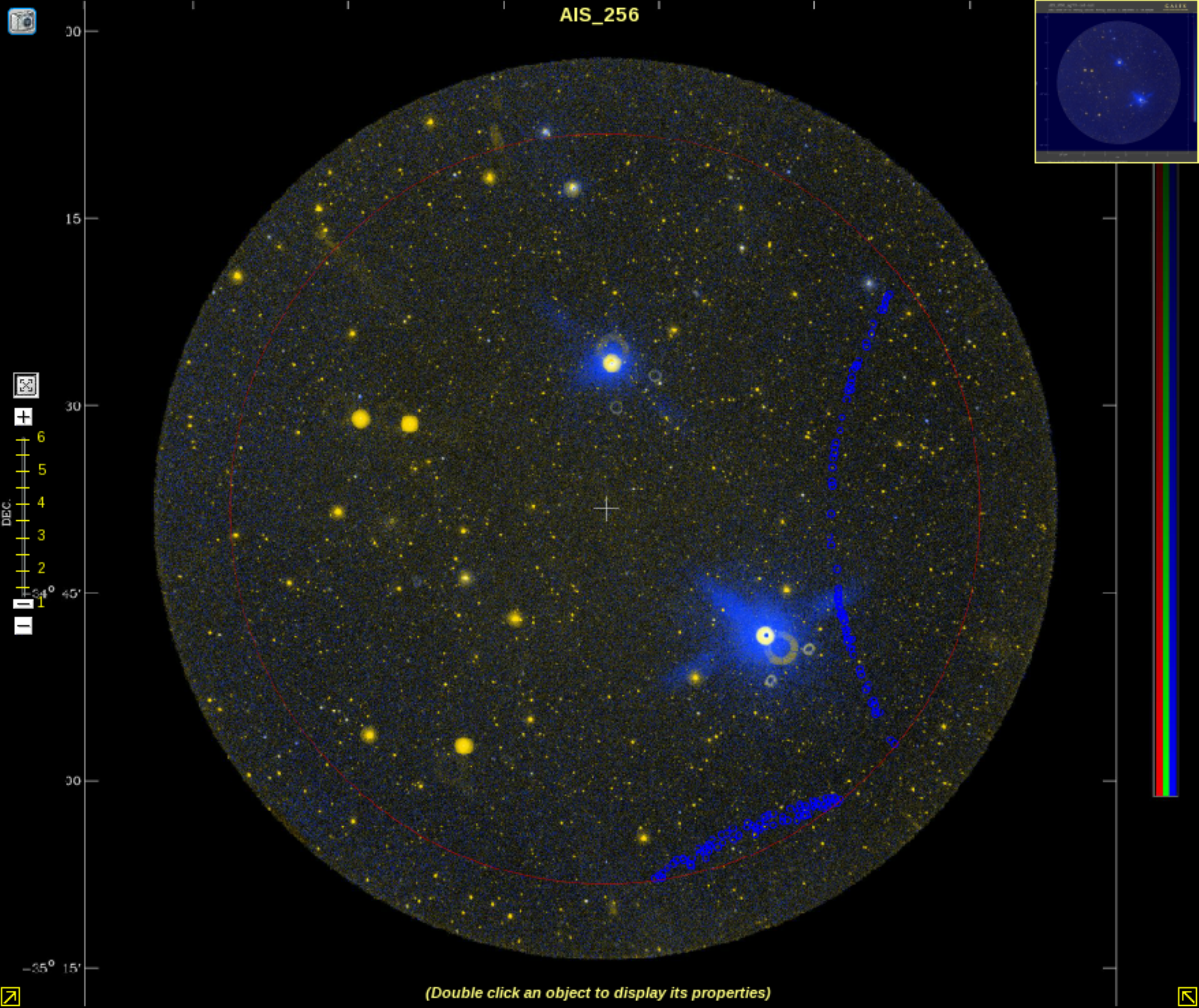}
\includegraphics[width=6.cm]{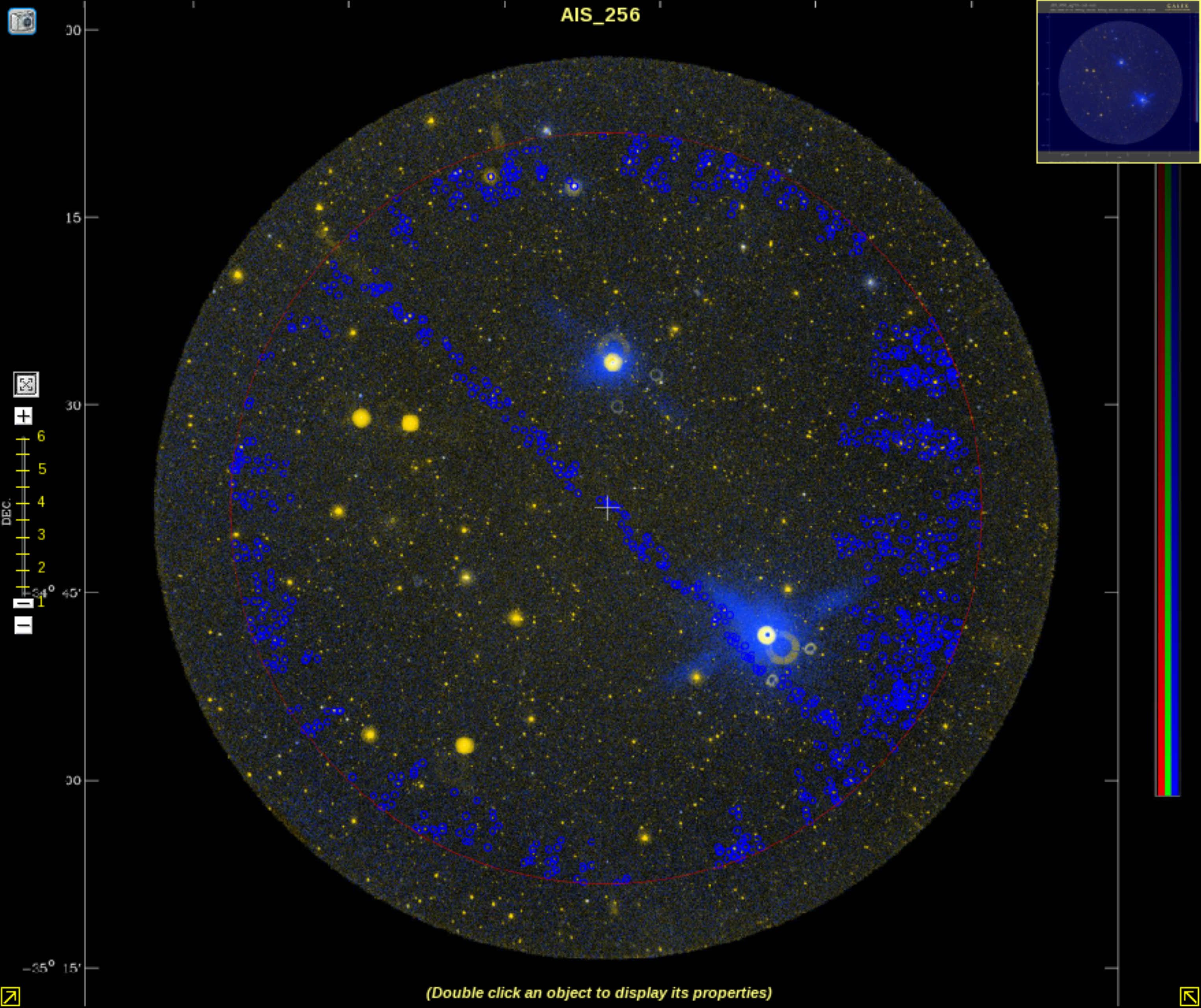}}
\vskip0.2cm
\centerline{
\includegraphics[width=6.cm]{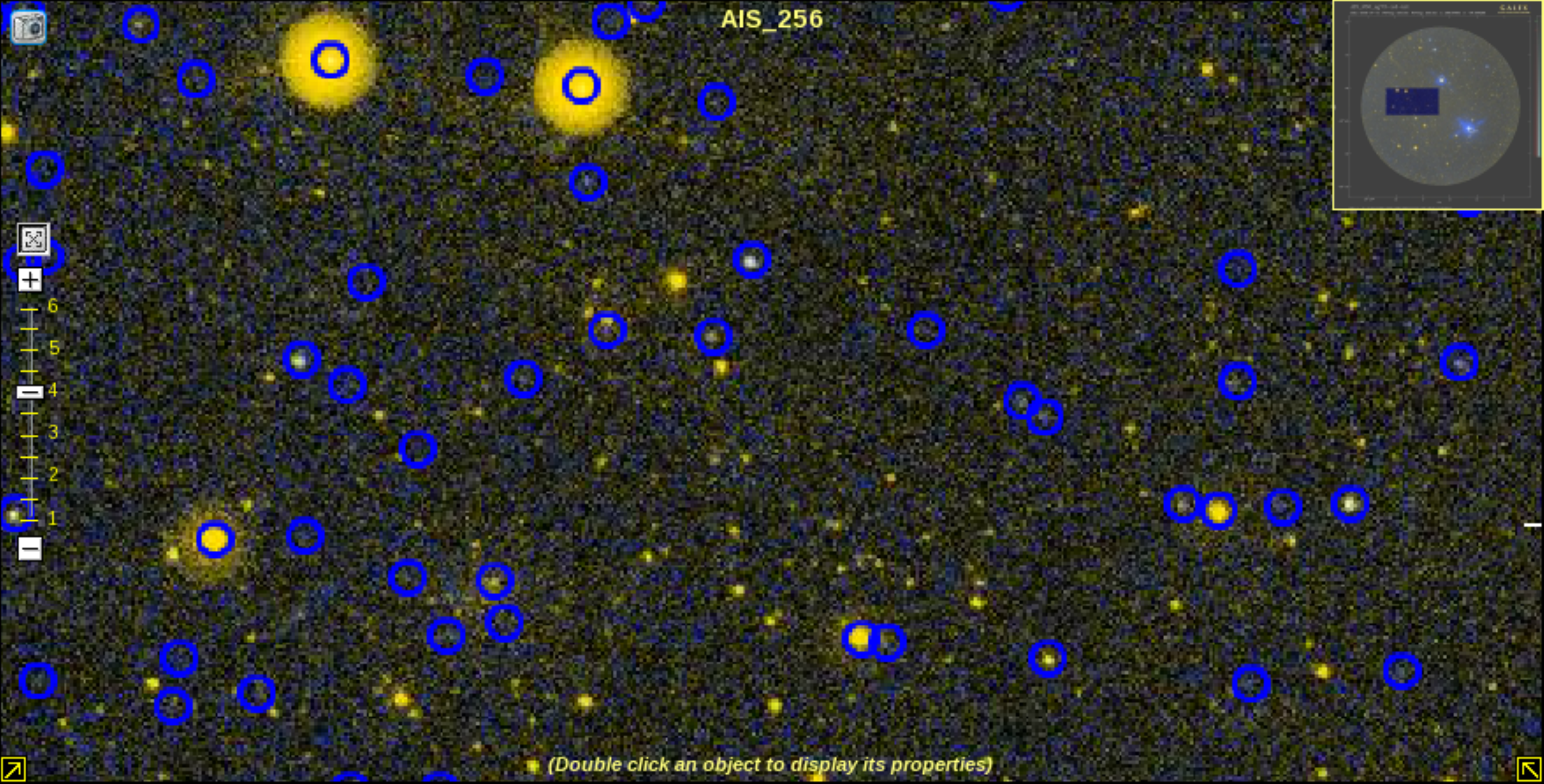}
\includegraphics[width=6.cm]{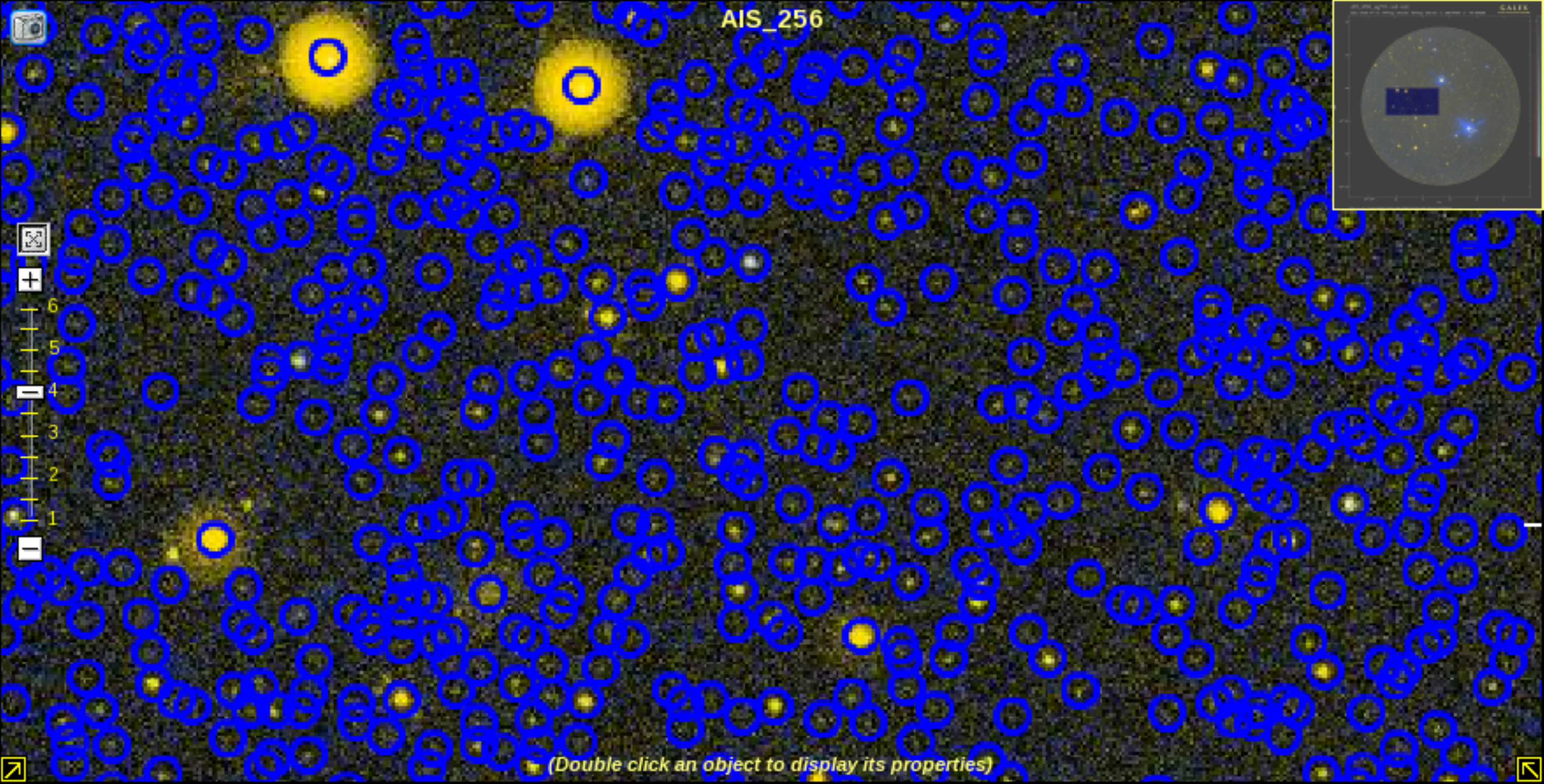}}
\caption{Continued from previous. Top: sources affected by NUV\_artifact=1 (left) and 256 (right: hot spots, taken care of by the pipeline); in the next row, left, sources with rim artifact (FUV\_artifact=32), these are not measured in this image, they come in the database from overlapping visits: they are not included in GUVcat; right: sources with NUV\_artifact=1 or 16. In the bottom row, we show separately sources detected in FUV (left) and NUV (right). As discussed in the text, this is one of five problematic fields, which represents an extreme example.}
\end{figure*}
\renewcommand{\thefigure}{\arabic{figure}}

\end{document}